\documentclass[aps,showpacs,nofootinbib,superscriptaddress,floatfix]{revtex4} 
\usepackage{amsmath}
\usepackage{graphicx,epsfig,wrapfig}
\usepackage{amsthm}
\usepackage{makeidx}
\usepackage{amsfonts}
\usepackage{amssymb}
\usepackage{natbib}
\usepackage{dsfont}
\usepackage{mathrsfs}
\usepackage{slashed}
\usepackage{epstopdf}
\usepackage{bm}
\usepackage{color}
\usepackage[normalem]{ulem} 
\renewcommand\sout{\bgroup \color[rgb]{0.55,0.00,0.99} \ULdepth=-.5ex \ULset}

\begin{document}
\newcommand{\red}[1]{{\color{red} #1}}

\newcommand{\mb}[1]{\mathbf{#1}}
\newcommand{\mc}[1]{\mathcal{#1}}
\newcommand{\non}{\nonumber}
\newcommand{\pep}[1]{\mathbf{#1}_{\perp}}
\newcommand{\pepgr}[1]{\bm{#1}_{\perp}}
\newcommand{\gdir}[1]{\gamma^{#1}}
\newcommand{\xk}{(x,\mathbf{k}_{\perp})}
\newcommand{\xkq}{(x,\mathbf{k}_{\perp}^{2})}
\newcommand{\pe}{(\mb{p}_{e})}
\newcommand{\xks}{(x,\pep{k};S)}
\newcommand{\pv}{\mathrm{P.V.}}
\newcommand{\xkl}{(x,\pep{k};\Lambda)}

\title{
Electron in three-dimensional momentum space
}

\author{Alessandro Bacchetta}
\email{alessandro.bacchetta@unipv.it}

\author{Luca Mantovani}
\email{luca.mantovani@pv.infn.it}

\author{Barbara Pasquini}
\email{barbara.pasquini@unipv.it}

\affiliation{Dipartimento di Fisica, Universit\`a degli Studi di Pavia, I-27100 Pavia, Italy}
\affiliation{Istituto Nazionale di Fisica Nucleare, Sezione di
  Pavia,  I-27100 Pavia, Italy}

\date{\today}


\allowdisplaybreaks[2]

\begin{abstract}
We study the electron as a system composed of an electron and a photon, using
lowest-order perturbation theory. We derive the leading-twist transverse-momentum-dependent distribution
functions for both the electron and photon in the dressed electron, thereby
offering a three-dimensional description of the dressed electron in momentum
space. 
To obtain the distribution functions, we apply
both the formalism of the light-front wave function overlap representation and
the diagrammatic approach. 
We perform the calculations both in light-cone gauge and Feynman gauge, 
and we present a detailed discussion of
the role of the Wilson lines to obtain gauge-independent results. We provide
numerical results and plots for many of the computed distributions. 
\end{abstract}

\maketitle

\section{Introduction}

The electron is an elementary field of quantum electrodynamics (QED) and, as
such, it is a structureless
object. 
When we  study the electron ``structure'', what we are in fact doing is
probing the quantum fluctuations of the theory. 
Due to the Heisenberg uncertainty principle, the electron can fluctuate into a
virtual electron-photon pair, carrying the same quantum number as the
electron,  
i.e. $e\rightarrow e\gamma\rightarrow e$. 
The virtual photons can themselves break up into pairs of virtual electrons
and positrons and, as a result, an isolated electron is in reality surrounded
by a composite virtual cloud of photons, electrons, and positrons in all
possible combinations, consistently with the quantum number of the
electron. Therefore, the ``bare'' electron effectively becomes a
``dressed'' electron. Photons, electrons, and positrons can be interpreted 
in this case as partons contained in the original electron, in analogy to
the partonic structure of hadrons.
  
If one of the virtual states of the electron cloud interacts
with a probe, the parton content of the electron is resolved and the
electron reveals its structure. 
Such behavior is an important aspect of quantum field theories, and similar phenomena arise in many different situations. 
For example, in QCD, the quark splitting to quarks and gluons drives the scaling violations in hadronic structure functions~\cite{Altarelli:1977zs}.
Analogously, the QED radiative corrections to the electron can be recast in the formalism of structure functions (see, e.g., Refs.~\cite{Kuraev:1985hb,Montagna:1991ku}).
Other possible descriptions of the parton structure of the electron can be
obtained in analogy to the QCD formalism of parton correlation functions,
which can be parametrized in terms of different types of distribution
functions (see, e.g., Refs.~\cite{Meissner:2009ww,Lorce:2013pza,Lorce:2011dv}) known as generalized transverse-momentum-dependent parton distributions (GTMDs), generalized parton distributions (GPDs),
transverse-momentum-dependent parton distributions (TMDs) and collinear parton
distribution functions (PDFs).  

These distribution functions allow us to view the dressed electron from a new
perspective. For instance, it is possible to introduce the notion of the shape
of the electron~\cite{Hoyer:2009sg,Miller:2014vla}, to analyze how
the spin of the electron is split into the spins and orbital
angular momenta (OAM) of the partons in the electron's
cloud~\cite{Brodsky:2000ii,Burkardt:2008ua,Liu:2014fxa}, and to compute all
possible correlations between spin, momentum and position of the
partons. Moreover, the study of QED parton distribution functions can help
to shed light on several formal aspects of their QCD analogues.

In this work, we will focus on the study of TMDs,
which describe the distribution of QED partons in the dressed electron 
as a function of their
longitudinal and transverse momentum, and therefore give access to the
three-dimensional picture of the 
dressed electron in momentum space.
At leading twist there are eight TMDs, for both the active electron and photon, 
characterizing the strength of
different spin-spin and spin-orbit correlations of the partons and electron
target. 
Three of them survive when integrated over the transverse momentum, giving
rise to 
the familiar collinear PDFs.
Formally, the transverse-momentum dependence in the distribution functions arises from the definition in terms of bilocal operator matrix elements off the light cone.
To comply with gauge invariance, one has to introduce nontrivial gauge links,
which compensate for the nonlocality of the
fields~\cite{Ji:2002aa,Belitsky:2002sm,Boer:2003cm,Ji:2004wu,Cherednikov:2014mua}. 
The Wilson line accounts for the (infinitely many, in principle) photons
that can be absorbed or emitted by the initial or final state, and it is the
source of interesting phenomena like single-spin
asymmetries~\cite{Brodsky:2002cx,Collins:2002kn}.

In a diagrammatic approach, the gauge-link contribution can be studied via a
perturbative expansion in the coupling constant.  
Thanks to its Abelian nature, QED is simpler to deal with compared to QCD, and
can therefore serve as a useful testing ground for full perturbative
calculations. 
In this article, we will work in the leading-order approximation, corresponding to the exchange of a single gauge boson
 between the active parton and the spectators.  
  In particular, we will use QED Feynman rules for the eikonal propagator, and we will explicitly show the equivalence of the resulting TMDs 
in different gauges, namely Feynman and light-cone gauge. 

Besides the diagrammatic approach, a useful framework to study parton
distributions is the representation in terms of overlap of light-front wave
functions (LFWFs)~\cite{Pasquini:2008ax}. 
The LFWFs of the electron corresponding to the minimal Fock state component of an electron-photon pair
are computable explicitly from time-ordered perturbation theory in the light-cone gauge.
 Several applications using the electron LFWFs  have been discussed, focusing in particular on the study of the distributions in impact-parameter space as obtained from the 
 GPDs~\cite{Hoyer:2009sg,Miller:2014vla,Brodsky:2000xy,Brodsky:2002cx,Chakrabarti:2004ci,Chakrabarti:2005zm,Brodsky:2006ku,Chakrabarti:2008mw,Kumar:2015tpa}.
 Here we will extend the study with LFWFs to the TMDs, of both the electron
 and photon partons.
 
 The diagrammatic approach  and the LFWF overlap representation are used as
 alternative, but equivalent, descriptions to emphasize different aspects  
 of the physical content of the TMDs. The diagrammatic approach allows us to
 unravel more explicitly the role of gauge invariance in the
 definition of the TMDs, while the representation in terms of LFWFs, being
 eigenstates of the parton light-front helicity and OAM  in the light-cone
 gauge, exposes more explicitly the information about the spin-spin and
 spin-orbit correlations  encoded in the TMDs.  
 
The outline of this work is as follows: In Sec.~\ref{definitions} we set up
the general framework for the TMDs of the electron, introducing the definition
of the correlator function and discussing the role of the Wilson line in
QED. In Sec.~\ref{tmdfg} we derive the leading-twist TMDs using the
diagrammatic approach in the Feynman gauge, at leading order in $\alpha$. In
Sec.~\ref{tmdslcg} we sketch the derivation of the LFWFs for the
electron-photon Fock state component and discuss the LFWF overlap
representation of the TMDs, showing the equivalence of the results within the
diagrammatic approach. In Sec.~\ref{tgl} we study in some detail the role of
the transverse components of the Wilson line in the light-cone gauge, using
different prescriptions for the regularization of the light-cone singularity
from the photon propagator. In Sec.~\ref{secphot} we turn our attention to the
photon contribution to the dressed-electron TMDs. 
In Sec.~\ref{results}, we present
the results for the electron and photon TMDs, discussing the role of the spin
and parton OAM in shaping the transverse-momentum dependence of the different
TMDs. Finally, we conclude with a section summarizing our findings.

 \section{General Framework}\label{definitions}

\begin{figure}[t]
\begin{center}
\includegraphics[height=3cm]{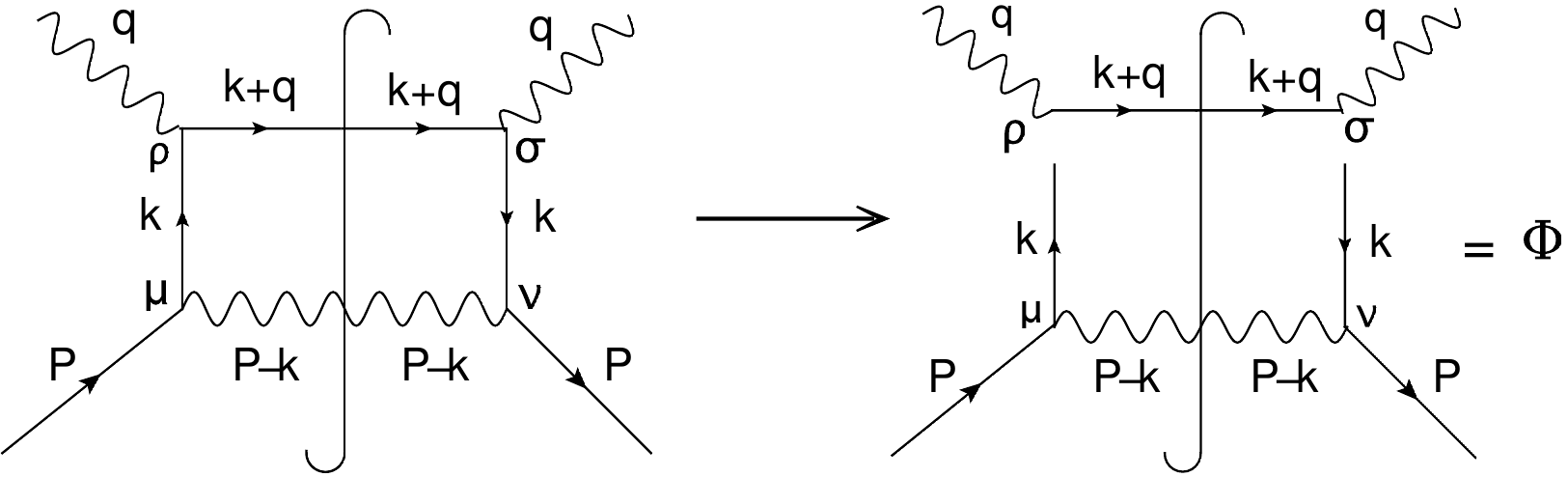}
\end{center}
\caption{\footnotesize{Handbag diagram for the dressed electron in QED at
    order $\alpha$ (left),
    separated in its upper and lower part (right), 
    the latter being the diagrammatic
    representation of the electron-electron correlator $\Phi$.}}
\label{handbag}
\end{figure}

Let us consider a dressed electron of mass $m$, spin $S$ and four-momentum $P$,
composed of a bare electron of momentum $k$ and a photon of momentum $P-k$. If
we write a generic vector $a^{\mu}$ in light-cone coordinates as
$a^{\mu}=(a^{+},a^{-},\pep{a})$, we set the momentum of the inner electron as
$k\equiv(xP^{+},k^{-},\pep{k})$, with the longitudinal momentum fraction $x$
defined as $x:\,=k^{+}/P^{+}$. We begin with the definitions related to the
distribution functions of the internal electron; those for the photon will be
discussed in Sec. \ref{secphot}. In analogy to the QCD case, if we probe the
internal structure of the electron via the scattering of a virtual photon (see
Fig.~\ref{handbag}), the most general expression for the correlator function
between initial and final electron states is given
by~\cite{Soper:1977jc,Collins:1981uw,Mulders:1995dh,Meissner:2009ww}  
\begin{equation} \Phi_{ij}(k;P,S):\,=\int\frac{d^{4}\xi}{(2\pi)^{4}}\,e^{ik\cdot\xi}\;\langle P,S\lvert \bar{\psi}_{j}(0)\psi_{i}(\xi)\rvert P,S\rangle \; . \label{corr}\end{equation}
In Eq.~\eqref{corr} $\psi(x)$ is the electron field, while the indices $i$ and $j$ refer to Dirac space and will be omitted from now on. 
If we integrate \eqref{corr}  over $k^{-},$ we obtain the transverse-momentum-dependent correlator, which enters the definition of the TMDs:
\begin{equation} 
\Phi(x,\mb{k}_{\perp};P,S)=\int
  dk^{-}\,\Phi(k;P,S)
=\int\frac{d\xi^{-}d^{2}\bm{\xi}_{\perp}}{(2\pi)^{3}}e^{ik\cdot\xi}\langle
  P,S\lvert \bar{\psi}(0)\psi(\xi)\rvert P,S\rangle \bigg \rvert_{\xi^{+}=0}\;
  . 
\label{corr2}
\end{equation}  
If we further integrate  \eqref{corr2} over $\pep{k}$, we reduce ourselves to a correlator function that depends only on the longitudinal momentum fraction $x$, giving 
access to the collinear PDFs:
\begin{equation} \Phi(x;P,S)=\int dk^{-}d^{2}\mb{k}_{\perp} \,\Phi(k;P,S)=\int\frac{d\xi^{-}}{2\pi}e^{ik\cdot\xi}\langle P,S\lvert \bar{\psi}(0)\psi(\xi)\rvert P,S\rangle \bigg \rvert_{{\xi^{+}=\bm{\xi}_{\perp}=0}} \;. \label{corr3}\end{equation}
 From a diagrammatic point of view, the correlator corresponds to the lower
 part of the handbag diagram in Fig.~\ref{handbag}.

In order for 
the bilocal products of field operators in the correlator functions to be
gauge invariant, 
we have to insert a Wilson line between the two electron 
fields~\cite{Collins:1981uw,Ji:2002aa,Belitsky:2002sm,
Bomhof:2004aw,Bomhof:2006dp}.
Indeed, the QED Lagrangian\footnote{Here, and in the following, we take $e>0$, meaning that the electron charge is equal to $-e$.}
\begin{equation}\mathscr{L}_{\rm QED}=\bar{\psi}(i\slashed{\partial}-m)\psi-\frac{1}{4}F^{\mu\nu}F_{\mu\nu}+e\bar{\psi}\gamma^{\mu}\psi
  A_{\mu}, \end{equation}
where the photon field tensor is 
$F^{\mu\nu}(\xi)=\partial^{\mu}A^{\nu}(\xi)-\partial^{\nu}A^{\mu}(\xi)$, 
is invariant under the following local gauge transformation of the fields:
\begin{equation} \psi(\xi)\rightarrow e^{ie\alpha(\xi)}\psi(\xi) \, , \quad A^{\mu}(\xi)\rightarrow A^{\mu}(\xi)+\partial^{\mu}\alpha(\xi) \; . \label{gtran} \end{equation} 
However, this transformation does not leave the correlator $\Phi$ invariant, since 
\begin{equation}\bar{\psi}(0)\psi(\xi)\rightarrow \bar{\psi}(0)\psi(\xi)\,e^{-ie[\alpha(0)-\alpha(\xi)]}\; .\label{gltrans} 
\end{equation} The standard way to ensure gauge invariance of the correlator
$\Phi$ is to insert in our definition a gauge-link operator $U_{(0,\xi)}$
connecting the fields evaluated at different points. Let us define, in general \cite{Cherednikov:2014mua},
\begin{equation}
  \mc{U}_{(\xi_{1},\xi_{2})}:\,=\exp\left[-ie\int_{\xi_{1}}^{\xi_{2}}d\eta^{\mu}A_{\mu}(\eta)\right]
  \; , 
\end{equation}
where the integral is meant to run along any path connecting $\xi_{1}$ to $\xi_{2}$. Acting with the gauge transformation \eqref{gtran}, the Wilson line will turn into
\begin{equation} 
\mc{U}_{(\xi_{1},\xi_{2})}\rightarrow
  \exp\left[-ie\int_{\xi_{1}}^{\xi_{2}}d\eta^{\mu}A_{\mu}(\eta)\right]\,\exp\left[-ie\int_{\xi_{1}}^{\xi_{2}}d\eta^{\mu}\partial_{\mu}\alpha(\eta)\right]=e^{ie[\alpha(\xi_{1})-\alpha(\xi_{2})]}\,
  \mc{U}_{(\xi_{1},\xi_{2})} \; , \label{wltran}
\end{equation} 
which is exactly the transformation that we need to render $\Phi$ gauge invariant.

\begin{figure}[b]
\begin{center}
\includegraphics[width=6cm]{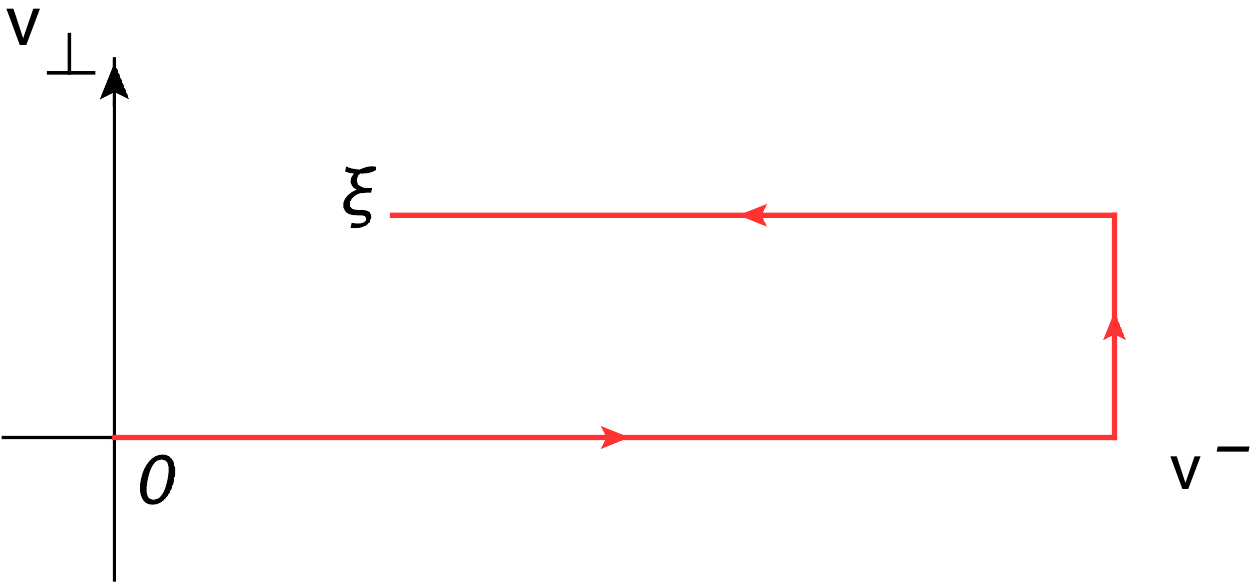}
\end{center}
\caption{\footnotesize{Wilson lines connecting two points at $0=(0^{-},\pep{0})$ and $\xi=(\xi^{-},\bm{\xi}_{\perp})$.}}
\label{path}
\end{figure}

Since different processes require different paths along which the Wilson line
must be evaluated \cite{Collins:2002kn,Boer:2003cm,Bomhof:2004aw,Bomhof:2006dp}, we are now
going to introduce the specific path that enters the definition of the
transverse-momentum-dependent correlator~\cite{Boer:2003cm}.
Consider a generic vector $v^{\mu}=(v^{+},v^{-},\mb{v}_{\perp})$ in light-cone
coordinates; for our purposes we can keep the plus component $v^{+}$ fixed at
$0$, since the fields in \eqref{corr2}  are taken along the light cone at
$\xi^{+}=0$.  
We move in the $v^{-}$-$\mb{v}_{\perp}$ plane and define Wilson lines
along the $v^{-}$ and $\mb{v}_{\perp}$ directions as follows: in order to
connect the points $(0,a^{-},\mb{c}_{\perp})$ and $(0,b^{-},\mb{c}_{\perp})$
along the minus direction, we introduce the longitudinal gauge link: 
\begin{equation}
  \mc{U}^{v^{-}}(a^{-},b^{-};\mb{c}_{\perp}):\,=\mathcal{P}\exp\left[-ie\int_{a^{-}}^{b^{-}}d\eta^{-}\,A^{+}(0,\eta^{-},\mb{c}_{\perp})\right]\;
  , \label{gll}\end{equation} 
while in order to connect the points $(0,c^{-},\mb{a}_{\perp})$ and $(0,c^{-},\mb{b}_{\perp})$, along the transverse direction, we use the transverse gauge link:
\begin{equation}
  \mc{U}^{\perp}(\mb{a}_{\perp},\mb{b}_{\perp};c^{-}):\,=\mathcal{P}\exp\left[ie\int_{\mb{a}_{\perp}}^{\mb{b}_{\perp}}d\pepgr{\eta}\cdot\pep{A}(0,c^{-},\pepgr{\eta})\right]\;
  . \label{glt}\end{equation} 
In order to link the points $0=(0^{-},\mb{0}_{\perp})$ and
$\xi=(\xi^{-},\bm{\xi}_{\perp})$ appearing in our correlator, we follow a path
along the $v^{-}$ direction all the way to infinity, then get to
$\bm{\xi}_{\perp}$ along the transverse direction and finally go back to
$\xi^{-}$, as shown in Fig.~\ref{path}. To this purpose, we define
\begin{align} \mc{U}_{(0,\infty)}:\,=\mc{U}^{v^{-}}(0^{-},\infty^{-};\pep{0})\,\mc{U}^{\perp}(\mb{0}_{\perp},\pep{\infty};\infty^{-}) \; , \label{u1} \\
\mc{U}_{(\infty,\xi)}:\,=\mc{U}^{\perp}(\pep{\infty},\bm{\xi}_{\perp};\infty^{-})\,\mc{U}^{v^{-}}(\infty^{-},\xi^{-};\bm{\xi}_{\perp})
\; . \label{u2} \end{align} 
We therefore define the gauge-invariant transverse-momentum-dependent
correlator as: 
\begin{equation}
\Phi(x,\mb{k}_{\perp};P,S) :\,=
\int\frac{d\xi^{-}d^{2}\bm{\xi}_{\perp}}{(2\pi)^{3}}e^{ik\cdot\xi}\langle
  P,S\lvert \bar{\psi}(0)\mc{U}_{(0,\infty)}\mc{U}_{(\infty,\xi)}\psi(\xi)\rvert P,S\rangle \bigg \rvert_{{\xi^{+}=0}}\;. 
\label{gicorr} \end{equation} 
It is clear that if we choose the light-cone gauge $A^{+}=0$, the longitudinal
gauge link $\mc{U}^{v^{-}}$ reduces to the identity; in this gauge, only the
transverse gauge link survives, and this  is crucial to maintain the gauge
independence under residual gauge transformations, which can be fixed by
imposing appropriate boundary conditions~\cite{Belitsky:2002sm}. 
In Feynman gauge, instead, the longitudinal gauge link is nontrivial;
however, the transverse gauge link does not give any contribution, because the
boundary condition for the transverse component of the photon field is
$\pep{A}(\infty)=0$. In the following, we will discuss the differences between
the two choices of the gauge in more detail.

\begin{figure}[t]
\centering
\epsfig{file=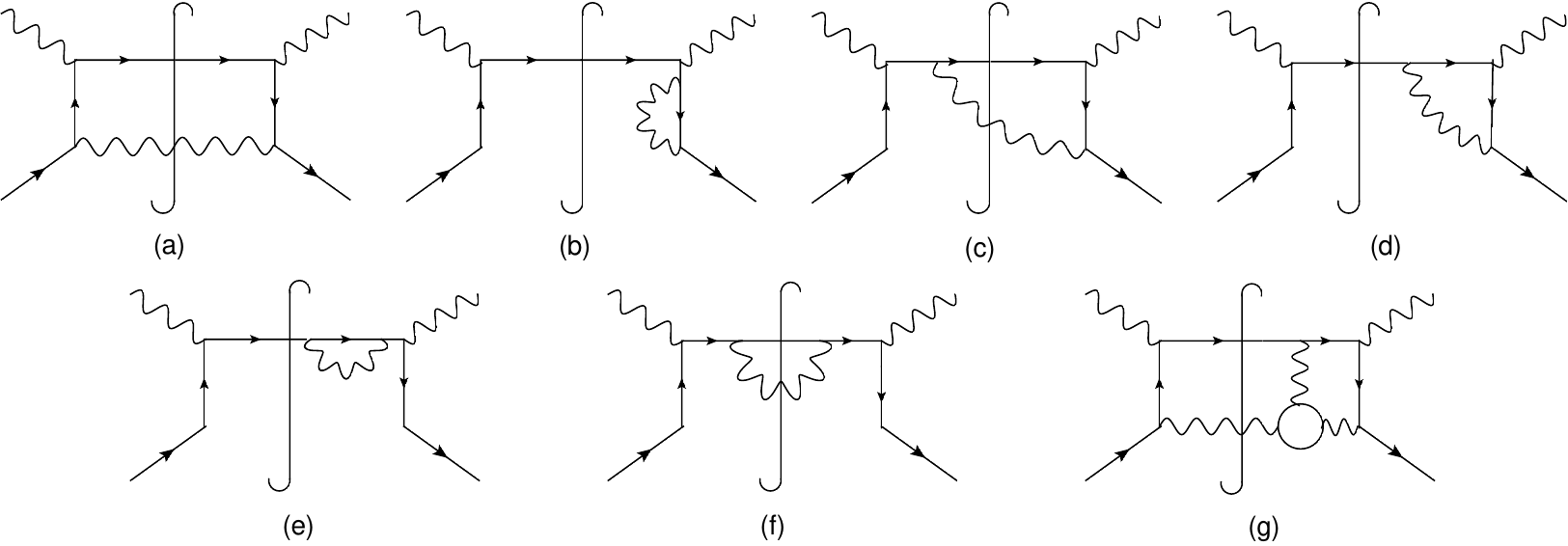, height=5cm}
\caption{\footnotesize{Contributions to the correlator at different orders in
    $\alpha$ and in the gauge link. Notice that also the Hermitean conjugates
    of all diagrams should be taken into account.}} 
\label{gaugephotons}
\end{figure}

The insertion of a gauge link  brings about some changes to the correlator
from a diagrammatic point of view~\cite{Boer:2003cm,Ji:2002aa}.
Wilson lines physically correspond to final- and initial-state
interactions, i.e. they account for the (infinitely many, in principle)
photons that can be absorbed or emitted by the incoming or outgoing
particles. Some examples of this kind of interaction are shown in
Fig.~\ref{gaugephotons}, where we gather all the $\alpha$-order contributions
to the correlator (diagrams (a) to (f)), along with a case of a higher-order
contribution (diagram (g)). Diagrams (a) and (b) are actually of zeroth order
in the gauge link, as they refer, respectively, to the handbag diagram we
already discussed and to the correction due to the self-energy of the bare
electron; diagrams (c) and (d) contain one gauge photon and they are hence
described using the first-order approximation of the Wilson line in the
coupling constant $e$, while (e) and (f) are of second order in the expansion
of the gauge link. We also notice that, in diagram (d), the photon loop is
responsible for the virtual vertex correction, which contributes  only at
the end point $x=1$, $\pep{k}=\pep{0}$; this is true also for the self-energy
in diagram (b). Diagrams (e) and (f) are instead proportional to $n_-^{2}$ [with $n_{-}^{\mu}=(0,1,\bm{0}_\perp)$] and can be neglected, unless the direction of the gauge link 
is {\em not} taken along the
light cone.

We remark here that a proper definition of TMDs must deal also with the
end points and the occurrence of infrared and rapidity
divergences, which can be regularized, e.g., by taking $n_-$ to be off the light
cone. These are critical issues in proving factorization theorems and
introducing well-defined TMDs, both for QCD and for QED. Careful investigations of
these issues can be found in several papers and books (see, e.g.,
\cite{Collins:1981uk,Ji:2004wu,Aybat:2011zv,GarciaEchevarria:2011rb,Collins:2011zzd,Echevarria:2012pw,Echevarria:2012js,Collins:2012uy}).

The correlator $\Phi(k;P,S)$ in \eqref{corr} can be written as a combination
of elements of a basis in Dirac space
$\mc{D}:\,=\left\{\mathds{1},\gamma^{\mu},\gamma^{5},\gamma^{5}\gamma^{\mu},i\sigma^{\mu\nu}\right\}$. Let
us consider from now on the case of the transverse-momentum-dependent
correlator $\Phi(x,\pep{k};P,S)$ defined in
\eqref{corr2}~\cite{Mulders:1995dh,Goeke:2005hb,Bacchetta:2006tn}.  
We can choose a reference frame where the incoming dressed electron has no
transverse momentum, i.e. $P=(P^{+},P^{-},\pep{0})$, with $P^{-}=m^{2}/2P^{+}$
for the on-shell condition. Moreover, since it must be $P\cdot S=0$,
$S^{2}=-1$, we can set $S\equiv(S_{z}P^{+}/m,-S_{z}m/2P^{+},\pep{S})$ and
therefore replace the dependence on the covariant four-vector $S$ with the
dependence on the three-vector $\mb{S}:\,=(\pep{S},S_z)$, with
$\mb{S}^{2}=1$. 

 Introducing also the light-like vector $n_{+}=(1,0,\mb{0}_\perp)$ and the
 antisymmetric tensor,
\begin{equation} \epsilon_{\perp}^{12}=-\epsilon_{\perp}^{21}=1 \; ,\quad \epsilon_{\perp}^{11}=\epsilon_{\perp}^{22}=0 \; ,  \label{antitens}\end{equation} one finds, in the leading-twist approximation,
\begin{align} 
\Phi\approx &\frac{1}{2}\bigg[ 
f_{1}^{e}\slashed{n}_{+}
-\frac{\epsilon_{\perp}^{ij}k_{\perp}^{i}S_{\perp}^{j}}{m}f_{1T}^{\perp e}\slashed{n}_{+}
+S_{z}g_{1L}^{e}\gamma^{5}\slashed{n}_{+}
+\frac{\pep{k}\cdot\pep{S}}{m}g_{1T}^{e}\gdir{5}\slashed{n}_{+}
+\frac{[\slashed{S}_\perp,\slashed{n}_{+}]}{2}\gdir{5}h_{1T}^{e}\non \\
&+S_{z}\frac{[\slashed{k}_\perp,\slashed{n}_{+}]}{2m}\gdir{5}h_{1L}^{\perp e}
+\frac{\pep{k}\cdot\pep{S}[\slashed{k}_\perp,\slashed{n}_{+}]}{2m^{2}}\gdir{5}h_{1T}^{\perp\,e}
+\frac{i[\slashed{k}_\perp,\slashed{n}_{+}]}{2m}h_{1}^{\perp e}\bigg] \, ,\label{phipar}\end{align}
where $i,j=1,2$ and $a_{\perp}^{\mu}=(0,0,\pep{a})$. In the above equation we
introduced the leading-twist TMDs of an electron inside the dressed electron, $f_{1}^{e}$, $f_{1T}^{\perp e}$, $g_{1L}^{e}$, $g_{1T}^{e}$, $h_{1T}^{e}$, $h_{1L}^{\perp e}$,
$h_{1T}^{\perp e}$, $h_{1}^{\perp e}$, that depend on $\xkq$. As far as the
notation is concerned, the letters $f$, $g$ and $h$ refer to unpolarized,
longitudinally polarized and transversely polarized internal electrons; 
$L$ ($T$) refers
to the spin of the dressed electron being along the longitudinal (transverse)
direction; the ``$\perp$'' symbols signal an explicit dependence on
transverse momenta with an uncontracted index. Out of the eight TMDs in
Eq.~\eqref{phipar},  
 the so-called Boer-Mulders function $h_{1}^{\perp e}$  \cite{PhysRevD.57.5780}
 and the Sivers function $f_{1T}^{\perp e}$ \cite{PhysRevD.41.83} are T-odd,
 i.e. they change sign under naive time reversal, which is defined as usual
 time reversal but without interchange of initial and final 
states, while the other six are T-even.

Let us take the trace of the correlator multiplied by an appropriate Dirac
operator and define 
\begin{equation} \Phi^{[\Gamma]}(x,\pep{k};\mb{S}):\,=\frac{1}{2}\mathrm{Tr}\left[\Phi(x,\pep{k};\mb{S})\Gamma\right] \; ,\end{equation}
with $\Gamma\in\mc{D}$. It turns out that, at the leading twist in $1/Q$, these Dirac space projections are given by \cite{Mulders:1995dh,Bacchetta:2006tn}
\begin{align} &\Phi^{[\gdir{+}]}(x,\pep{k};\mb{S})=f_{1}^{e}-\frac{\epsilon_{\perp}^{ij}k_{\perp}^{i}S_{\perp}^{j}}{m}f_{1T}^{\perp e} \; , \label{tmd1}\\
&\Phi^{[\gdir{+}\gdir{5}]}(x,\pep{k};\mb{S})=S_{z}g_{1L}^{e}+\frac{\pep{k}\cdot\pep{S}}{m}g_{1T}^{e} \; , \label{tmd2} \\
&\Phi^{[i\sigma^{j+}\gdir{5}]}(x,\pep{k};\mb{S})=S_{\perp}^{j}h_{1}^{e}+S_{z}\frac{k^{j}_{\perp}}{m}h_{1L}^{\perp e}+S^{i}_{\perp}\frac{2k_{\perp}^{i}k_{\perp}^{j}-\pep{k}^{2}\delta^{ij}}{2m^{2}}h_{1T}^{\perp e} +\frac{\epsilon_{\perp}^{ji}k_{\perp}^{i}}{m}h_{1}^{\perp e}\; , \label{tmd3}\end{align}
where 
\begin{equation} h_{1}^{e}:\,=h_{1T}^{e}+\frac{\pep{k}^{2}}{2m^{2}}h_{1T}^{\perp e} \;. 
\end{equation} The different TMDs in Eqs.~\eqref{tmd1}-\eqref{tmd3} can be
isolated by taking proper spin configurations of the dressed electron, while
the Dirac structure $\Gamma$ in the field correlator projects on different
polarization states of the internal bare electron. 

\section{Calculation of TMDs in Feynman gauge} \label{tmdfg}

In this section, we derive explicit analytic expressions for
the TMDs.  We stress
that, from now on, we will exclude the end point $x=1\, ,\pep{k}=\pep{0}$ from
our analysis, since it would require a proper regularization procedure
which is beyond the scope of the present work. 
This implies that we will neglect in our analysis all self-energy and virtual
vertex corrections discussed in section \ref{definitions} (diagrams (b) and
(d) in Fig.~\ref{gaugephotons}). 

An important statement can be made straightforwardly about the T-odd TMDs
$f_{1T}^{\perp e}$ and $h_{1}^{\perp e}$. Had we tried to derive them starting
from the correlator without the gauge link, i.e. the one in Eq.~\eqref{corr2},
we would have found that they are vanishing. This, eventually, can be
explained by the fact that only diagrams containing a loop at one side
of the cut (such as diagrams (d) to (g) in Fig.~\ref{gaugephotons}) can
potentially  give a
non zero contributions to T-odd TMDs \cite{Boer:2003cm}. The only possibility
to have such a loop is either to consider diagrams of higher perturbative
order, or consider the end point $x=1\, ,\pep{k}=\pep{0}$. Therefore,
already at this stage we are allowed to conclude that \textsl{at order
  $\alpha$} and \textsl{for} $x\neq 1$, $\pep{k}\neq\pep{0}$, we have: 
\begin{align} h_{1}^{\perp e}(x,\pep{k}^{2})=0 \; ,\\
f_{1T}^{\perp e}(x,\pep{k}^{2})=0 \; . \end{align}

\begin{figure}[t]
\centering
\includegraphics[height=3cm]{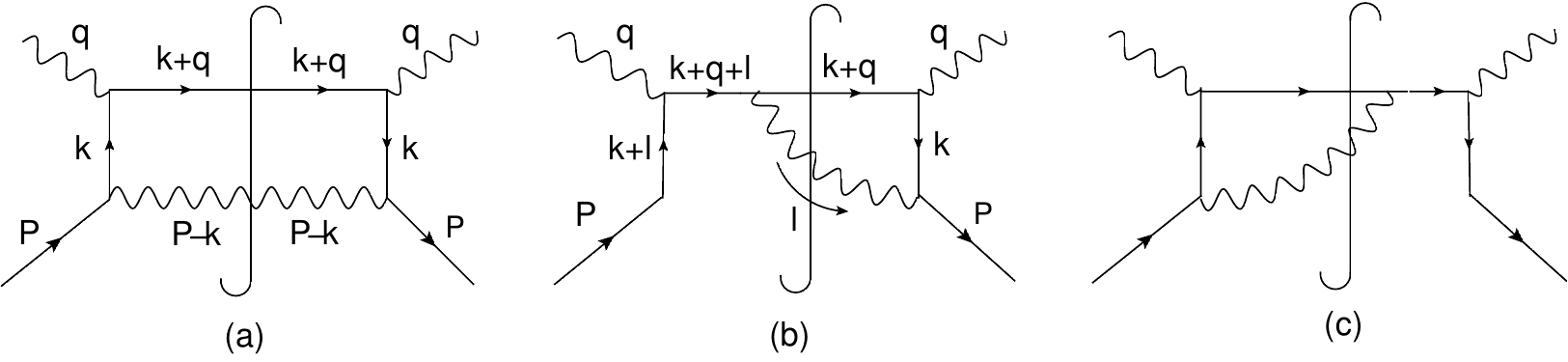}
\caption{\footnotesize{Diagrams contributing to the TMDs in Feynman gauge, for $x\neq 1$ and $\pep{k}\neq\pep{0}$.}
}
\label{feyngdiag}
\end{figure} 

As we mentioned, from a diagrammatic point of view, the correlator
\eqref{corr} corresponds to the diagram shown in the right-hand side of
Fig.~\ref{handbag}; it is possible, then, to evaluate it with usual Feynman
rules and take its trace with a proper matrix $\Gamma$, in order to recover
the analytic expressions of the TMDs. 
Introducing for greater clarity the Dirac indices, the Feynman rules give for diagram (a) in Fig.~\ref{feyngdiag}
\begin{equation} \lvert\mc{M}\rvert^{2}=
  \frac{1}{(2\pi)^{3}}\delta\left((P-k)^{2}\right)\delta\left((k+q)^{2}-m^{2}\right)
  M^{\mathrm{low}}_{ij}M^{\mathrm{up}}_{ji} \;
  , \label{amplitude} 
\end{equation} 
where we factor out the transition matrix corresponding to the lower part of the diagram,
\begin{equation} 
\mc{M}^{\mathrm{low}}_{ij}:\,=e^{2}d^{\mu\nu}(P-k)\bar{u}_{\mb{S},k}(\mb{P})\left(\gamma_{\nu}\frac{\slashed{k}+m}{k^{2}-m^{2}-i\epsilon}\right)_{kj}\left(\frac{\slashed{k}+m}{k^{2}-m^{2}+i\epsilon}\gamma_{\mu}\right)_{il}u_{\mb{S},l}(\mb{P}) \; ,\label{soft}
\end{equation} 
from the one corresponding to the upper part, where the interaction between
the parton electron and the virtual photon takes place,
\[ \mc{M}^{\mathrm{up}}_{ij}:\,=e^{2}\Big(\slashed{\epsilon}^{*}_{\lambda}(\mb{q})(\slashed{k}+\slashed{q}+m)\slashed{\epsilon}_{\lambda}(\mb{q})\Big)_{ij} \; . \]
We can henceforth write for the correlator, setting the Dirac indices
according to \eqref{corr}
\begin{equation} 
\Phi(k;\bm{S})_{ij}=\frac{e^{2}}{(2\pi)^{3}}\delta\left((P-k)^{2}\right)\mc{M}^{\mathrm{low}}_{ij} \; . \label{corlow}
\end{equation}
The sum over photon polarization vectors $d_{\mu\nu}$ takes different forms depending on the gauge. We choose the Feynman gauge first, and hence use
\begin{equation} d^{\mu\nu}(p):\,=\sum_{\lambda}\varepsilon_{\lambda}^{\mu\;^{*}}(\bm{p})\varepsilon_{\lambda}^{\nu}(\bm{p})=-g^{\mu\nu}. \label{polsumfeyn}
\end{equation} 

As an example, we would like to show the result for the TMD $f_{1}^{e}\xkq$, which
is the distribution for unpolarized electrons in an unpolarized dressed
electron. From Eq.~\eqref{tmd1}, since it is $f_{1T}^{\perp e}=0$, we see that
$f_{1}^{e}$ can be recovered by taking the trace of $\Phi\gamma^{+}$ (with a
factor $1/2$) and including also an average over the spin state of the target
electron. We rewrite
also: \[\delta\left((P-k)^{2}\right)=\frac{1}{2(P-k)^{+}}\delta\left(k^{-}-P^{-}+\frac{\pep{k}^{2}}{2(P-k)^{+}}\right)
\; ,\] where, from the on-shell condition, 
$P^{-}={m^{2}/(2P^{+})}$ (working in a frame of reference where
  $\pep{P}=\pep{0}$). 
Hence, it is easy to find that the contribution of diagram (a) of Fig.~\ref{feyngdiag} to the TMD $f_{1}^{e}$ is
\begin{align}f_{1}^{a}\xkq&=\sum_{S}\int dk^{-}\frac{1}{8(2\pi)^{3}(P-k)^{+}}\delta\left(k^{-}-P^{-}+\frac{\pep{k}^{2}}{2(P-k)^{+}}\right)\frac{(-e^{2})}{(k^{2}-m^{2})^{2}}\left[\bar{u}_{\bm{S}}(\mb{P})\gamma_{\mu}(\slashed{k}+m)\gamma^{+}(\slashed{k}+m)\gamma^{\mu}u_{\bm{S}}(\mb{P})\right]\non \\
&= \frac{1}{(2\pi)^{3}}\left[\frac{\pep{k}^{2}+m^{2}(1-4x+x^{2})}{x}\right]\varphi^{2}\xkq \; , \label{f1A}\end{align}
 where we drop the imaginary part of the electron propagator since the internal electron  cannot be on-shell, and we define the function $\varphi\xkq$ as
\begin{equation}  \varphi(x,\mathbf{k}_{\perp}^{2}):\,=\frac{e\,\sqrt{x(1-x)}}{[(1-x)^{2}m^{2}+\mathbf{k}_{\perp}^{2}]} \; . \label{varphi}
\end{equation}

\begin{figure}[t]
\centering
\includegraphics[width=8cm]{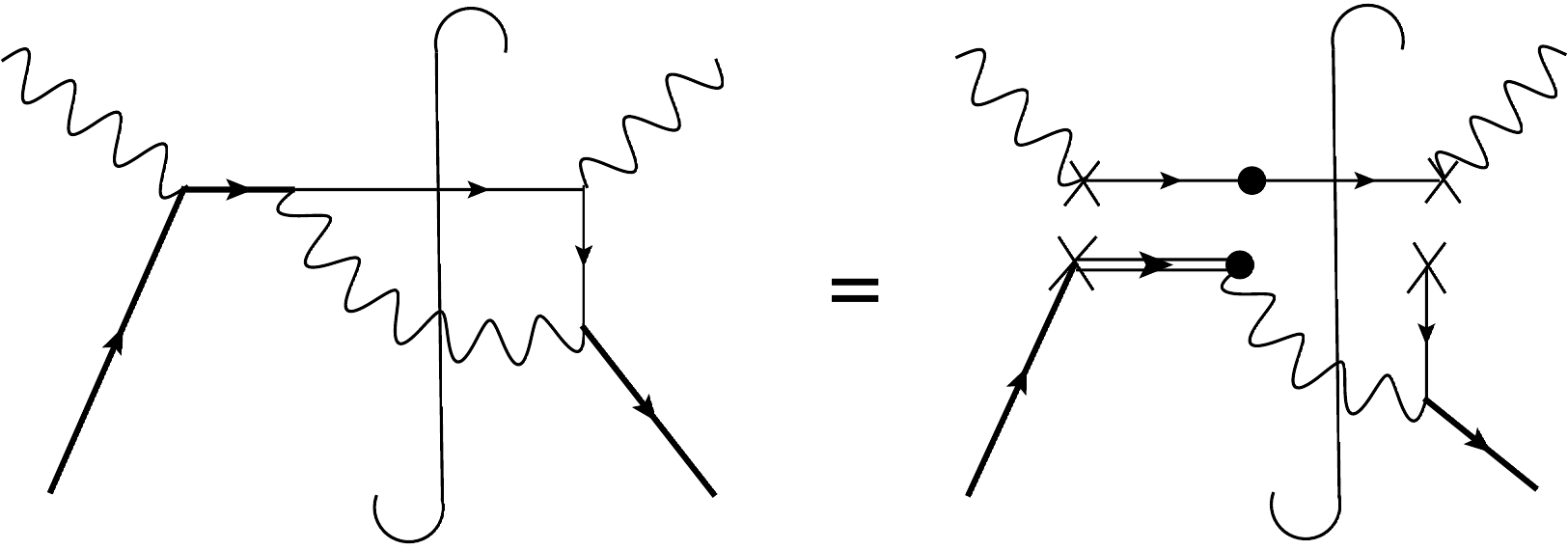}
\caption{\footnotesize{Handbag diagram with a photon loop separated into its upper and lower part, with the latter corresponding to the correlator.
The two factorized subdiagrams are meant to be ``attached'' by joining the crosses with the crosses and the circles with circles.}}
\label{eikonal}
\end{figure} 
This is not actually the final result, since we have not yet considered the
longitudinal gauge link, which is not vanishing in the Feynman gauge and
brings about two more diagrams to be evaluated, namely diagram (b) and (c) of
Fig.~\ref{feyngdiag}. Applying the Feynman rules to diagram (b), one finds\footnote{For convenience, we set the momentum of the dressed electron as
  $P=k+l$.} 
\begin{equation}
  \lvert\mc{M}\rvert^{2}
=-e^{4}\,\delta\left(l^2\right)\,\bar{u}_{\bm{S}}(\mb{P})\gamma_{\nu}g^{\mu\nu}
\frac{\slashed{k}+m}{k^{2}-m^{2}-i\epsilon}
\slashed{\epsilon}^{*}_{\lambda}(\mb{q})(\slashed{k}+\slashed{q}+m)\gamma_{\mu}
\frac{\slashed{k}+\slashed{q}+\slashed{l}+m}{(k+q+l)^{2}-m^{2}+i\epsilon}
\slashed{\epsilon}_{\lambda}(\mb{q})u_{\bm{S}}(\mb{P})
  \; . \label{m2} \end{equation} 

\begin{figure}[h]
\centering
\begin{align*}
 \raisebox{-5mm}{ \includegraphics[height=1.2cm]{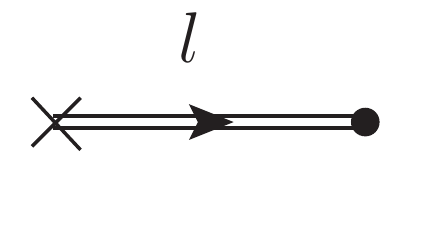}} &= \frac{i}{l^+ +i \epsilon},
&
   \raisebox{-4mm}{\includegraphics[height=1.2cm]{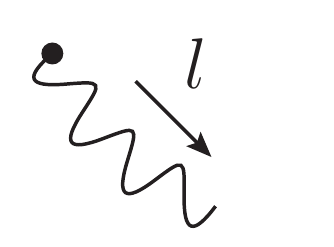}} &= - i e \, n_-^{\mu}.
\end{align*} 
\caption{\footnotesize{Feynman rules for the eikonal propagator (left) and
    vertex (right), when appearing on the left-hand side of cut diagrams. 
Notice that the momentum in the propagator is equal to the momentum of the
photon attached to vertex.}}
\label{feynrules}
\end{figure}

In order to separate the term corresponding to the lower part, this time we
need to apply the so-called {\em eikonal approximation} \cite{Collins:1981uw,Collins:1981uk},  which consists of considering as
relevant only the ``$-$'' component of the momentum of the fermion after the
interaction with the virtual photon; moreover, it is assumed that the emission
or the absorption of a ``soft'' photon, with small momentum $l$, does not
alter the fermion momentum. Under this approximation
the propagator of the struck fermion becomes
\begin{equation} \frac{\slashed{k}+\slashed{q}+\slashed{l}+m}{(k+q+l)^{2}-m^{2}+i\epsilon}\simeq \frac{\gdir{+}}{2l^{+}+i\epsilon} \; , \label{eikprop}\end{equation}
where we use
 \[(k+q+l)^{-}\gg(k+q+l)^{+} \; , \quad (k+q+l)^{-}\gg k_{\perp i}+q_{\perp i}+l_{\perp i} \; ,\quad (k+q+l)^{-} \gg m \; , \quad  (k+q)^{-}\gg l^{-} \; , \]
along with the on-shell conditions $(k+q)^2=m^2$, $l^2=0$. In this way we are able to factor out a lower and an upper part in the
amplitude, since
\begin{equation} (\slashed{k}+\slashed{q}+m)\gamma^{\rho}\gamma^{+}\simeq
  (k+q)^{-}\gamma^{+}\gamma^{\rho}\gamma^{+}\delta^{-}_{\rho}=2(k+q)^{-}\gamma^{+}\simeq
  2(\slashed{k}+\slashed{q}+m) \; , \end{equation} 
where in the first and in the last step we applied again the eikonal
approximation. The modified propagator described in Eq.~\eqref{eikprop} is
shown in the right-hand side of Fig.~\ref{eikonal}; the eikonal approximation
is taken into account by using the 
modified Feynman rules for the eikonal
propagator and vertex described in Fig.~\ref{feynrules}. 

The resulting transition matrix for the lower part of the diagram in Fig.~\ref{eikonal} is now 
\begin{equation} \mc{M}^{\mathrm{low}}_{ij}=-\frac{e^{2}}{l^{+}+i\epsilon}g^{\mu\nu}n_{-\mu}\,\bar{u}_{\mb{S},k}(\mb{P})\left(\gamma_{\nu}\frac{\slashed{k}+m}{k^{2}-m^{2}-i\epsilon}\right)_{kj}u_{\mb{S},i}(\mb{P})  \; . \label{mprime}\end{equation} 
Let us focus our attention on the denominators first. 
As we already noticed, the imaginary part of the electron propagator can be dropped, and we are left with
\begin{align} 
\frac{1}{l^{+}+i\epsilon}\frac{1}{k^{2}-m^{2}}&=\frac{1}{l^{+}(k^{2}-m^{2})} -i\pi\frac{1}{k^{2}-m^{2}}\delta\left(l^{+}\right),
\label{propagators}
\end{align}
where the term with $\delta\left(l^{+}\right)$ will be canceled by the contribution of
diagram (c) of Fig.~\ref{feyngdiag}, since the imaginary unit from the photon
vertex changes sign, being on the right-hand side of the cut. Therefore,  we are allowed to drop the $i\epsilon$ also in the photon propagator.
If we derive the correlator corresponding to the amplitude \eqref{mprime}, as we did
in \eqref{corlow} starting from \eqref{soft}, we obtain that the
contribution of diagram (b) to the 
TMD $f_{1}^{e}$ is
\begin{equation}
  f_{1}^{b}\xkq=\frac{1}{(2\pi)^{3}}\left[\frac{\pep{k}^{2}+m^{2}(1-x)^{2}}{(1-x)^{2}}\right]
  \,\varphi^{2}\xkq \; . \label{f1b} \end{equation} 

A completely equivalent result is found by computing the contribution $f_{1}^{c}$ to $f_{1}^{e}$ from diagram (c); henceforth we have $f_{1}^{c}=f_{1}^{b}$, and we can finally write as the final result
\begin{equation}f_{1}^{e}\xkq=f_{1}^{a}\xkq+f_{1}^{b}\xkq+f_{1}^{c}\xkq=\frac{1}{(2\pi)^{3}}\left[\frac{(1+x^{2})\pep{k}^{2}+m^{2}(1-x)^{4}}{x(1-x)^{2}}\right]
  \varphi^{2}\xkq \; . \end{equation} 
The same procedure can be applied to all the TMDs; as a result, we find that
the  leading-twist TMDs, at $\xk\neq (1,\mb{0})$, are
\begin{align}
&f_{1}^e(x,\mathbf{k}_{\perp})=\frac{1}{(2\pi)^{3}}\left[\frac{\mathbf{k}_{\perp}^{2}(1+x^{2})+m^{2}(1-x)^{4}}{x(1-x)^{2}}\right]\varphi^{2}(x,\mathbf{k}_{\perp}^{2})=\frac{e^{2}}{(2\pi)^{3}}\frac{\pep{k}^{2}(1+x^{2})+m^{2}(1-x)^{4}}{(1-x)\left[\pep{k}^{2}+m^{2}(1-x)^{2}\right]^2} \, , \label{Tmd1}\\
&g_{1L}^{e}(x,\mathbf{k}_{\perp})=\frac{1}{(2\pi)^{3}}\left[\frac{\mathbf{k}_{\perp}^{2}(1+x^{2})-m^{2}(1-x)^{4}}{x(1-x)^{2}}\right]\varphi^{2}(x,\mathbf{k}_{\perp}^{2})=\frac{e^{2}}{(2\pi)^{3}}\frac{\pep{k}^{2}(1+x^{2})-m^{2}(1-x)^{4}}{(1-x)\left[\pep{k}^{2}+m^{2}(1-x)^{2}\right]^2} \, ,\label{TMD2} \\
&g_{1T}^{e}\xkq=-\frac{2m^{2}}{(2\pi)^{3}}\,\varphi^{2}(x,\mathbf{k}_{\perp}^{2})=-\frac{2e^{2}}{(2\pi)^{3}}m^{2}\frac{x(1-x)}{\left[\pep{k}^{2}+m^{2}(1-x)^{2}\right]^{2}} \; ,\label{TMD3} \\
&h_{1L}^{\perp e}\xkq=\frac{2m^{2}}{(2\pi)^{3}x}\,\varphi^{2}(x,\mathbf{k}_{\perp}^{2})=\frac{2e^{2}}{(2\pi)^{3}}m^{2}\frac{(1-x)}{\left[\pep{k}^{2}+m^{2}(1-x)^{2}\right]^{2}} \; ,\label{TMD4} \\
&h_{1T}^{\perp e}\xkq=0 \; , \\
&h_{1}^{e}\xkq=\frac{2}{(2\pi)^{3}}\frac{\mathbf{k}_{\perp}^{2}}{(1-x)^{2}}\,\varphi^{2}(x,\mathbf{k}_{\perp}^{2})= \frac{2e^{2}}{(2\pi)^{3}}\frac{x\pep{k}^{2}}{(1-x)\left[\pep{k}^{2}+m^{2}(1-x)^{2}\right]^{2}}\; ,\label{TMD6} \\
&h_{1}^{\perp e}\xkq=f_{1T}^{\perp e}\xkq=0 \; . \label{Tmd8}
\end{align}

Note that the results for the T-even TMDs in Eqs.~\eqref{Tmd1}-\eqref{TMD6} correspond
to the leading-order predictions in the quark-target model of Ref.~\cite{Meissner:2007rx},
up to the QCD color factor. The Abelian nature of QED shows up in the
results for the T-odd TMDs, which are vanishing at variance with the
quark-target model.

\section{Calculation of TMDs in light-cone gauge}\label{tmdslcg}
\subsection{Overlap representation}\label{sec:overlap}

We now switch to the light-cone gauge, aiming to recover the same results that
we found in the Feynman gauge. We will also apply a different approach,
adopting the formalism of light-cone quantization and thus obtaining an
overlap representation of TMDs in terms of LFWFs
\cite{Brodsky:2000ii,Brodsky:1997de,brodskynot}.   

We already remarked that the longitudinal gauge link is simply an identity in
the light-cone gauge; we also neglect, at least at this level, the
contribution coming from the transverse gauge link, that we will study
separately later. Therefore, the expression we should start with is
\eqref{corr2}. Taking the trace of a generic Dirac operator $\Gamma$ is
equivalent to rewriting \cite{Mulders:1995dh}
\begin{equation}
  \Phi^{[\Gamma]}(x,\pep{k};\mb{S})=\int\frac{d\xi^{-}d^{2}\xi_{\perp}}{2(2\pi)^{3}}\,e^{-ik\xi}\langle
  P,\mb{S}|\bar{\psi}(\xi)\Gamma\psi(0)|P,\mb{S}\rangle \bigg \rvert_{{\xi^{+}=0}}
  \; . \label{traced} \end{equation} 

We expand the eigenstate $\lvert P,\mb{S}\rangle$ of the electron
in the light-cone Fock space \cite{Brodsky:1997de} truncating the expansion to
the two-particle Fock state of a photon and an electron. Since we are going to
use light-cone quantization, we need to decompose the dressed electron states
in terms of light-front helicity states $\lvert P,\Lambda\rangle$.
The relation between the eigenstates of canonical spin $\mb{S}$ 
and the eigenstates of light-front helicity $\Lambda$ can be found, for instance, in Ref.~\cite{Lorce:2011zta}.
%
In this representation, we then have
\begin{equation} |P,\Lambda\rangle= Z|e\rangle +|e\gamma\rangle \; , \label{fockdec} \end{equation}
where $\Lambda=\pm$ stands for the values of light-front helicity $\pm 1/2$.
In the above equation $Z$ is a renormalization constant that we may simply
take to be $1$ for our purposes (it would become relevant only when
considering the end point $x=1,\,\pep{k}=\pep{0}$).  The one-particle state is
simply: 
\begin{equation} 
|e\rangle=b^{\dag}_{\Lambda}(\mb{P})|0\rangle \; ,\label{estate}
 \end{equation} 
 with $b^\dagger_\Lambda(\mb{P})$ the creation operator of an electron with
 light-front helicity $\Lambda$ and momentum
 $\mb{P}=(P^+,\mb{P}_\perp)$.  

 In order to find the exact expression of the multiparticle Fock states, one
 should solve the eigenvalue equation for the full QED Hamiltonian. However,
 by considering the interaction potential small with respect to the free
 Hamiltonian, one is allowed to apply time-ordered perturbation theory and,
 for the two-particle state at first order in the coupling constant $e$, one  
 finds~\cite{brodskynot}
\begin{equation}|e\gamma\rangle=
  \int\frac{dx\,d^{2}\mathbf{k}_{\perp}}{2(2\pi)^{3}\sqrt{x(1-x)}}\sum_{\lambda,\lambda_{\gamma}}
  \Psi_{\lambda,\lambda_{\gamma}}^{\Lambda}(x,\mathbf{k}_{\perp})|e \, \gamma;
  \, \mb{p}_{e},\lambda; \,\mb{p}_{\gamma},\lambda_{\gamma}\rangle \;
  . \label{2bod} \end{equation} 
Here $\lambda=\pm,\,\lambda_{\gamma}=\pm1$ stand for the internal electron light-front helicity $\pm 1/2$ and photon polarization, respectively, and $\mb{p}_{e}, \mb{p}_{\gamma}$ refer to the momenta of the partons\footnote{One should pay attention not confusing the $|e\gamma\rangle$ state, which describes the dressed electron as a 2-body system in Eq.~\eqref{fockdec}, with the state of two free quanta $|e\gamma; \, \mb{p}_{e},\lambda; \,\mb{p}_{\gamma},\lambda_{\gamma}\rangle$ .}:
\begin{equation} 
\mb{p}_{e}=(xP^{+},\pep{k}) \; , \quad \mb{p}_{\gamma}=\left( (1-x)P^{+},-\pep{k}\right) \; .\label{param}
\end{equation} 
The wave functions $\Psi_{\lambda,\lambda_{\gamma}}^{\Lambda}(x,\mathbf{k}_{\perp})$ appearing in \eqref{2bod} are given by
\begin{equation} \Psi^{\Lambda}_{\lambda,\lambda_{\gamma}}(x,\mathbf{k}_{\perp})=\varphi(x,\mathbf{k}_{\perp}^{2})\, \bar{u}_{\lambda}(\mathbf{p_{e}})\gamma^{\mu}\varepsilon^{*}_{\lambda_{\gamma},\mu}(\mb{p}_{\gamma})u_{\Lambda}(\mathbf{P}) \, , \label{lfwfex} 
\end{equation} 
with the function $\varphi\xkq$ defined in Eq.~\eqref{varphi}. 
In the light-cone gauge $A^{+}=0$, they are the LFWFs for the $\lvert e\gamma\rangle$ component of the dressed electron. For a dressed electron with light-front helicity $\Lambda=+$, one has explicitly\footnote{There is a sign difference with respect to Ref. \cite{Brodsky:2000ii}, due to a different choice of the sign of the charge $e$.} \cite{Brodsky:2000ii}
\begin{align}
&\Psi^{+}_{+,+1}(x,\mathbf{k}_{\perp})=\sqrt{\frac{2}{x}}\frac{(k_{x}-ik_{y})}{(1-x)}\, \varphi(x,\mathbf{k}_{\perp}^{2}) \; , \label{lcwf1} \\
&\Psi^{+}_{+,-1}(x,\mathbf{k}_{\perp})=-\sqrt{2x}\frac{(k_{x}+ik_{y})}{(1-x)} \,\varphi(x,\mathbf{k}_{\perp}^{2}) \; , \label{lcwf2} \\
&\Psi^{+}_{-,+1}(x,\mathbf{k}_{\perp})= \sqrt{\frac{2}{x}}m(1-x) \,\varphi(x,\mathbf{k}_{\perp}^{2}) \; , \label{lcwf3}\\
&\Psi^{+}_{-,-1}(x,\mathbf{k}_{\perp})=0 \; .\label{lcwf4}
\end{align}

Assuming instead $\Lambda=-$, one has
\begin{align}
&\Psi^{-}_{+,+1}(x,\mathbf{k}_{\perp})=0 \; ,\label{lcwf5}\\
&\Psi^{-}_{+,-1}(x,\mathbf{k}_{\perp})= \sqrt{\frac{2}{x}}m(1-x)\, \varphi(x,\mathbf{k}_{\perp}^{2}) \; ,  \\
&\Psi^{-}_{-,+1}(x,\mathbf{k}_{\perp})= \sqrt{2x}\frac{(k_{x}-ik_{y})}{(1-x)}\, \varphi(x,\mathbf{k}_{\perp}^{2})\; ,\\
&\Psi^{-}_{-,-1}(x,\mathbf{k}_{\perp})= -\sqrt{\frac{2}{x}}\frac{(k_{x}+ik_{y})}{(1-x)}\, \varphi(x,\mathbf{k}_{\perp}^{2}) \; \label{lcwf8}.
\end{align}
Notice that the LFWFs in Eqs.~\eqref{lcwf1}-\eqref{lcwf8} are related by
\begin{equation}
\Psi^\Lambda_{\lambda,\lambda_{\gamma}}\xk=(-1)^{(\Lambda-\lambda)/2+\lambda_{\gamma}}
\Psi^{-\Lambda\;^*}_{-\lambda,-\lambda_{\gamma}}\xk\,.\label{eq:relation}
\end{equation}

If we use Eq.~\eqref{fockdec} in the correlator \eqref{traced}, and take the proper projections for each $\Gamma$ matrix, we can obtain the LFWF overlap representation of the correlator, which reads
\begin{equation} \Phi^{[\Gamma]}\xkl=\frac{1}{4(2\pi)^{3}xP^{+}}\sum_{\lambda,\lambda',\lambda_{\gamma}} \, \Psi^{\Lambda\;^{*}}_{\lambda',\lambda_{\gamma}}(x,\pep{k})\bar{u}_{\lambda'}(\mb{p}_{e})\Gamma u_{\lambda}(\mb{p}_{e})\Psi^{\Lambda}_{\lambda,\lambda_{\gamma}}(x,\pep{k}) \,  . \label{phii} \end{equation} 

An alternative, yet equivalent, way to derive the TMDs explicitly is to represent the correlator in the basis where we consider the light-front helicity of both the dressed and  the internal electron, and we treat them symmetrically. We can therefore define the light-front helicity amplitudes \cite{Lorce:2011zta}:
\begin{equation} 
\Phi_{\Lambda',\lambda';\Lambda,\lambda}(x,\pep{k}):\,=\frac{1}{N}\langle P,\Lambda'|b_{\lambda'}^{\dag}(x,\pep{k})b_{\lambda}(x,\pep{k})|P,\Lambda\rangle \; , 
\label{eq:helicity-ampl}
\end{equation}
with \[N:\,=[2(2\pi)^{3}]^2 x\delta^{(3)}(\mb{0}) \; . \] 
Inserting in \eqref{eq:helicity-ampl} the Fock-state expansion \eqref{fockdec}, we obtain
\begin{equation} 
\Phi_{\Lambda',\lambda';\Lambda,\lambda}(x,\pep{k})=\frac{1}{2(2\pi)^{3}}\sum_{\lambda_{\gamma}}\Psi^{\Lambda'\;^{*}}_{\lambda',\lambda_{\gamma}}(x,\pep{k})\Psi_{\lambda,\lambda_{\gamma}}^{\Lambda}\xk \; . \label{phiss} 
\end{equation} 
The light-front helicity amplitudes are parametrized by the following combinations of TMDs~\cite{Bacchetta:1999kz,Lorce:2011zta}: 
\begin{equation} \Phi=\begin{pmatrix} \frac{1}{2}(f_{1}^{e}+g_{1L}^{e}) & -\frac{k_{R}}{2m}(ih_{1}^{\perp e}-h_{1L}^{\perp e}) & \frac{k_{L}}{2m}(if_{1T}^{\perp e}+g_{1T}^{e}) & h_{1}^{e} \\
\frac{k_{L}}{2m}(ih_{1}^{\perp e}+h_{1L}^{\perp e}) & \frac{1}{2}(f_{1}^{e}-g_{1L}^{e}) & \frac{k_{L}^{2}}{2m^{2}}h_{1T}^{\perp e} & \frac{k_{L}}{2m}(if_{1T}^{\perp e}-g_{1T}^{e}) \\
-\frac{k_{R}}{2m}(if_{1T}^{\perp\, e}-g_{1T}^{e}) & \frac{k_{R}^{2}}{2m^{2}}h_{1T}^{\perp e} & \frac{1}{2}(f_{1}^{e}-g_{1L}^{e}) & -\frac{k_{R}}{2m}(ih_{1}^{\perp e}+h_{1L}^{\perp e}) \\
h_{1}^{e} & -\frac{k_{R}}{2m}(if_{1T}^{\perp e}+g_{1T}^{e}) & \frac{k_{L}}{2m}(ih_{1}^{\perp e}-h_{1L}^{\perp e}) & \frac{1}{2}(f_{1}^{e}+g_{1L}^{e}) \; ,\label{matrix}\end{pmatrix}
\end{equation} where $k_{R,L}:\,=k_{x}\pm ik_{y}$. The row entries are $(\Lambda',\lambda')=(++),(+-),(-+),(--)$, while the column entries are $(\Lambda,\lambda)=(++),(+-),(-+)$, $(--)$. 

Using Eq.~\eqref{phii} and the projections \eqref{tmd1}-\eqref{tmd3}, or
equivalently Eq.~\eqref{phiss} along with \eqref{matrix}, one obtains the
following results for the LFWF overlap representation of the contribution of
the internal electron to the T-even TMDs:
\begin{align} 
&f_{1}^{e}\xkq =\frac{1}{4(2\pi)^{3}}\sum_{\Lambda,\lambda,\lambda_{\gamma}}|\Psi^{\Lambda}_{\lambda,\lambda_{\gamma}}\xk|^{2}=\frac{1}{2(2\pi)^{3}}\left[|\Psi^{+}_{+,+1}\xk|^{2}+|\Psi^{+}_{+,-1}\xk|^{2}+|\Psi^{+}_{-,+1}\xk|^{2}\right] \; ,\label{overlap-1} \\
&g_{1L}^{e}\xkq=\frac{1}{2(2\pi)^{3}}\left[|\Psi^{+}_{+,+1}\xk|^{2}+|\Psi^{+}_{+,-1}\xk|^{2}-|\Psi^{+}_{-,+1}\xk|^{2}\right] \, ,\label{overlap-2} \\
&g_{1T}^{e}\xkq=\frac{m}{4(2\pi)^{3}\pep{k}^{2}}\sum_{\lambda_{\gamma}}\left[k_{R}\, \Psi^{+\;^{*}}_{+,\lambda_{\gamma}}\xk\Psi_{+,\lambda_{\gamma}}^{-}\xk+k_{L}\,\Psi^{-\;^{*}}_{+,\lambda_{\gamma}}\xk\Psi_{+,\lambda_{\gamma}}^{+}\xk\right] \, , \label{overlap-3}\\
&h_{1L}^{\perp e}\xkq=\frac{m}{2(2\pi)^{3}\pep{k}^{2}}\sum_{\lambda_{\gamma}}\left[k_{R}\,\Psi_{-,\lambda_{\gamma}}^{+\;^{*}}\xk\Psi_{+,\lambda_{\gamma}}^{+}\xk+k_{L}\,\Psi_{+,\lambda_{\gamma}}^{+\;^{*}}\xk\Psi_{-,\lambda_{\gamma}}^{+}\xk\right] \; , \label{overlap-4} \\
&h_{1T}^{\perp e}\xkq=\frac{m^{2}}{2(2\pi)^{3}\pep{k}^{2}}\sum_{\lambda_{\gamma}}\left[k_{R}^{2}\,\Psi^{+\;^{*}}_{-,\lambda_{\gamma}}\xk\Psi^{-}_{+,\lambda_{\gamma}}\xk+k_{L}^{2}\,\Psi^{-\;^{*}}_{+,\lambda_{\gamma}}\xk\Psi^{+}_{-,\lambda_{\gamma}}\xk\right] \; ,  \label{overlap-5}\\
&h_{1}^{e}\xkq=\frac{1}{2(2\pi)^{3}}\left[\Psi^{+\;^{*}}_{+,+1}\xk\Psi^{-}_{-,+1}\xk+\Psi^{+\;^{*}}_{+,-1}\xk\Psi_{-,-1}^{-}\xk\right]\; . \label{overlap-6} \end{align}
The analytic results found by inserting the explicit expressions
\eqref{lcwf1}-\eqref{lcwf8} of the LFWFs indeed coincide with the results
\eqref{Tmd1}-\eqref{Tmd8} obtained in Feynman gauge and using Feynman diagrams. 

The LFWFs are eigenstates of the total spin of the partons in the $z$ direction, $J_z=\lambda/2+\lambda_\gamma$, and of the 
total OAM  $L_z=\Lambda/2-J_z$, as required by angular momentum conservation. 
For the two-body LFWFs of the electron, one can have only $L_z=0$, and $L_z=\pm 1$. They are commonly labeled as S and P waves, respectively, although 
this is an abuse of language as partial waves should refer to $L$ and not $L_z$.
The LFWF overlap representation in Eqs.~\eqref{overlap-1}-\eqref{overlap-6} allows one to disclose the different contributions to the TMDs from the spin and OAM configurations of the partons.
In particular, $f_{1}^{e}$, $g_{1L}^{e}$ and $h_{1}^{e}$ are all diagonal in the OAM, 
while the remaining TMDs contain different interference terms between S and P waves, and therefore involve a
transfer of OAM  between
the initial and final electron states. The interplay between the different partial waves in the TMDs will be discussed in more detail in Sec.~\ref{results}.

 \subsection{Feynman diagram approach}

It is interesting to follow the diagrammatic approach in light-cone gauge
also, in order to explicitly discuss the gauge invariance of TMDs. In this case,
since the longitudinal gauge link is trivial and we are neglecting for the
moment the transverse gauge link, we only need to evaluate diagram (a) of
Fig.~\ref{feyngdiag}; the amplitude is still given by \eqref{amplitude}, but
this time we take the following sum over the photon polarization vectors\footnote{Notice that the third term of the polarization sum, proportional to $p^2$, is omitted since in our case the photon is on shell.}: 
\begin{equation} 
d^{\mu\nu}(p):\,=\sum_{\lambda}\varepsilon_{\lambda}^{\mu\ast}(\mb{p})\varepsilon_{\lambda}^{\nu}(\mb{p})=-g^{\mu\nu}+\frac{1}{p^{+}}\left(p^{\nu}n^{\mu}_{-}+p^{\mu}n^{\nu}_{-}\right) \; . \label{polsum}\end{equation}  
If we factorize the amplitude separating the upper and lower parts, we actually get three contributions for the lower part of the diagram:
\[ \mc{M}^{\mathrm{low}}\equiv A+B+C \; , \] with
\begin{align}
&A:\,=-e^{2}\bar{u}_{\mb{S}}(\mb{P})\gamma_{\mu}\frac{\slashed{k}+m}{k^{2}-m^{2}-i\epsilon}\frac{\slashed{k}+m}{k^{2}-m^{2}+i\epsilon}\gamma^{\mu}u_{\mb{S}}(\mb{P})\; , \label{A} \\
&B:\,=\frac{e^{2}}{(P-k)^{+}}\bar{u}_{\mb{S}}(\mb{P})\gdir{+}\frac{\slashed{k}+m}{k^{2}-m^{2}-i\epsilon}\frac{\slashed{k}+m}{k^{2}-m^{2}+i\epsilon}(\slashed{P}-\slashed{k})u_{\mb{S}}(\mb{P}) \; , \label{B}\\
&C:\,=\frac{e^{2}}{(P-k)^{+}}\bar{u}_{\mb{S}}(\mb{P})(\slashed{P}-\slashed{k})\frac{\slashed{k}+m}{k^{2}-m^{2}-i\epsilon}\frac{\slashed{k}+m}{k^{2}-m^{2}+i\epsilon}\gdir{+}u_{\mb{S}}(\mb{P}) \; . \label{C}\end{align}
It is now easy to recognize that the term $A$ is equivalent to the
contribution of Eq.~\eqref{soft} evaluated in the Feynman gauge. Let
us focus on  $B$; using the Dirac equation
$(\slashed{P}-m)u_{\mb{S}}(\mb{P})=0$, we can rewrite 
\[
\frac{\slashed{k}+m}{k^{2}-m^{2}}\,(\slashed{P}-\slashed{k})u_{\mb{S}}(\mb{P})=\frac{\slashed{k}+m}{k^{2}-m^{2}}(m-\slashed{k})u_{\mb{S}}(\mb{P})=-u_{\mb{S}}(\mb{P})
\; . \] 
Hence we find that  term $B$, which comes from the extra term in the
polarization sum in light-cone gauge, actually coincides with the contribution
of Eq.~\eqref{mprime}, treated with the eikonal
approximation. The same applies also to   term $C$, which in turn corresponds to
diagram (c) of Fig.~\ref{feyngdiag} in Feynman gauge after eikonal
approximation.   

We can now also easily check that the term of the correlator that is related to diagram (b) and (c) of Fig.~\ref{feyngdiag} are actually vanishing in the light-cone gauge. If we still use the eikonal approximation, but substitute the sum \eqref{polsum} instead of \eqref{polsumfeyn} for the polarization vectors in \eqref{mprime}, the latter becomes: 
\begin{equation} \mc{M}^{\mathrm{low}}=-\frac{e^{2}}{(P-k)^{+}+i\epsilon}\bar{u}_{S}(\mb{P})\gamma_{\nu}\left[-g^{\mu\nu}+\frac{1}{(P-k)^{+}}
\left((P-k)^{\mu}n^{\nu}_{-}+(P-k)^{\nu}
n^{\mu}_{-}\right)\right]\frac{\slashed{k}+m}{k^{2}-m^{2}-i\epsilon}n_{-\mu} u_{\mb{S}}(\mb{P}) \; , \end{equation}
which contains the Dirac structure,
\[ -\gamma^{+}+\frac{1}{(P-k)^{+}}\gamma_{\mu}n^{\mu}_{-}(P-k)^{\nu}n_{-\nu}+\frac{1}{(P-k)^{+}}(\slashed{P}-\slashed{k})n^{2}_{-}=0 \; . \]
Within the diagrammatic approach in the light-cone gauge we recover the same results for the TMDs we obtained in the Feynman gauge derivation of Sec.~\ref{tmdfg}.

\section{Transverse gauge link} \label{tgl}

In the previous sections we performed the derivation of the TMDs for the
electron first in Feynman gauge including the contribution from the
longitudinal gauge link, then in light-cone gauge but neglecting the Wilson
line; as a result, we found perfect agreement between the two gauges, as it
should be, since the TMDs are gauge-invariant objects. However, one may wonder
if there is a contribution of the transverse gauge link. 
We actually neglected a correction to our results that is illustrative to
analyze, although it will show up only at the end point $x=1$.
 
As it was noticed by Ji \textsl{et al}.~\cite{Ji:2002aa,Belitsky:2002sm}, the
transverse gauge link does have an important role in taking into account the
final-state interactions when working in  light-cone gauge. The point is that
the four-momentum of the intermediate photon must be integrated over, according
to cut-diagram rules. This means that, when we use the sum over the
polarization vectors \eqref{polsum}, we actually should regularize the
singularity due to the $p^{+}$ factor in the denominator. We can in fact do
this in different ways, because we have the freedom of choosing a certain
prescription for the regularization of the denominator; the following choices
are possible: 
\begin{align} &\mathrm{Retarded:} \quad \frac{1}{p^{+}} \; \; \longrightarrow \; \; \frac{1}{p^{+}+i\epsilon}  \; , \label{ret}\\
&\mathrm{Advanced:} \quad \frac{1}{p^{+}} \; \; \longrightarrow \; \; \frac{1}{p^{+}-i\epsilon}  \; , \label{adv}\\
&\mathrm{Principal\;value:} \quad \frac{1}{p^{+}} \; \; \longrightarrow \; \; \frac{1}{2}\left[\frac{1}{p^{+}+i\epsilon}+\frac{1}{p^{+}-i\epsilon}\right]  \; . \label{pv}
\end{align}
It is obvious that
our final result must be independent of the choice of a specific prescription,
and it must also be the same we found in  Feynman gauge, where the problem
of regularizing the denominators does not arise and the transverse gauge link
does not contribute. Nonetheless, if we come back to our evaluation of the TMD
$f_{1}^{e}\xkq$ from Feynman diagrams, we see that it is not the
case if we make the replacement \eqref{ret}-\eqref{pv} in \eqref{polsum} and
then put the latter in the amplitude \eqref{soft}.   

Let us first take the retarded prescription as an example: term $A$ in
\eqref{A} would be unchanged, but the denominator of terms $B$ and $C$ in
\eqref{B} and \eqref{C} would be modified according to \eqref{ret}, giving
\begin{equation} 
B=\frac{e^{2}}{(P-k)^{+}+i\epsilon}\bar{u}_{S}(\mb{P})\gdir{+}\frac{\slashed{k}+m}{k^{2}-m^{2}-i\epsilon}\frac{\slashed{k}+m}{k^{2}-m^{2}+i\epsilon}(\slashed{P}-\slashed{k})u_{\mb{S}}(\mb{P}) \; ,
\end{equation} 
and similarly for $C$. We rewrite, as usual:
\begin{equation} 
\frac{1}{(P-k)^{+}+i\epsilon}=\pv\left(\frac{1}{(P-k)^{+}}\right)-i\pi\delta\left((P-k)^{+}\right) \; . 
\end{equation}
This leads us to the final result for the TMD $f_{1}^{e}$ (including all three terms):
\begin{align} f^{\mathrm{ret}}_{1}\xkq &=\frac{1}{(2\pi)^{3}}\left[\mathbf{k}_{\perp}^{2}\frac{(1+x^{2})}{x(1-x)^{2}}+m^{2}\frac{(1-x)^{2}}{x}\right]\varphi^{2}(x,\mathbf{k}_{\perp}^{2})-\frac{i}{(2\pi)^{2}}\left[\frac{\pep{k}^{2}}{1-x}+m^{2}(1-x)\right]\varphi^{2}\xkq\, \delta(1-x)\non \\
&= f_{1}^{e}\xkq-i\frac{e^{2}}{(2\pi)^{2}}\frac{1}{\pep{k}^{2}}\,\delta(1-x) \; . \label{f1ret} \end{align}
Similarly, we obtain for the advanced prescription:
\begin{equation}f^{\mathrm{adv}}_{1}\xkq=f_{1}^{e}\xkq+i\frac{e^{2}}{(2\pi)^{2}}\frac{1}{\pep{k}^{2}}\,\delta(1-x)
  \; . \label{f1adv} \end{equation} 
In the principal-value prescription, instead, the denominators of $B$ and $C$ remain unchanged because:
\[ \frac{1}{2}\left[\frac{1}{p^{+}+i\epsilon}+\frac{1}{p^{+}-i\epsilon}\right]
=\frac{1}{2}\left[\pv\left(\frac{1}{p^{+}}\right)-i\pi\delta(p^{+})+\pv\left(\frac{1}{p^{+}}\right)+i\pi\delta(p^{+})\right]\equiv
\frac{1}{p^{+}}\] 
and hence the final result coincides with the one found in Feynman gauge:
\begin{equation}f^{\mathrm{pv}}_{1}\xkq\equiv f_{1}^{e}\xkq \; . \label{f1pv} \end{equation}
We expect all results to be the same when adding also the transverse gauge-link contribution. As we will show, the extra terms proportional to
$\delta(1-x)$ in the retarded and advanced prescriptions are compensated by
the contribution of the transverse gauge link, while the latter is vanishing
in the principal-value prescription. We remark that this is not what one finds
usually in the case of the nucleon (see e.g.~\cite{Ji:2002aa}), since the
retarded prescription is the one corresponding to having $\pep{A}(\infty)=0$
and hence the transverse gauge link should vanish in that
prescription. However, we stress that in the present discussion we have not
taken into account all the corrections to the distributions at the end point
$x=1$, and this might explain the difference with the existing calculations in
QCD.

We now turn to the evaluation of the transverse gauge link in the LFWF overlap
representation. We start again with the integrated version of the correlator,
with the states in the light-front helicity basis: 
\begin{equation} \Phi(x,\pep{k};\Lambda):\,=\int\frac{d\xi^{-}d^{2}\pepgr{\xi}}{(2\pi)^{3}}\,e^{ik\cdot\xi}\;\langle P,\Lambda|\bar{\psi}(0)\mc{U}_{(0,\infty)}\mc{U}_{(\infty,\xi)}\psi(\xi)|P,\Lambda\rangle \bigg \rvert_{{\xi^{+}=0}} \; . \label{tgl2} \end{equation} 
We express the Wilson line in the light-cone gauge, using \eqref{glt}
approximated according to
\begin{equation} \mc{U}_{(0,\infty)}\mc{U}_{(\infty,\xi)}\simeq \mathds{1} +ie\int_{\pep{0}}^{\pepgr{\xi}}d\pepgr{\eta}\cdot\pep{A}(0,\pepgr{\eta},\infty) =\mathds{1} +ie\int_{0}^{1}dt\,\pep{A}(0,t\pepgr{\xi},\infty)\cdot\pepgr{\xi} \; . \label{tglappr}\end{equation}
We neglect the identity in \eqref{tglappr} and we focus on the gauge-link terms. 
If we switch again to light-cone quantization and replace the dressed electron
state $|P,\Lambda\rangle$ according to \eqref{fockdec}, in principle 
we will get four different combinations of states to consider. However, the term
containing the gauge link between two $|e\gamma\rangle$ states is vanishing,
because the photon coming from the final- (or initial-) state interaction cannot
couple to the spectator photon. 
The 
only relevant contributions will then come
from the terms where the matrix element is between the states $|e\rangle$ and
$|e\gamma\rangle$; in this situation, the gauge photon which interacts with
the bare particle in the initial (or final) state coincides with the photon in
the dressed electron. Therefore, we are left with the following two terms,
contributing to the correlator: 
\begin{equation} I:\,=\frac{ie}{2}\int\frac{d\xi^{-}d^{2}\pepgr{\xi}}{(2\pi)^{3}}\,e^{ik\cdot\xi}\;\langle e|\bar{\psi}(0)\left(\int_{0}^{1}dt\,\pep{A}(0,t\pepgr{\xi},\infty)\cdot\pepgr{\xi}\right)\psi(0)|e\gamma\rangle\bigg \rvert_{\xi^{+}=0} \; , \end{equation}
\begin{equation} 
II:\,=ie\int\frac{d\xi^{-}d^{2}\pepgr{\xi}}{(2\pi)^{3}}\,e^{ik\cdot\xi}\;\langle e\gamma|\bar{\psi}(0)\left(\int_{0}^{1}dt\,\pep{A}(0,t\pepgr{\xi},\infty)\cdot\pepgr{\xi}\right)\psi(0)|e\rangle\bigg \rvert_{{\xi^{+}=0}} \; .
 \end{equation} 
Let us now restrict to the case of the TMD $f_{1}^{e}$; we will begin with the evaluation of $I$. After some manipulations and resolving the integral over $t$, it can be split into two terms, $I=I_{A}+I_{B}$, with
\begin{align}
I_{A}:\,=&-\frac{e}{2}\int\frac{d\xi^{-}d^{2}\pepgr{\xi}}{2(2\pi)^{3}}\int\frac{dx'd^{2}\pep{k'}}{2(2\pi)^{3}\sqrt{x'(1-x')}}e^{ik\cdot\xi}e^{-ip_{e}\cdot\xi}e^{-i p_{\gamma}^{+}\infty}\non \\
&\times\sum_{\Lambda,\lambda,\lambda_{\gamma}}\frac{\bm{\varepsilon}_{\perp_{\lambda_{\gamma}}}(\mb{p}_{\gamma})\cdot\pepgr{\xi}}{\pep{p_{\gamma}}\cdot\pepgr{\xi}} \, \bar{u}_{\Lambda}(\mb{P})\gamma^{+}u_{\lambda}(\mb{p}_{e})\Psi^{\Lambda}_{\lambda,\lambda_{\gamma}}(x',\pep{k'})\bigg \rvert_{{\xi^{+}=0}} \; , \label{1a}\\
I_{B}:\,=&\frac{e}{2}\int\frac{d\xi^{-}d^{2}\pepgr{\xi}}{2(2\pi)^{3}}\int\frac{dx'd^{2}\pep{k'}}{2(2\pi)^{3}\sqrt{x'(1-x')}}e^{ik\cdot\xi}e^{-ip_{e}\cdot\xi}e^{-i p_{\gamma}^{+}\infty} e^{i\pepgr{\xi}\cdot\pep{p_{\gamma}}}\non \\
&\times\sum_{\Lambda,\lambda,\lambda_{\gamma}}\frac{\bm{\varepsilon}_{\perp_{\lambda_{\gamma}}}(\mb{p}_{\gamma})\cdot\pepgr{\xi}}{\pep{p_{\gamma}}\cdot\pepgr{\xi}} \, \bar{u}_{\Lambda}(\mb{P})\gamma^{+}u_{\lambda}(\mb{p}_{e})\Psi^{\Lambda}_{\lambda,\lambda_{\gamma}}(x',\pep{k'})\bigg \rvert_{{\xi^{+}=0}} \; ,
 \end{align}
where $\mb{p}_{e}=(x' P^+,\mb{k}'_\perp)$ and $\mb{p}_\gamma=((1-x')P^+,-\mb{k}'_\perp)$.
The sum appearing in both $I_{A}$ and $I_{B}$ is
\begin{equation}
T:\,=\sum_{\Lambda,\lambda,\lambda_{\gamma}}\bar{u}_{\Lambda}(\mb{P})\gamma^{+}u_{\lambda}(\mb{p}_{e})\bar{u}_{\lambda}(\mb{p}_{e})\gamma_{\mu}u_{\Lambda}(\mb{P})\varepsilon^{\mu\;^{*}}_{\lambda_{\gamma}}(\mb{p}_{\gamma})\bm{\varepsilon}_{\perp_{\lambda_{\gamma}}}(\mb{p}_{\gamma})\cdot\pepgr{\xi}=-4P^{+}\pep{p_{e}}\cdot\pepgr{\xi}\left(1-2\frac{p^{+}_e}{p_{\gamma}^{+}}\right) \; , 
\end{equation}
where we used the explicit expression of the LFWFs in terms of the spinors and polarization vectors. Equation \eqref{1a} consequently becomes
\begin{equation}
I_{A}=-2eP^{+}\int\frac{d\xi^{-}d^{2}\pepgr{\xi}}{2(2\pi)^{3}}\int\frac{dx'd^{2}\pep{k'}}{2(2\pi)^{3}\sqrt{x'(1-x')}} e^{ik\cdot\xi}e^{-ip_{e}\cdot\xi}\left(-e^{-i \infty p_{\gamma}^{+}}+2x'P^{+}\frac{e^{-i \infty p_{\gamma}^{+}}}{p_{\gamma}^{+}}\right) 
\varphi(x',\bm{k}_{\perp}'^{2})\bigg\rvert_{{\xi^{+}=0}}\; .
\label{IAc}
\end{equation}
We isolated the term in brackets in the above equation because we need to specify how to handle the exponential at infinity \cite{Belitsky:2002sm,Ji:2002aa}: in the sense of principal-value distribution, we can write
\begin{equation}
 \frac{e^{-i\infty p_{\gamma}^{+}}}{p_{\gamma}^{+}}\equiv\lim_{L\rightarrow\infty}\frac{e^{-iLp_{\gamma}^{+}}}{p_{\gamma}^{+}}=2\pi i \chi\delta(p_{\gamma}^{+}) \; ,\label{limit} 
 \end{equation}
where $\chi$ is a constant that takes different values, according to the prescription that we adopt for the denominator:
\begin{equation} \chi= \begin{cases}
-1 & \text{Retarded },\\
\phantom{-}0 & \text{Advanced }, \\
-\frac{1}{2} & \text{Principal value }.
\end{cases} 
\label{cases} \end{equation}

Again in the sense of principal-value distribution, instead, we have
\begin{equation} e^{-i \infty p_{\gamma}^{+}}=0 \; , \end{equation} which
allows us to drop the first term in the bracket of~\eqref{IAc}. This is not surprising, because that term comes from the $-g^{\mu\nu}$ in the polarization sum, which appears in Feynman gauge also; then, this is consistent with the fact that the transverse gauge link does not contribute in the Feynman gauge.  

Now we can easily perform the integration over $\xi^-, \,\pepgr{\xi}^2$, and then over $x'$, $\pep{k'}$, to arrive at
\begin{equation} I_{A}=-\frac{i\chi e^{2}}{(2\pi)^{2}}\frac{xP^{+}}{\left[(1-x)M^{2}+\pep{k}^{2}\right]} \,\delta(p_{\gamma}^{+}) \; . \label{1aa}\end{equation}
As for term $I_{B}$, which is now
\begin{equation}I_{B}=-eP^{+^{2}}\int\frac{d\xi^{-}d^{2}\pepgr{\xi}}{(2\pi)^{3}}\int\frac{dx'd^{2}\pep{k'}}{(2\pi)^{3}\sqrt{x'(1-x')}} e^{ik\cdot\xi}e^{-ip_{e}\cdot\xi}\frac{e^{-i \infty p_{\gamma}^{+}}}{p_{\gamma}^{+}}\, x' \varphi(x',\bm{k}_{\perp}'^{2})\bigg\rvert_{{\xi^{+}=0}}\; ,\end{equation}
we notice that in any prescription (except for the advanced, where it is vanishing), it would become proportional to $\delta(1-x)$, thanks to \eqref{limit}; furthermore, the integration over $\pepgr{\xi}^{2}$ would now bring a $\delta^{(2)}(\pep{k})$. Then, we can conclude that this term would be relevant only in the end point $(x=1,\pep{k}=\pep{0})$,  and for this reason we will not consider it, consistently with what we did previously in the Feynman gauge.  

A similar argument can be applied to the term $II$. It can in turn be split into the sum of two contributions, of which the first one becomes relevant only at the end point, while the second one reads
\begin{equation} II_{A}=-\frac{eP^{+}}{(2\pi)^{3}}\int dx'
  \frac{x'}{\sqrt{x'(1-x')}}\frac{e^{i\infty
      p_{\gamma}^{+}}}{p_{\gamma}^{+}}\,\varphi(x',\pep{k}^{2})\delta(1-x) \;
  . \end{equation} In the sense of principal-value distribution, also in the
above equation we can write
\begin{equation} \frac{e^{i\infty p_{\gamma}^{+}}}{p_{\gamma}^{+}}=2\pi i \zeta\delta(p_{\gamma}^{+}) \; , \label{2a}\end{equation} where in this case, according to the different prescriptions, the constant $\zeta $ takes the values
\begin{equation} \zeta= \begin{cases}
0 & \text{Retarded },\\
1 & \text{Advanced }, \\
\frac{1}{2} & \text{Principal value }.
\end{cases} 
\end{equation}
We now need to fix a prescription in order to proceed further. In the retarded prescription $II_{A}$ is vanishing, while $I_{A}$ becomes, from \eqref{1aa}
\begin{equation}I_{A}=\frac{ie^{2}}{(2\pi)^{2}}\frac{xP^{+}}{\left[(1-x)M^{2}+\pep{k}^{2}\right]}\frac{\delta(1-x)}{P^{+}}\equiv \frac{ie^{2}}{(2\pi)^{2}}\frac{1}{\pep{k}^{2}}\delta(1-x) \; . \end{equation} If we were to fix the advanced prescription, instead, we would find that $I_{A}$ is zero, while $II_{A}$ would read
\begin{equation} II_{A}= \frac{-ieP^{+}}{(2\pi)^{2}}\int dx' \frac{x'}{\left[(1-x')^{2}M^{2}+\pep{k}^{2}\right]}\frac{\delta(1-x')}{P^{+}}\delta(1-x)=-\frac{ie}{(2\pi)^{2}}\frac{1}{\pep{k}^{2}}\delta(1-x) \; . \end{equation}
In the principal-value prescription, instead, the sum of $I_{A}$ and $II_{A}$ gives zero.  

We may summarize our results by stating that the contribution of the transverse gauge link to the TMD $f_{1}^{e}\xkq$, in the different prescriptions, is
\begin{align} &\text{Retarded: } \quad \frac{ie^{2}}{(2\pi)^{2}}\frac{1}{\pep{k}^{2}}\delta(1-x) \; , \non \\
&\text{Advanced: } \quad -\frac{ie^2}{(2\pi)^{2}}\frac{1}{\pep{k}^{2}}\delta(1-x) \; , \non \\
&\text{Principal value: } \quad 0 \; . \label{tglresult}\end{align}
Adding the contributions in \eqref{tglresult} to the results \eqref{f1ret}, \eqref{f1adv} and \eqref{f1pv} in the corresponding prescriptions, we can see that the transverse gauge link allows one to compensate for the extra terms that come from specific choices of a prescription for the regularization of the pole in the polarization sum, thus recovering the results obtained in the Feynman gauge. Similar results can be derived also for the other TMDs. We conclude that the transverse gauge link makes the evaluation of TMDs prescription independent in the light-cone gauge.

\section{Photon TMDs} \label{secphot}

\begin{figure}[t]
\includegraphics[height=3.5cm]{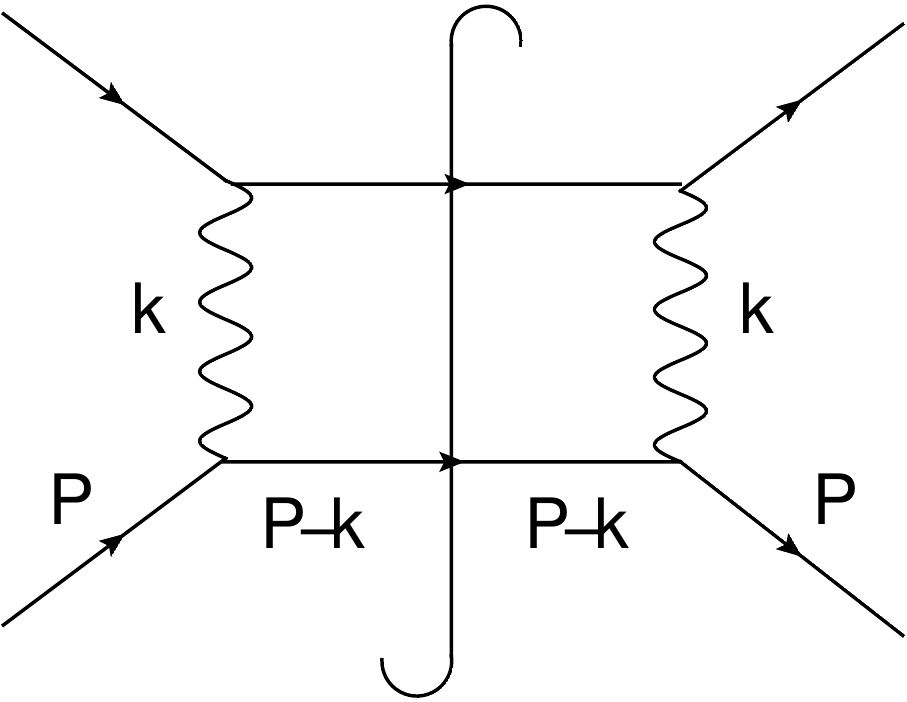}
\caption{\footnotesize{Handbag diagrams related to the TMDs of the photon.
}}
\label{handph}
\end{figure}

We now focus on the derivation of the TMDs for the photon inside the dressed electron. We still consider the scattering of the electron off a lepton, with the situation this time described by the handbag diagram in Fig.~\ref{handph}. 
We will now indicate with $k=(xP^{+},k^{-},\pep{k})$ the four-momentum carried
by the photon. In analogy with gluon TMDs in the QCD case
\cite{Mulders:2000sh,Goeke:2006ef,Meissner:2007rx,Collins:2008sg},  the relevant correlator between initial and final electron states for photon TMDs is\footnote{Indices $i$ and $j$ here refer to transverse 
Lorentz components.} 
\begin{equation} \Phi^{\gamma[ij]}(k;P,S)
  :\,=\frac{1}{xP^{+}}\int\frac{d\xi^{-}d^{2}\pepgr{\xi}}{(2\pi)^{3}}\,e^{ik\cdot\xi}\langle
  P,S\lvert F^{+j}(0)F^{+i}(\xi)\rvert P,S\rangle \bigg \rvert_{{\xi^{+}=0}} \; . \label{phcorr}\end{equation}

The field tensor $F^{\mu\nu}$ in QED is already invariant under the gauge
transformation \eqref{gtran}; therefore in this case a Wilson line is not
required. The situation is different in QCD, where the field tensor is not
gauge invariant and gluon self-interactions are present. 

If we introduce the symmetry operator $\hat{\mb{S}}$, whose action on an
operator $O^{ij}$ is defined by
\begin{equation} 
\hat{\mb{S}}O^{ij}=\frac{1}{2}\left(O^{ij}+O^{ji}-\delta_{i,j}\sum_{m}O^{mm}\right) \; , 
\end{equation}
we have the following decomposition of the correlator \eqref{phcorr} in terms of the leading-twist  TMDs of a photon inside the dressed electron, which are obviously functions of $\xkq$ \cite{Meissner:2007rx}:
\begin{align}
\sum_{i=1}^{2}\Phi^{\gamma[ii]}\xks=&f_{1}^{\gamma}-\frac{\epsilon_{\perp}^{ij}k_{\perp}^{i}S_{\perp}^{j}}{m}f_{1T}^{\perp\gamma} \; \label{tmdp1} ,\\
i\epsilon_{\perp}^{ij}\Phi^{\gamma[ij]}\xks=&S_{z}g_{1L}^{\gamma}+\frac{\pep{k}\cdot\pep{S}}{m}g_{1T}^{\gamma} \; , \label{tmdp2}\\
-\hat{\mb{S}}\Phi^{\gamma[ij]}\xks=&-\frac{\hat{\mb{S}}k_{\perp}^{i}k_{\perp}^{j}}{2m^{2}}h_{1}^{\perp\gamma}+S_{z}\frac{\hat{\mb{S}}k_{\perp}^{i}\epsilon_{\perp}^{jk}k_{\perp}^{k}}{2m^{2}}h_{1L}^{\perp\gamma}+\frac{\hat{\mb{S}}k_{\perp}^{i}\epsilon_{\perp}^{jk}S_{\perp}^{k}}{2m}\left(h_{1T}^{\gamma}+\frac{\pep{k}^{2}}{2m^{2}}h_{1T}^{\perp\gamma}\right) \non \\
&+\frac{\hat{\mb{S}}k_{\perp}^{i}\epsilon_{\perp}^{jk}(2k_{\perp}^{k}\pep{k}\cdot\pep{S}-S_{\perp}^{k}\pep{k}^{2})}{4m^{3}}h_{1T}^{\perp\gamma} \; , \label{tmdp3}
\end{align}
where the antisymmetric tensor $\epsilon_{\perp}^{ij}$ was defined in
\eqref{antitens}. For the photon TMDs we adopt  the nomenclature of
Ref.~\cite{Meissner:2007rx}: the letters $f$, $g$ and $h$ refer to the photon
being unpolarized, circularly polarized or linearly polarized, respectively,
while $L$ and $T$ still refer to the dressed electron being longitudinally or
transversely polarized. Out of the eight photon TMDs implicitly defined in
equations \eqref{tmdp1}-\eqref{tmdp3}, four are T-even
($f_{1}^{\gamma},g_{1L}^{\gamma},g_{1T}^{\gamma}$ and $h_{1}^{\perp\gamma}$)
and four T-odd ($f_{1T}^{\perp\gamma},h_{1L}^{\perp\gamma},h_{1T}^{\gamma}$
and $h_{1T}^{\perp\gamma}$).

If we work in the light-cone gauge, the components of the field tensor
entering the correlator \eqref{phcorr} 
reduce to $F^{+i}=\partial^{+}A^{i}$. If we
also switch to the light-front helicity basis and apply the LFWF overlap
representation method, we come up with
\begin{equation} \Phi^{\gamma[ij]}(x,\pep{k},\Lambda)=\frac{1}{2(2\pi)^{3}}\sum_{\lambda,\lambda_{\gamma},\lambda_{\gamma}'}\Psi^{\Lambda\;^{*}}_{\lambda,\lambda_{\gamma}'}\left((1-x),-\pep{k}\right)\Psi^{\Lambda}_{\lambda,\lambda_{\gamma}}\left((1-x),-\pep{k}\right)\varepsilon^{j\;^{*}}_{\lambda_{\gamma}'}(k)\varepsilon^{i}_{\lambda_{\gamma}}(k) \; . \label{phcorr2} \end{equation} 
Further development of \eqref{phcorr2} yields\footnote{Notice that the denominator from the photon propagator does not need to be regularized, since we are not integrating over the photon momentum.}
\begin{equation}
 \Phi^{\gamma[ij]}(x,\pep{k},\Lambda)=\frac{\lvert\varphi\left((1-x),\pep{k}^{2}\right)\rvert^{2}}{2(2\pi )^{3}} \bar{u}_{\Lambda}(\mb{P})\left(-\gamma^{j}+\frac{k_{\perp}^{j}}{xP^{+}}\gamma^{+}\right)(\slashed{p}_{e}+m)\left(-\gamma^{i}+\frac{k_{\perp}^{i}}{xP^{+}}\gamma^{+}\right)u_{\Lambda}(\mb{P}) \; , \label{phcorr3} 
\end{equation} where the momentum of the inner electron is now $\mb{p}_{e}=((1-x)P^{+},-\pep{k})$. The Feynman gauge result is completely equivalent, the difference in the polarization vectors being compensated by the extra $\partial_{i}A^{+}$ term in the field tensor. 

If one takes the proper combinations of helicities for the dressed electron and transverse indices, it is easy to recover from equations \eqref{tmdp1}-\eqref{tmdp3} the following results for the leading-twist photon TMDs:
\begin{align}
&f_{1}^{\gamma}\xkq=\frac{e^{2}}{(2\pi)^{3}}\frac{\pep{k}^{2}\big[1+(1-x)^{2}\big]+m^{2}x^{4}}{x\left[\pep{k}^{2}+m^{2}x^{2}\right]^2} \, , \label{Tmd1f}\\
&g_{1L}^{\gamma}\xkq=\frac{e^{2}}{(2\pi)^{3}}\frac{\pep{k}^{2}(2-x)+m^{2}x^{3}}{\left[\pep{k}^{2}+m^{2}x^{2}\right]^2} \, , \\
&g_{1T}^{\gamma}\xkq=-\frac{2e^{2}}{(2\pi)^{3}}m^{2}\frac{x(1-x)}{\left[\pep{k}^{2}+m^{2}x^{2}\right]^{2}} \; , \\
&h_{1}^{\perp\gamma}\xkq=\frac{2e^{2}}{(2\pi)^{3}}m^{2}\frac{(1-x)}{x\left[\pep{k}^{2}+m^{2}x^{2}\right]^{2}} \; , \\ 
&f_{1T}^{\perp\gamma}\xkq=h_{1L}^{\perp\gamma}\xkq= h_{1T}^{\perp\gamma}\xkq=h_{1T}^{\gamma}\xkq=0 \; . \label{Tmd8f} 
\end{align}

Notice that the T-odd functions are vanishing: this is not surprising, since
there are no contributions that can be interpreted as final-state
interactions. Even if we go up to order $\alpha^{2}$, we find that 
T-odd TMDs still vanish.
This is probably true to all orders, since it is related to the absence of a
gauge link in the correlator.  
The vanishing of the Sivers function up to order $\alpha^2$ 
is also consistent with the sum rule derived by Burkardt \cite{Burkardt:2004ur,Burkardt:2003yg,Goeke:2006ef}, 
since we found that the Sivers
  effect for the electron vanishes both at order $\alpha$ and
  at $\alpha^{2}$. 

\section{Results} \label{results}

\begin{figure}[b]
\centering
\includegraphics[width=0.45\textwidth, keepaspectratio]{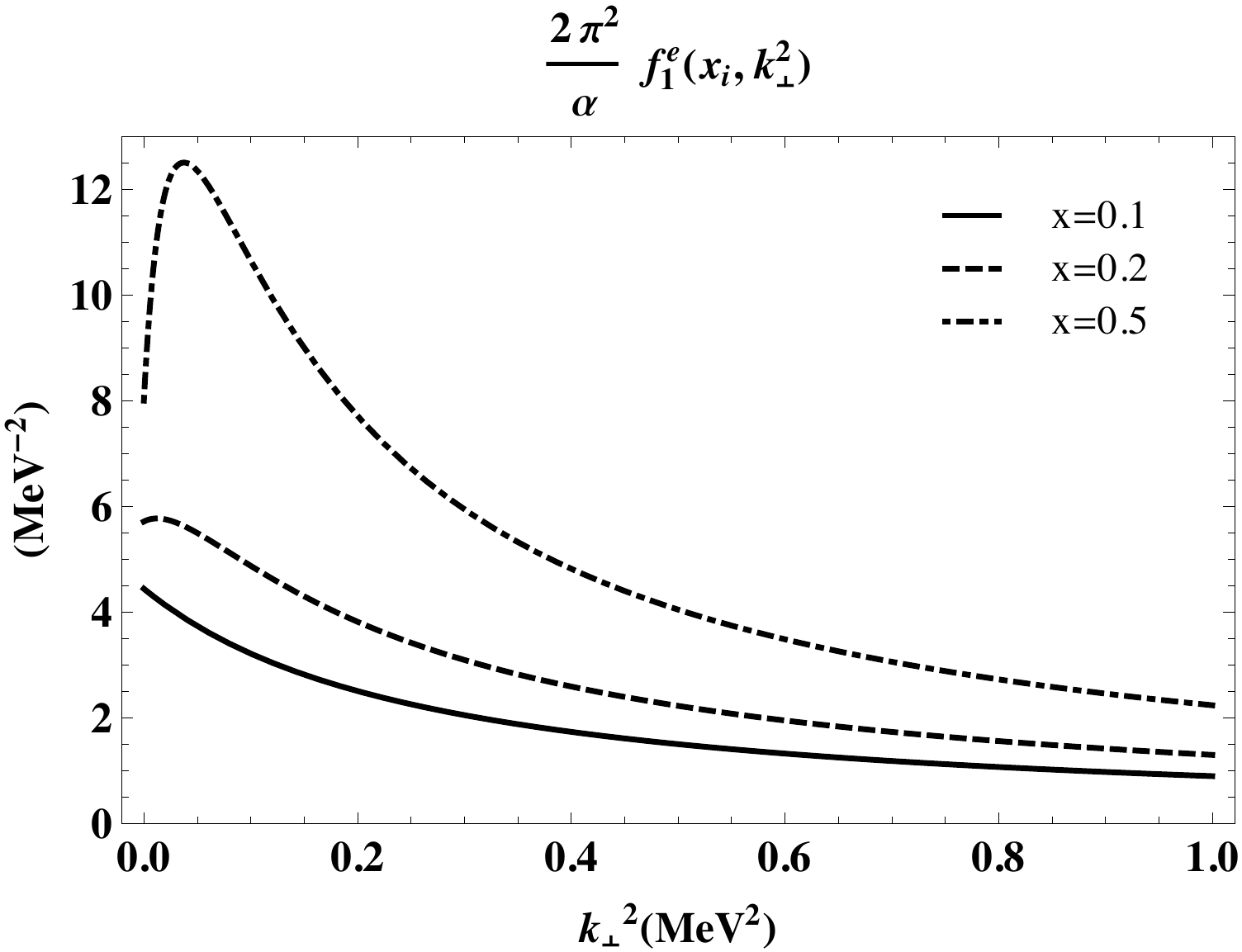}
\hspace{0.5cm}
\includegraphics[width=0.45\textwidth, keepaspectratio]{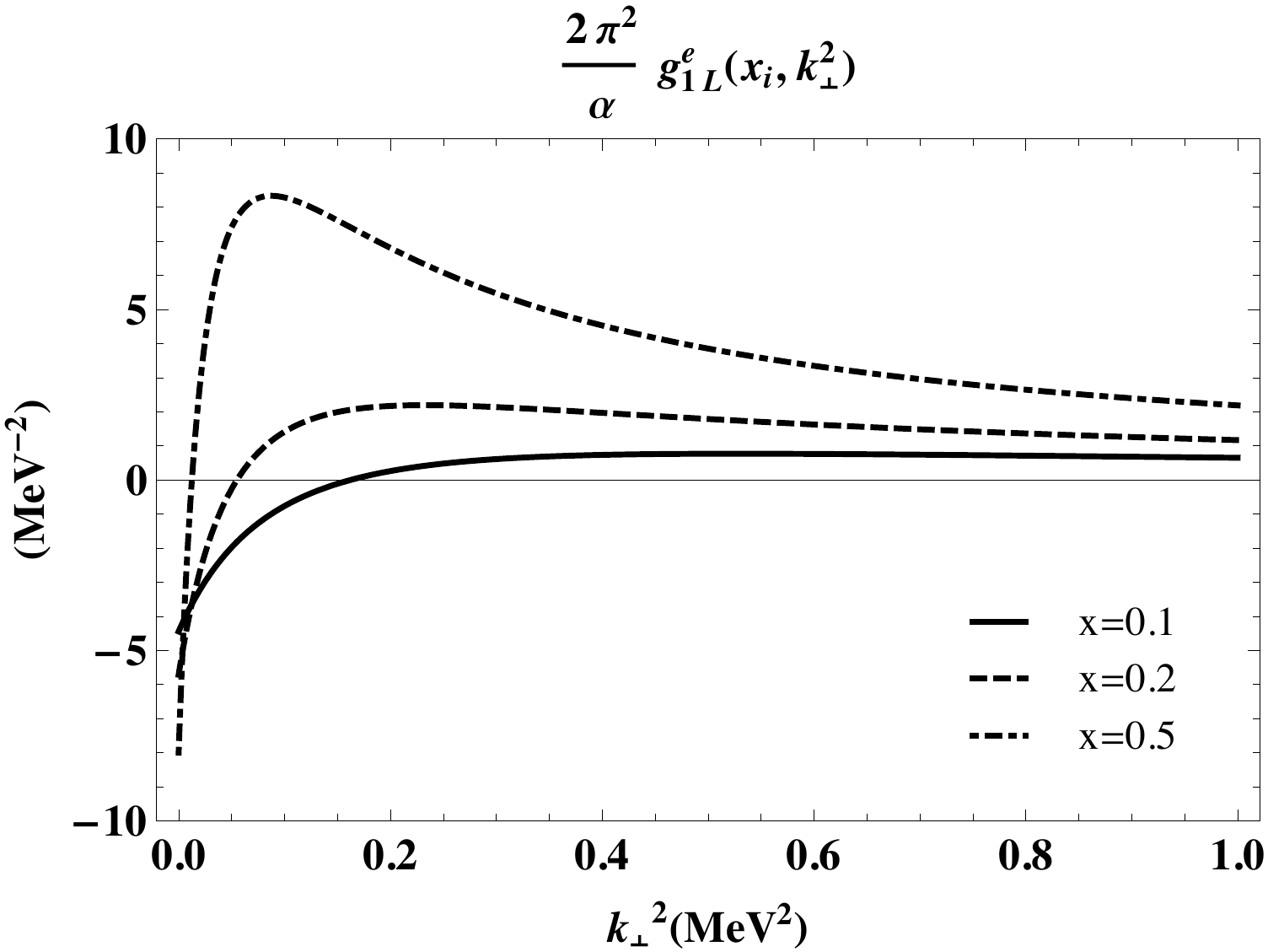}\\
\vspace{0.7cm}
\includegraphics[width=0.45\textwidth, keepaspectratio]{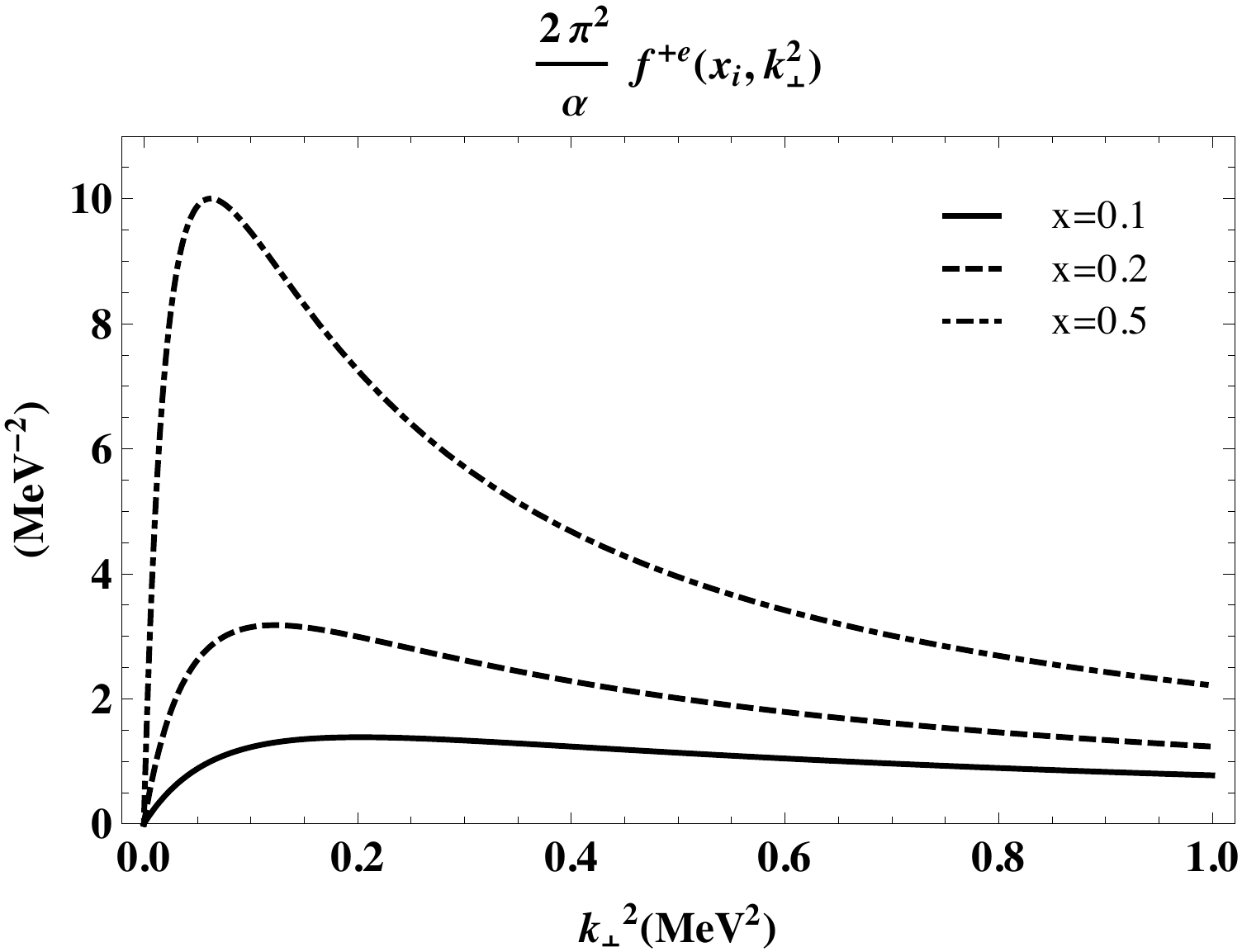}
\hspace{0.5cm}
\includegraphics[width=0.45\textwidth, keepaspectratio]{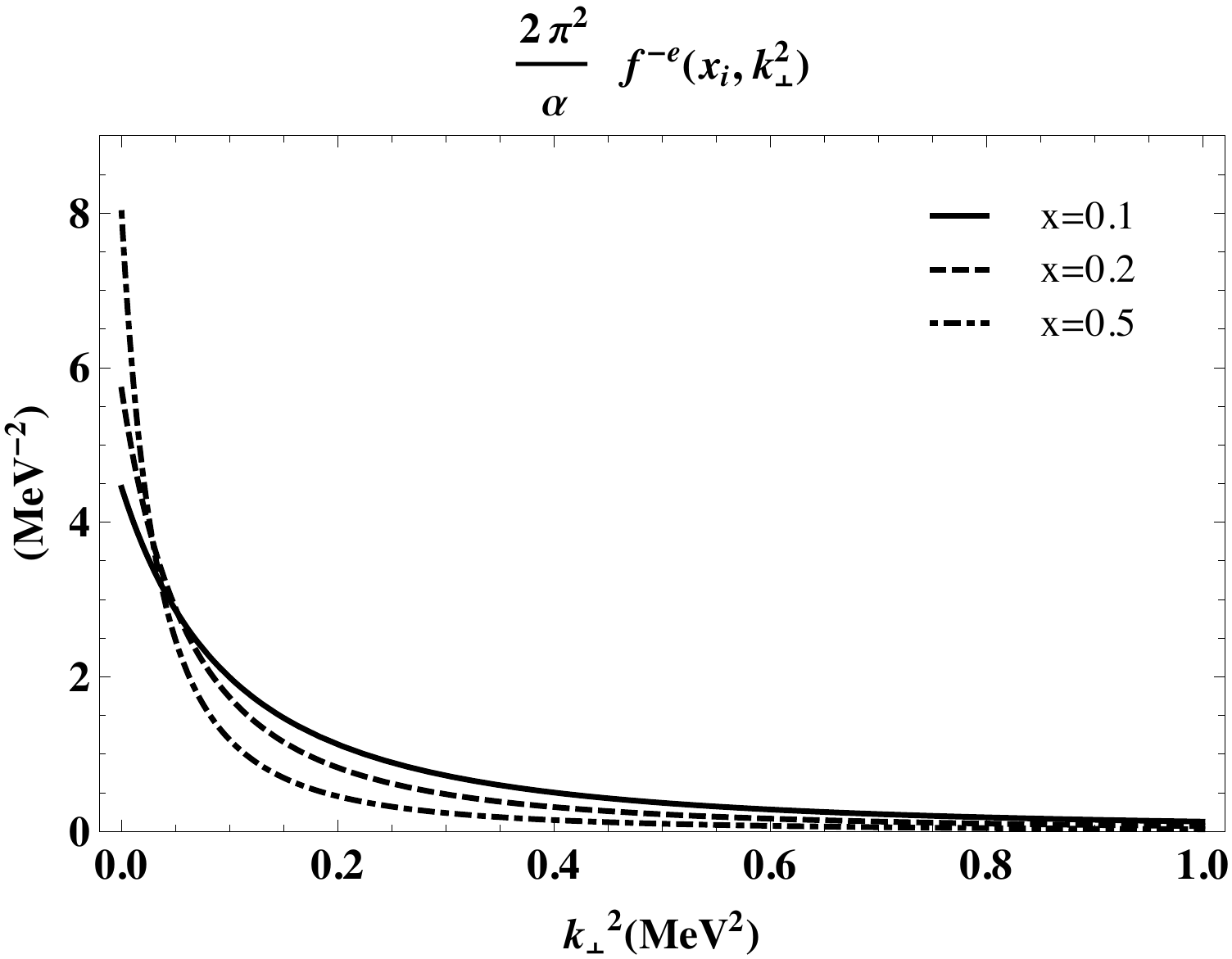}
\caption{\footnotesize{TMDs $f_{1}^{e}\xkq$ (upper-left panel), $g_{1L}^{e}\xkq$ (upper-right panel), along with their sum (lower-left panel) and their difference (lower-right panel). All distributions are rescaled by a factor $2\pi^2/\alpha$, and shown as function of  $\pep{k}^2$, and for different values of the longitudinal momentum fraction $x$: $x=0.1$ (solid curves); $x=0.2$ (dashed curves), and $x=0.5$ (dashed-dotted curves).}}
\label{f1g1plot}
\end{figure}

In this section we discuss the results of our calculation for the electron and photon TMDs at ${\cal O}(\alpha)$, which 
is strictly valid at $\xk\neq (1,\mb{0})$.
In  Fig.~\ref{f1g1plot}, we present the unpolarized TMD $f_{1}^{e}$ (upper-left
panel), and the TMD $g_{1L}^{e}$ (upper-right panel) for longitudinally polarized
electrons in a longitudinally polarized electron, as functions 
of $\bm{k}_{\perp}^{2}$ and for different values of $x$.
In the lower panels of Fig.~\ref{f1g1plot}, we show the corresponding results
for the combinations $f^{+ e}=(f_{1}^{e}+g_{1L}^{e})/2$  and $f^{- e}=(f_{1}^{e}-g_{1L}^{e})/2$,
describing the probability to find the internal electron with spin aligned
and antialigned to the spin of the parent electron, respectively. 
As discussed in Sec.~\ref{sec:overlap}, $f_{1}^e$ and $g_{1L}^{e}$ are given by
different combinations of the squares of S- and P-wave components. 
In particular, $f_{1}^{e}$ is obtained from  the sum of the  contribution from the
two partial waves, while in $g_{1L}^{e}$ one has the difference between the S- and
P-wave contributions. 
Correspondingly, with the combination $f^{+ e}$ we isolate the contribution from P
waves to both $f_{1}^{e}$ and $g_{1L}^{e}$, while  
$f^{- e}$ gives the S-wave contribution which enters with a positive sign in $f_{1}^{e}$
and a negative sign in $g_{1L}^{e}$. 
We notice that for $\bm{k}_{\perp}^{2}\rightarrow 0$, the P-wave contribution
is vanishing, while the S-wave contribution reaches its maximum, with larger
values at increasing  $x$. 
On the other side,  the falloff in $\bm{k}_{\perp}^{2}$ of the S-wave
contribution is very steep, and the contribution of the P wave takes over  at
increasing transverse momentum.  
This also means that, for vanishing transverse momentum, the configuration with the spin of the internal electron being antialigned with the spin of the target electron 
and the spin of the photon being aligned with the spin of the target electron is favored,  and at larger $\bm{k}_{\perp}^{2}$ the spin of the internal electron is most likely to flip in the  direction of the spin of the parent electron. The spin-flip of the active parton at low transverse momentum is also
responsible for the sign change in $g_{1L}^{e}$, which occurs faster in
$\bm{k}_{\perp}^{2}$ at larger values of $x$.

\begin{figure}[t]
\centering
\includegraphics[width=0.45\textwidth, keepaspectratio]{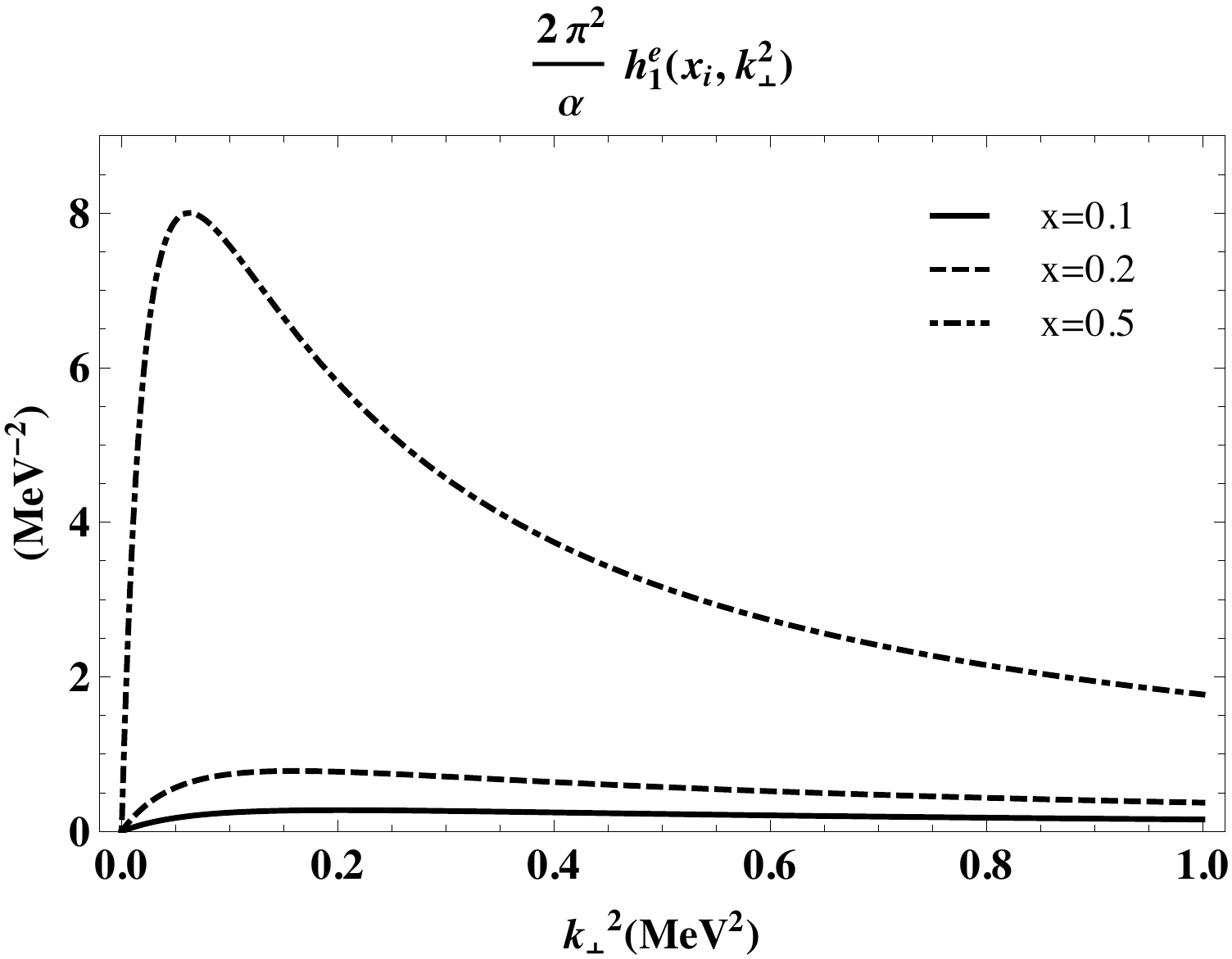}
\\
\vspace{0.7 cm}
\includegraphics[width=0.45\textwidth, keepaspectratio]{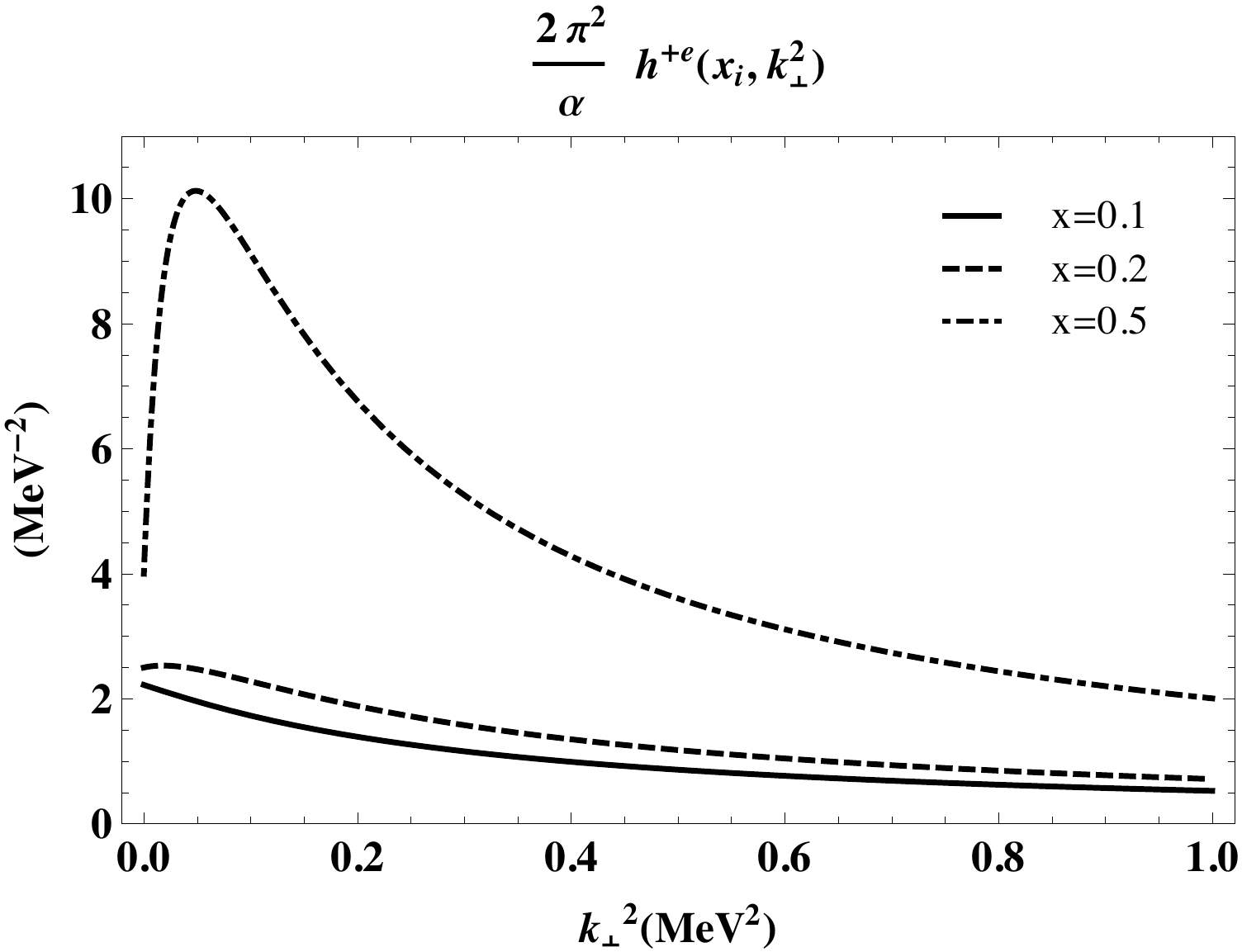}
\hspace{0.5cm}
\includegraphics[width=0.45\textwidth, keepaspectratio]{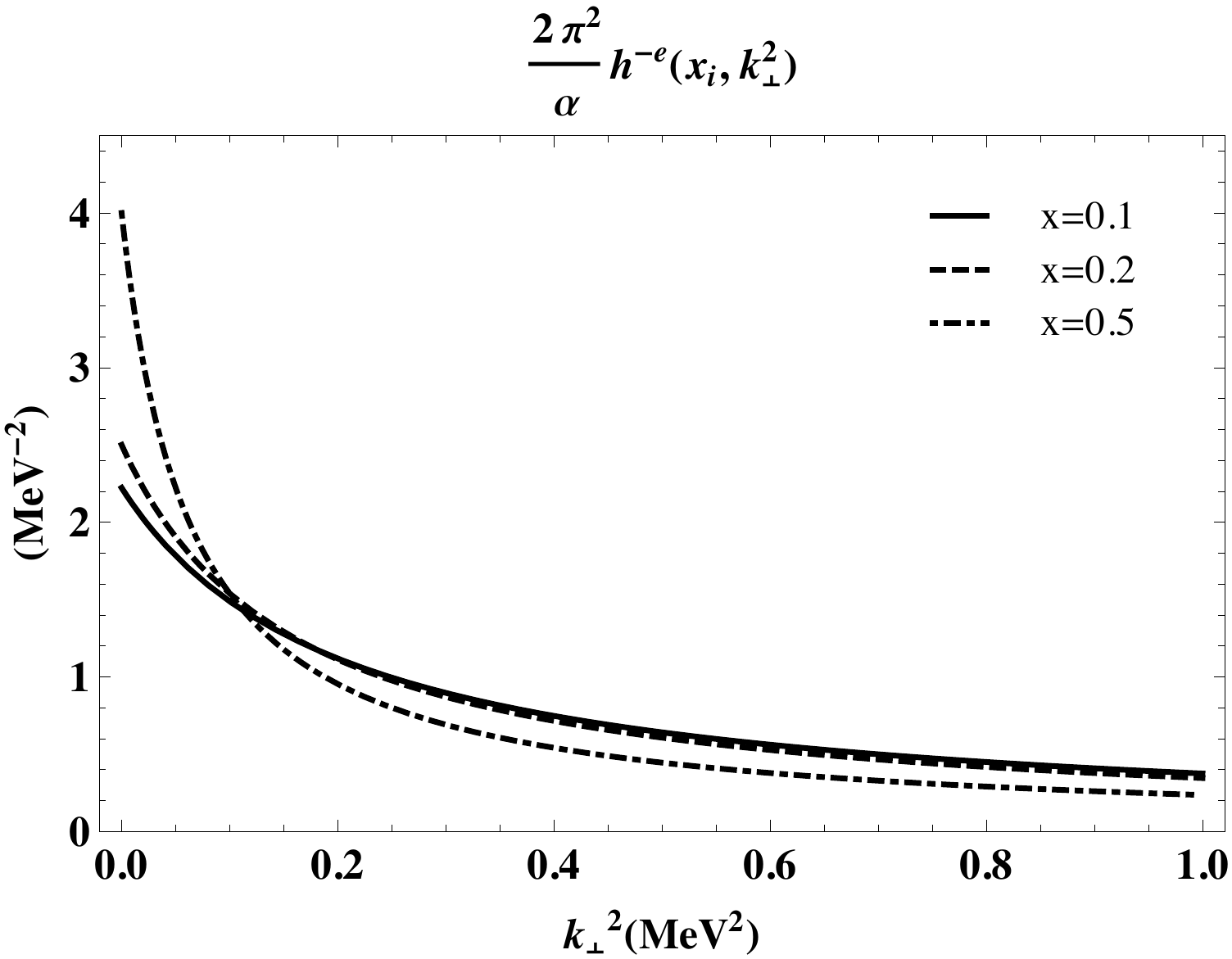}
\caption{\footnotesize{TMDs $h_{1}^{e}\xkq$ (upper panel),  
$h^{+ e}\xkq$ (lower left panel) and $h^{- e}\xkq$ (lower right panel), rescaled by a factor $2\pi^2/\alpha$, as function of  $\pep{k}^2$, and for different values of the longitudinal momentum fraction $x$: $x=0.1$ (solid curves); $x=0.2$ (dashed curves), and $x=0.5$ (dashed-dotted curves).}}
\label{figh1}
\end{figure}

In Fig.~\ref{figh1} we show the results for the transversity TMD $h_{1}^{e}$, and
the combinations $h^{+ e}=(f_{1}^{e}+h_{1}^{e})/2$ and $h^{- e}=(f_{1}^{e}-h_{1}^{e})/2$, describing
the probability of finding the internal electron with transverse spin aligned
and antialigned to the transverse spin of the parent electron, respectively. 
Being chiral odd, $h_{1}^{e}$  involves a helicity-flip of the internal electron
from the initial to the final state, which is compensated by a helicity flip
of the parent electron in same direction. 
Since total helicity is the same in initial and final states, 
$h_{1}^{e}$ is diagonal in OAM, and,
from the LFWF overlap
representation in Eq.~\eqref{overlap-6}, we note that it is given in terms of
the  partial waves with $L_z=\pm 1$ only, without contributions from the
S-wave components.  
The P-waves enter in $h_{1}^{e}$  with different combinations with respect to the
contribution in $f_{1}^{e}$ and $g_{1L}^{e}$, as given by $f^{+ e}$. 
In particular, from the LFWF overlap representation in Eqs.~\eqref{overlap-1},
\eqref{overlap-2}, and \eqref{overlap-6}, we find that $f^{+ e}$  contains the sum
of the square of the $L_z=+1$ and $L_z=-1$ components, while $h_{1}^{e}$ is given by
the interference between the two P waves, i.e. 
\begin{align} 
&f^{+ e}\xkq =\frac{1}{2(2\pi)^{3}}\left[|\Psi^{+}_{+,+1}\xk|^{2}+|\Psi^{+}_{+,-1}\xk|^{2}\right] \; ,\label{overlap-1bis} \\
&h_{1}^{e}\xkq=\frac{1}{(2\pi)^{3}}{\rm Re}\left[\Psi^{+\;^{*}}_{+,+1}\xk\Psi^{-}_{-,+1}\xk\right]\; , \label{overlap-6bis} 
\end{align}
where we used the property~\eqref{eq:relation} of the two-body component of the electron LFWFs.
From the analytical expressions in Eqs.~\eqref{Tmd1}, \eqref{TMD2} and \eqref{TMD6}, one also finds
\begin{equation}
h_{1}^{e}\xkq=\frac{2x}{1+x^2}f^{+e}\xkq\;.
\end{equation}
As a result, $h_{1}^{e}$ has the same dependence on   $\bm{k}_{\perp}^{2}$ as $f^{+ e}$,
but its overall size is always smaller, particularly 
at smaller values of $x$.  
In the combination $h^{+ e}$, the positive contribution from the S-wave
components of $f_{1}^{e}$ is evident at vanishing transverse momentum, while at
higher values of $\bm{k}_{\perp}^{2}$ the contributions of the P waves from
$f_{1}^{e}$ and $h_{1}^{e}$ become dominant.  
In the case of $h^{- e}$  the P-wave contributions from $f_{1}^{e}$ and $h_{1}^{e}$
partially cancel, especially at higher values of $x$, and we find that the low-$\bm{k}_{\perp}^{2}$ behavior is dominated by the S-wave contribution of
$f_{1}^{e}$, while the P waves still control  the  tail at higher transverse
momentum. 
As a consequence, at $x=0.5$ 
the transverse spin of the internal electron tends to be
aligned with the transverse spin of the  parent electron in
the full $\bm{k}_{\perp}^{2}$ range, while for smaller values of  $x$ there is
no marked preference between the parallel or antiparallel configuration of the
two transverse spins. 

\begin{figure}[t]
\centering
\includegraphics[width=0.45\textwidth, keepaspectratio]{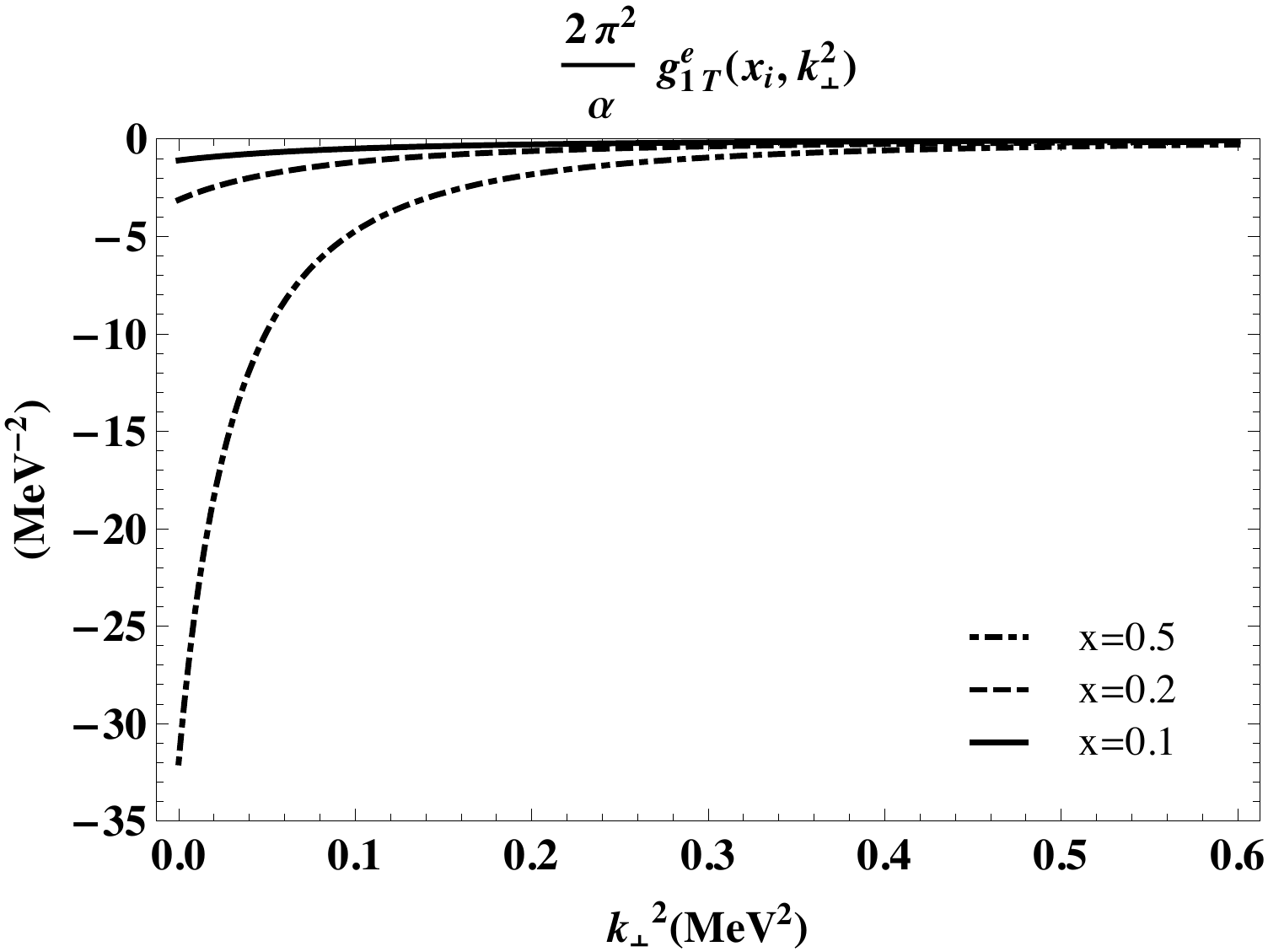}
\hspace{0.5cm}
\includegraphics[width=0.45\textwidth, keepaspectratio]{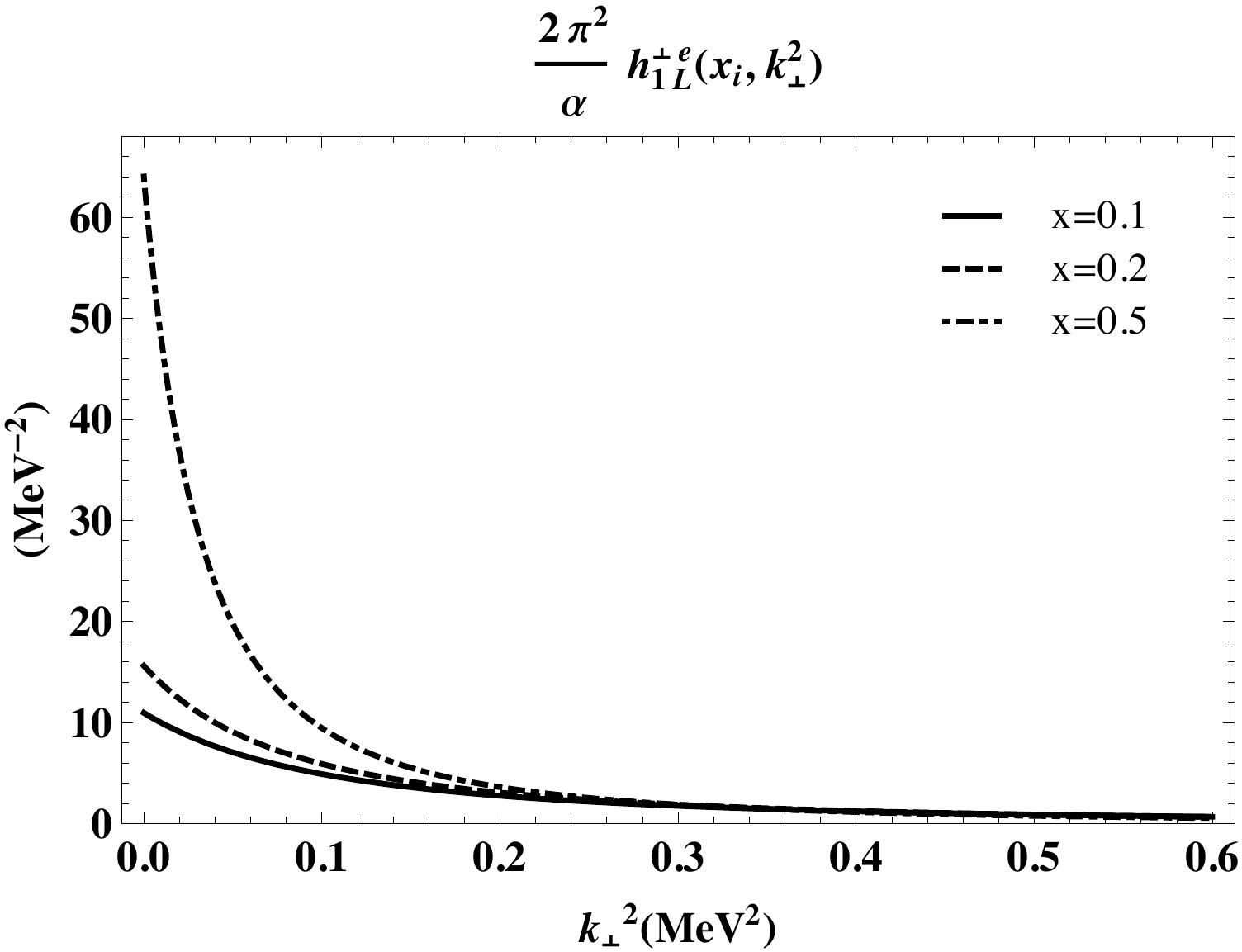}
\caption{\footnotesize{TMDs $g_{1T}^{e}\xkq$ (left panel), $h_{1L}^{e}\xkq$ (right panel),  rescaled by a factor $2\pi^2/\alpha$, as function of \ $\pep{k}^2$, and for different values of the longitudinal momentum fraction $x$: $x=0.1$ (solid curves); $x=0.2$ (dashed curves), and $x=0.5$ (dashed-dotted curves).}}
\label{figg1thil}
\label{tplot}
\end{figure}

In Fig.~\ref{figg1thil} we show the TMD $g_{1T}^{e}$ for longitudinally polarized
electrons in a transversely polarized electrons target in comparison with
$h_{1L}^{\perp e}$, describing the momentum distribution  for transversely
polarized electron in a longitudinally polarized electron target.  
These distributions are characteristic effects due to transverse
momentum, since they are the only ones which have no analog in the spin
densities related to the GPDs in the impact parameter
space~\cite{Diehl:2005jf}, and vanish after integration over $\bm{k}_{\perp}$.  
Since the  LFWFs $\Psi^{+}_{-,-1}$ and $\Psi^{-}_{+,+1}$ vanish, 
the LFWF overlap representation in Eqs.~\eqref{overlap-3} and \eqref{overlap-4}
can be rewritten as   
\begin{align}
&
g_{1T}^{e}\xkq=\frac{m}{2(2\pi)^{3}\pep{k}^{2}}
{\rm Re}
\left[k_{R}\, \Psi^{+\;^{*}}_{+,-1}\xk\Psi_{+,-1}^{-}\xk\right] \, , \label{overlap-3bis}\\
&
h_{1L}^{\perp e}\xkq=\frac{m}{(2\pi)^{3}\pep{k}^{2}}{\rm Re}\left[k_{R}\,\Psi_{-,+1}^{+\;^{*}}\xk\Psi_{+,+1}^{+}\xk\right] \; .\label{overlap-4bis} 
\end{align}
We notice that $g_{1T}^{e}$ requires a helicity flip of the  electron target that is not compensated by a change of the helicity of the internal electron,
and vice versa $h_{1L}^{\perp e}$ involves a helicity flip of the internal electron, but is diagonal in the helicity of the electron target. As a result, 
the two distributions require a transfer of OAM between the initial and final states: in the case of $g_{1T}^{e}$, this comes from the interference of the S wave and the P wave with $L_z=+1$, while 
for $h_{1L}^{\perp e}$ it is driven from the S wave and the P wave with $L_z=-1$.
The difference between the two P waves in Eqs.~\eqref{lcwf1} and \eqref{lcwf2}
leads to the following relations for the TMDs:
\begin{equation}
g_{1T}^{e}\xkq=-x h_{1L}^{\perp e}\xkq\; ,
\end{equation}
which explains the suppression of $g_{1T}^{e}$, in absolute value, with respect to $h_{1L}^{\perp e}$, especially at lower values of $x$.

From Eqs.~\eqref{tmd1}-\eqref{tmd3} and omitting  the contributions of $f_{1T}^{\perp e}$, $h_{1}^{\perp e}$ and $h_{1T}^{\perp e}$, which are vanishing at leading order in $\alpha$,
we can form the following  densities for electrons of definite longitudinal or transverse polarization:
\begin{eqnarray}
\rho(x,k_x,k_y,(\lambda,\bm{s}_\perp),(\Lambda,\bm{S}_\perp))
 = \frac{1}{2} \Bigg[ f_{1}^{e}
   + \lambda \Lambda\, g_{1L}^{e}
   + \lambda\, S_\perp^i k_\perp^i \frac{1}{m}\, g_{1T}^{e} 
   +  \Lambda\, s^i_\perp k_\perp^i \frac{1}{m}\, h_{1L}^{\perp e}
    + s^i_\perp \, S_\perp^i \, h_{1}^{e}
   \Bigg]\; ,
   \label{eq:rho}
   \end{eqnarray}
where we denote the transverse spin of the internal electron with  $s_\perp$.

\begin{figure}[t]
\centering
\includegraphics[width=0.45\textwidth, keepaspectratio]{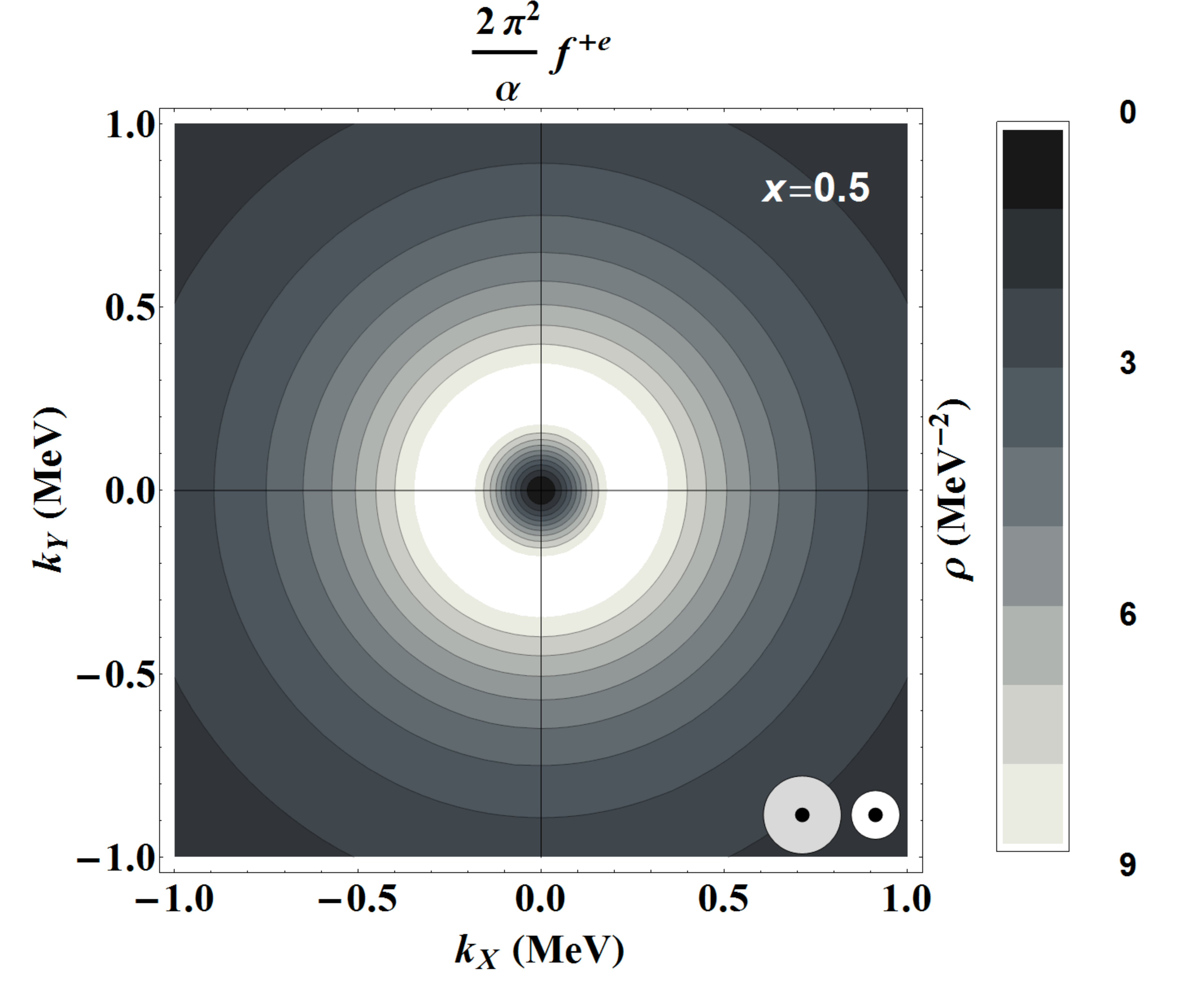}
\vspace{0.5cm}
\includegraphics[width=0.45\textwidth, keepaspectratio]{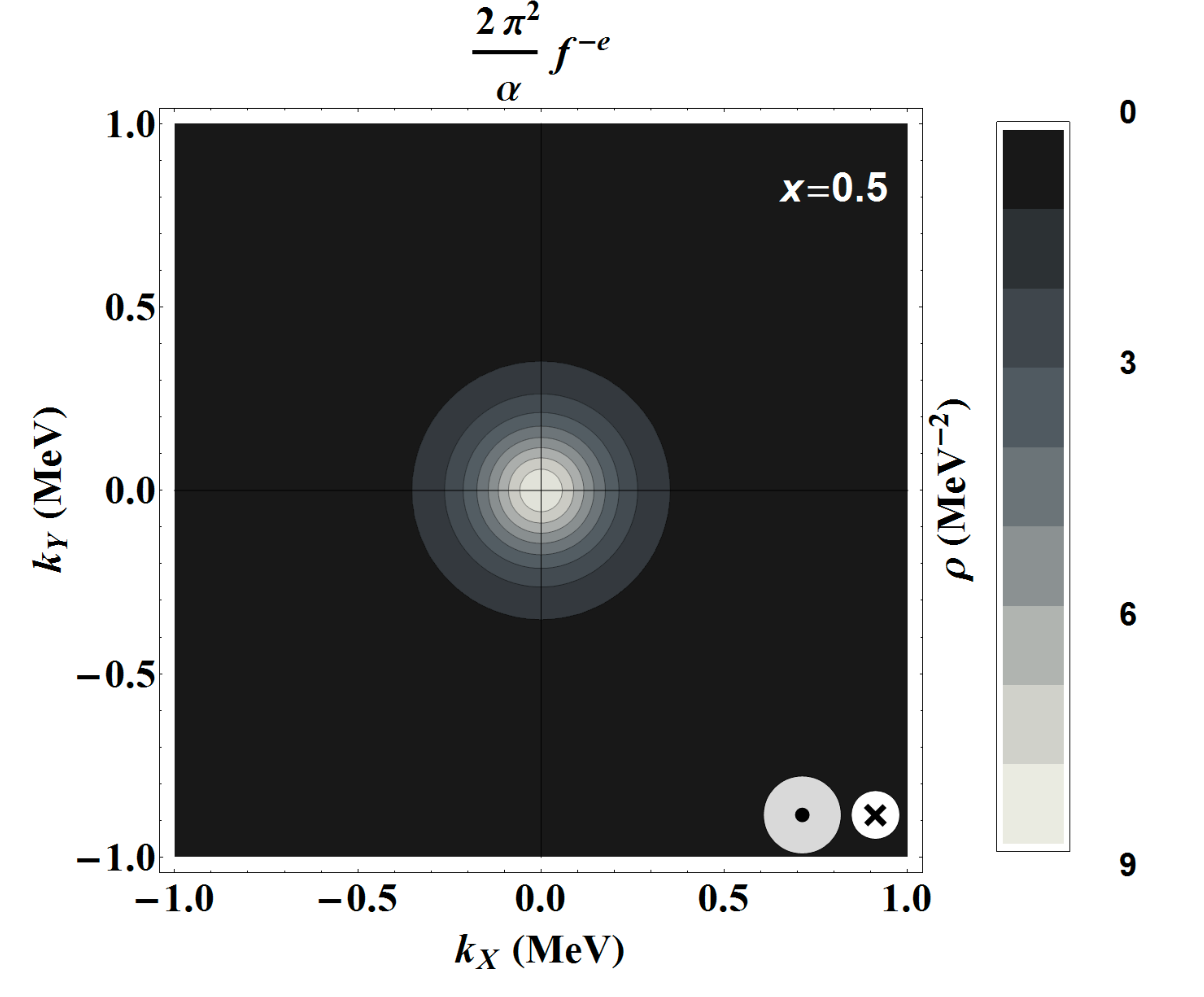}
\vspace{0.2cm}
\includegraphics[width=0.45\textwidth, keepaspectratio]{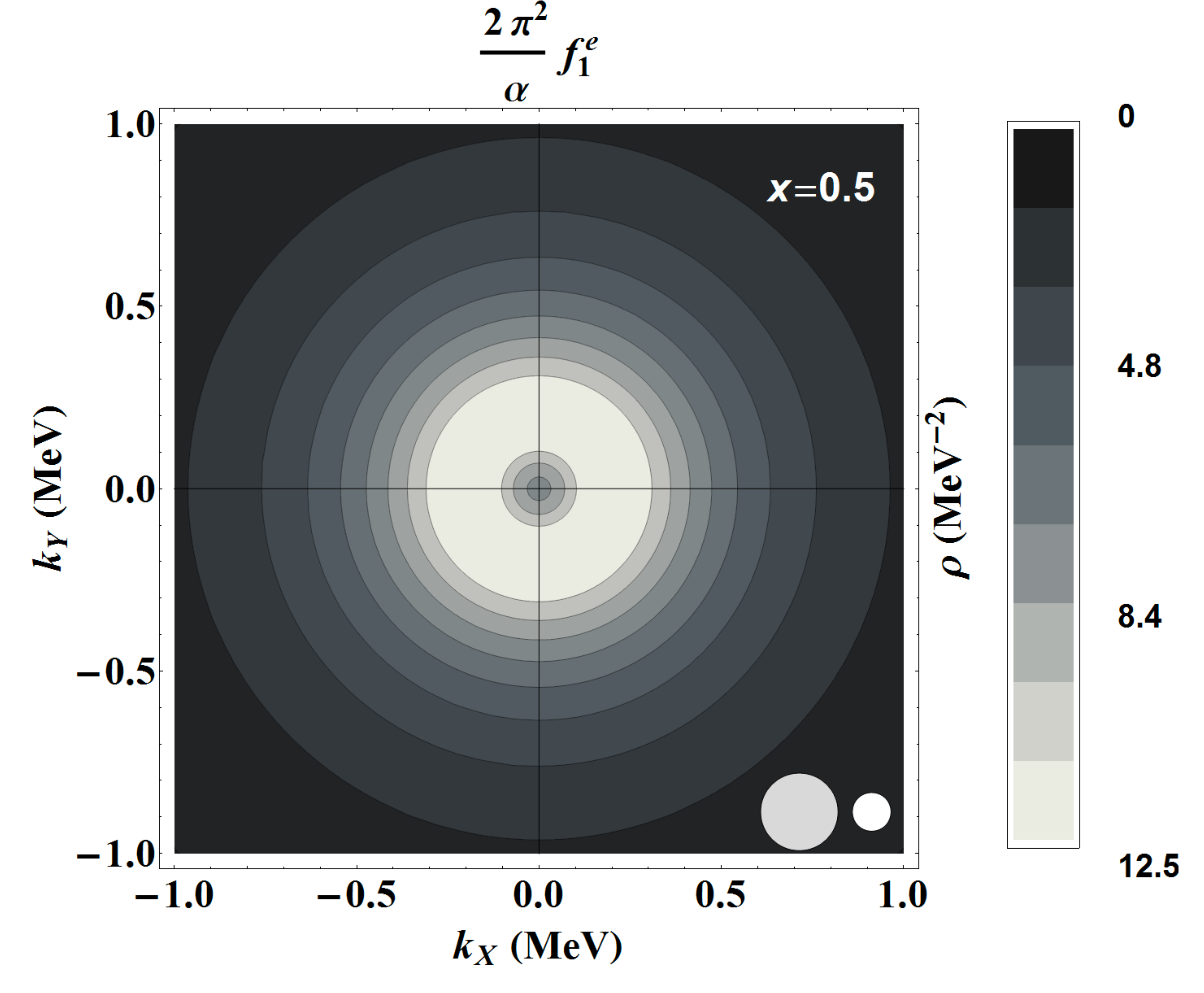}
\caption{\footnotesize{Density plots in the transverse momentum plane $k_{x}-k_{y}$ and at fixed value of $x=0.5$ for $f^{+ e}$ (upper left panel), $f^{- e}$ (upper right panel)  and $f_{1}^{e}$ (lower panel), rescaled by a factor $2\pi^2/\alpha$.
  The legend in the bottom-right corner of each panel indicates the corresponding spin configurations: the grey and white empty discs refer to the unpolarized dressed and unpolarized bare electron, respectively; 
the  circle and cross inside the discs stand for polarization along the longitudinal axes, in opposite directions. }}
\label{density-f1}
\end{figure}

In Fig.~\ref{density-f1} we show the densities in the transverse momentum plane and at fixed $x=0.5$ for a longitudinally polarized electron target and longitudinally polarized internal electron, 
with helicity in the same ($f^{+ e}$) or opposite ($f^{- e}$) direction, along with the corresponding results for $f_{1}^{e}$. 
As discussed above, the behavior of $f^{+e}$  is determined from the P waves,
which rapidly increase starting from the dip in the center, reach a
maximum at $|\bm{k}_{\perp}|\approx 200$ keV and smoothly decrease at larger
values. 
On the other hand, the S-wave contribution  to $f^{- e}$ is maximum at the centre
and rapidly falls off towards the periphery of the $k_x$-$k_y$ plane.
The interplay between the P-and S-wave contributions determines the pattern of
$f_{1}^{e}=f^{+ e}+f^{- e}$ from the center to higher values of $\bm{k}_\perp$. In summary,
the density of electrons in a dressed electron in momentum space, averaged
over all polarizations, at $x=0.5$, looks like a ring-shaped image, with a
radius of about 200 keV. 

\begin{figure}[h]
\centering
\includegraphics[width=0.45\textwidth, keepaspectratio]{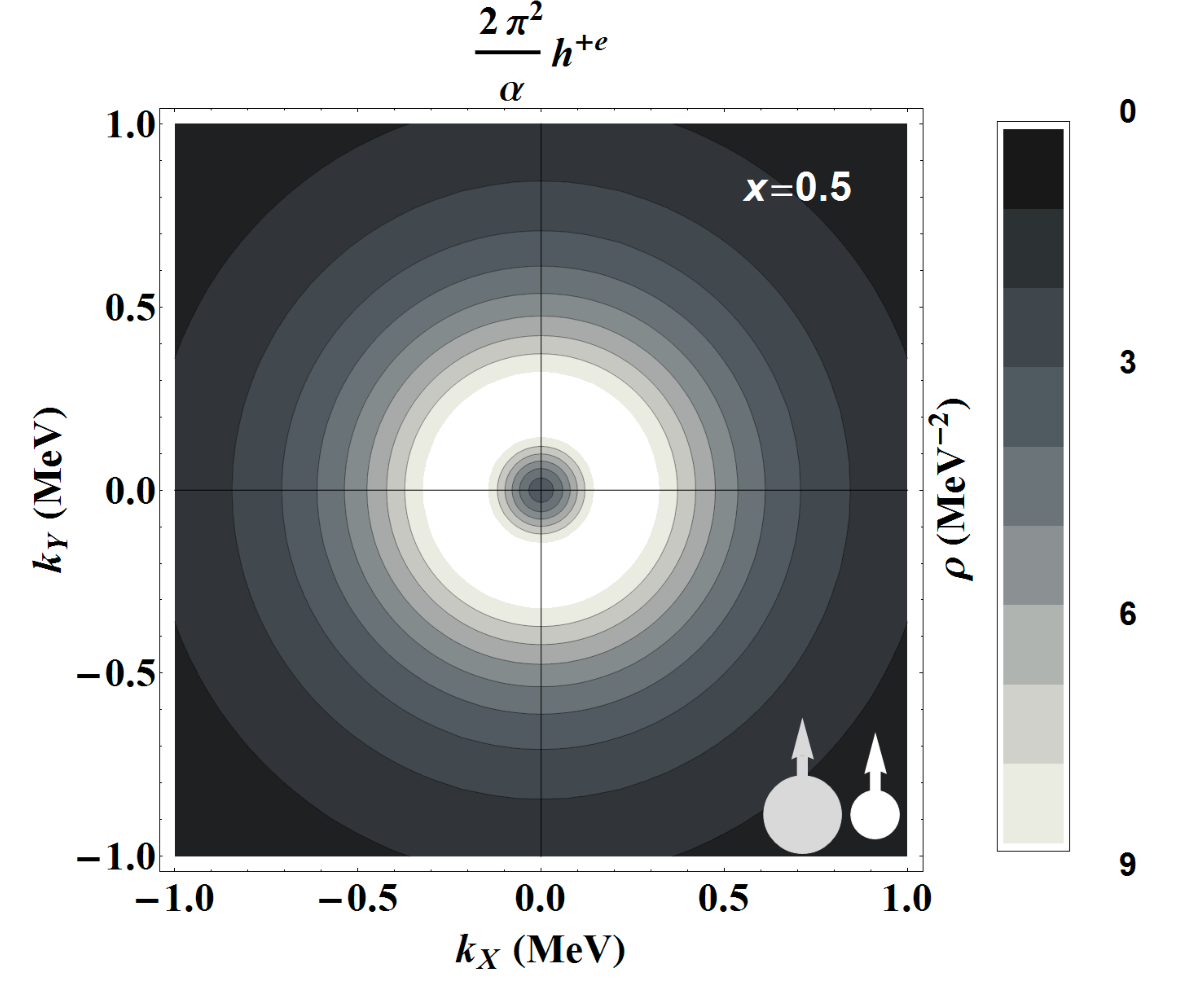}
\hspace{0.5cm}
\includegraphics[width=0.45\textwidth, keepaspectratio]{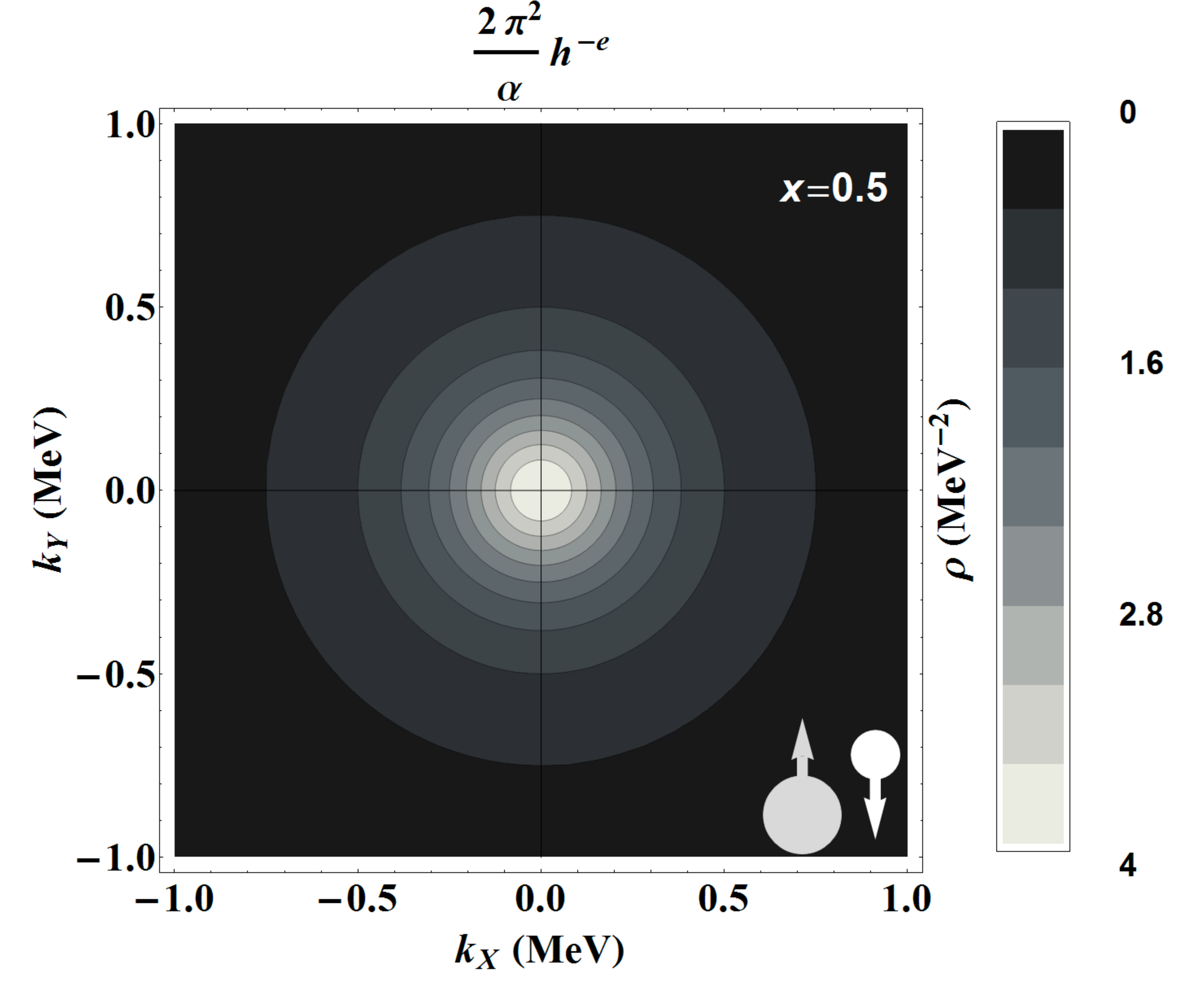}
\caption{\footnotesize{Density plots in the transverse momentum plane $k_{x}-k_{y}$ and at fixed value of $x=0.5$ for $h^{+ e}$ (left panel) and $h^{-e}$ (right panel), rescaled by a factor $2\pi^2/\alpha$.
  The legend in the bottom-right corner of each panel indicates the corresponding spin configurations: the grey and white discs refer to the  dressed and  bare electron, respectively;  the arrows indicate polarization along the $y$ direction.}}
\label{density-h1}
\end{figure}

In Fig.~\ref{density-h1} we show the transverse polarization densities $h^{+ e}$
and $h^{- e}$. Qualitatively, they look similar to their longitudinal
counterparts, but they are different in the details. The P-wave contribution
is more pronounced in $h^{+ e}$, but the S wave is also present, and the
density at $\bm{k}_{\perp}^{2}=0$ is not zero. The S wave dominates in $h^{- e}$,
but a P-wave contribution is also present, even though it is suppressed. 
Note that, in principle, the transverse polarization density could
in principle be not cylindrically symmetric. However, the fact that
$h_{1T}^{\perp e}$ and $h_{1}^{\perp e}$ vanish in QED at this order makes the density
symmetric.  

\begin{figure}[t]
\centering
\includegraphics[width=0.45\textwidth, keepaspectratio]{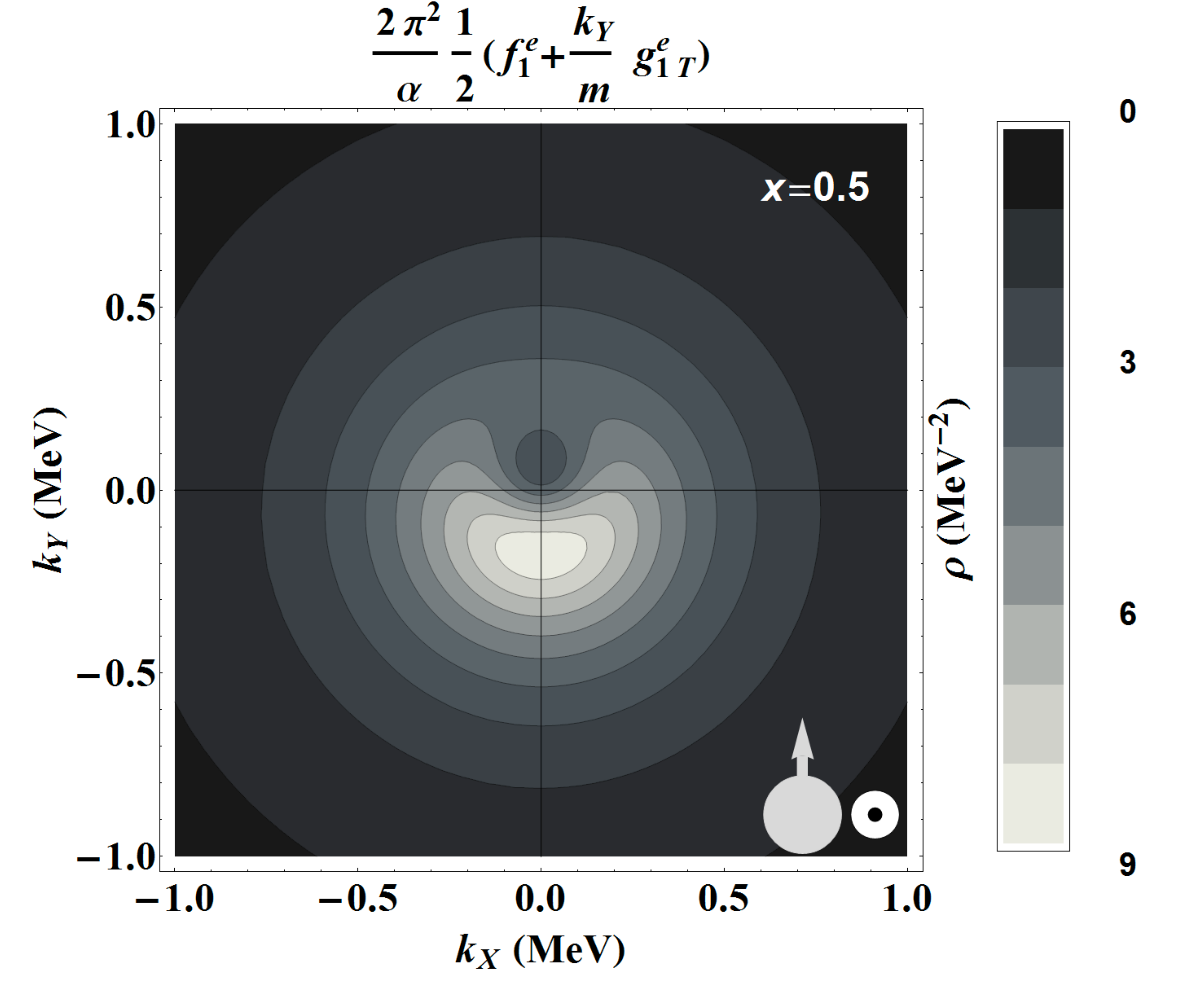}
\hspace{0.5cm}
\includegraphics[width=0.45\textwidth, keepaspectratio]{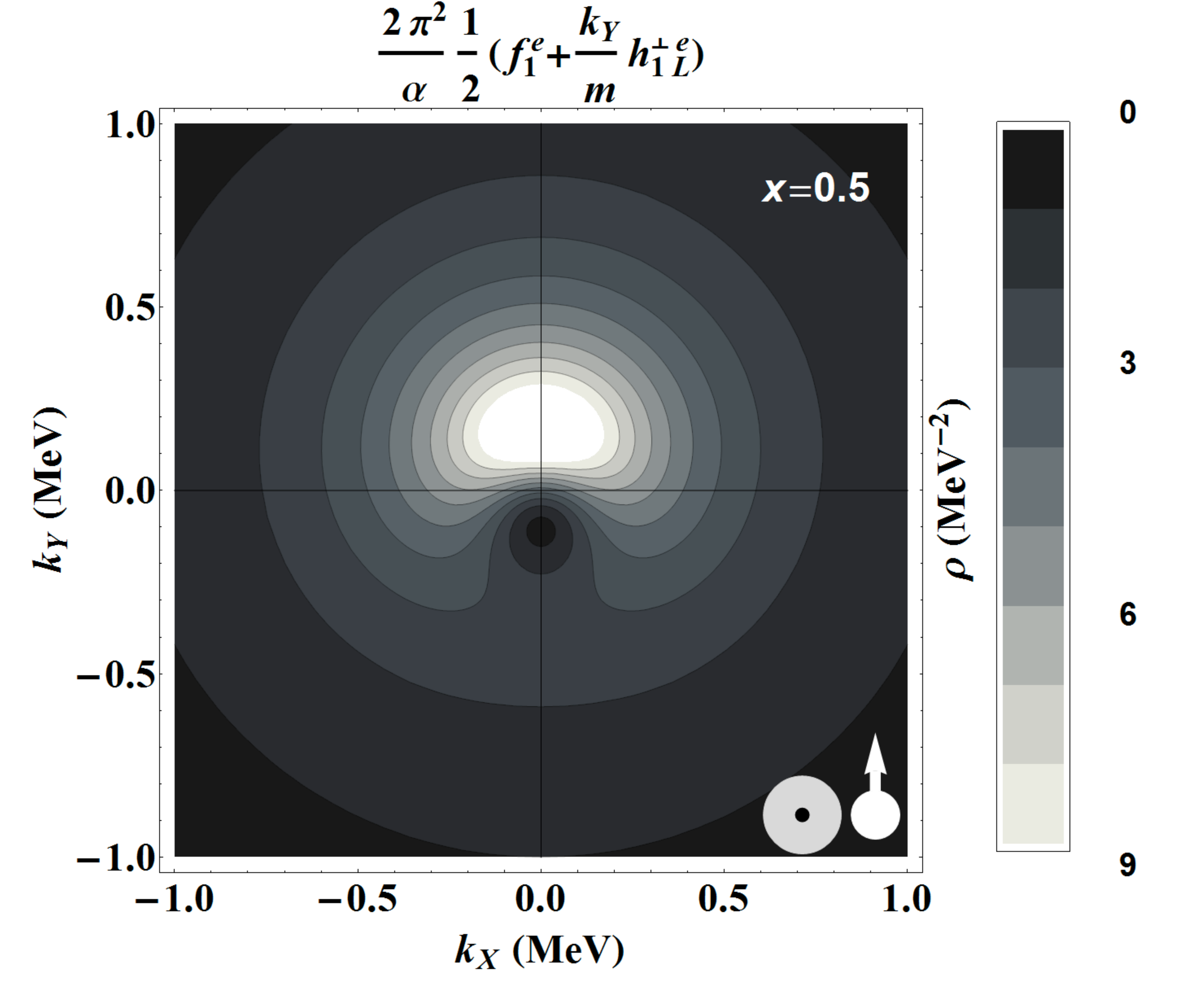}
\caption{\footnotesize{Density plots  in the transverse momentum plane $k_{x}-k_{y}$ and at fixed value of $x=0.5$ for the sum $(f_{1}^{e}+\tfrac{k_y}{m}\,g_{1T}^{e})/2$ (left panel) and $(f_{1}^{e}+\tfrac{k_y}{m}\,h_{1L}^{\perp e})/2$ (right panel), rescaled by a factor $2\pi^2/\alpha$. The legend in the bottom-right corner of each panel indicates the corresponding spin configurations: the grey and white discs refer to the  dressed and  bare electron, respectively; 
the  circle inside the discs stands for polarization along the longitudinal axes; the arrows indicate polarization along the $y$ direction. }}
\label{density-g1t}
\end{figure}

In Fig.~\ref{density-g1t} the distorting effect induced by the polarization is
shown. In the left panel, we present the density in the transverse momentum
plane and at fixed $x=0.5$ for longitudinally polarized electrons in an
electron target transversely polarized along the $y$ direction. This is
obtained by adding the dipole deformation due to the term $  \tfrac{k_y}{m}\,
g_{1T}^{e} $ in Eq.~\eqref{eq:rho} to the monopole
distribution of $f_{1}^{e}$.  
The term in  $g_{1T}^{e}$ features a significant dipole deformation along the
direction of the spin of the parent electron, and is responsible for the
sizable downward shift along $y$. 
An analogous observation can be made for transversely polarized electrons in a
longitudinally polarized electron target, as displayed in the right panel of
Fig.~\ref{density-g1t}. 
In this case, the effect of the distortion is even more pronounced and in the
opposite direction, since the strength of the deformation is  due to the term
$ \tfrac{k_y}{m}\, h_{1L}^{\perp e} $, with $h_{1L}^{\perp e}=-2 g_{1T}^{e}$ at $x=0.5$.

\begin{figure}[h]
\centering
\includegraphics[width=0.45\textwidth, keepaspectratio]{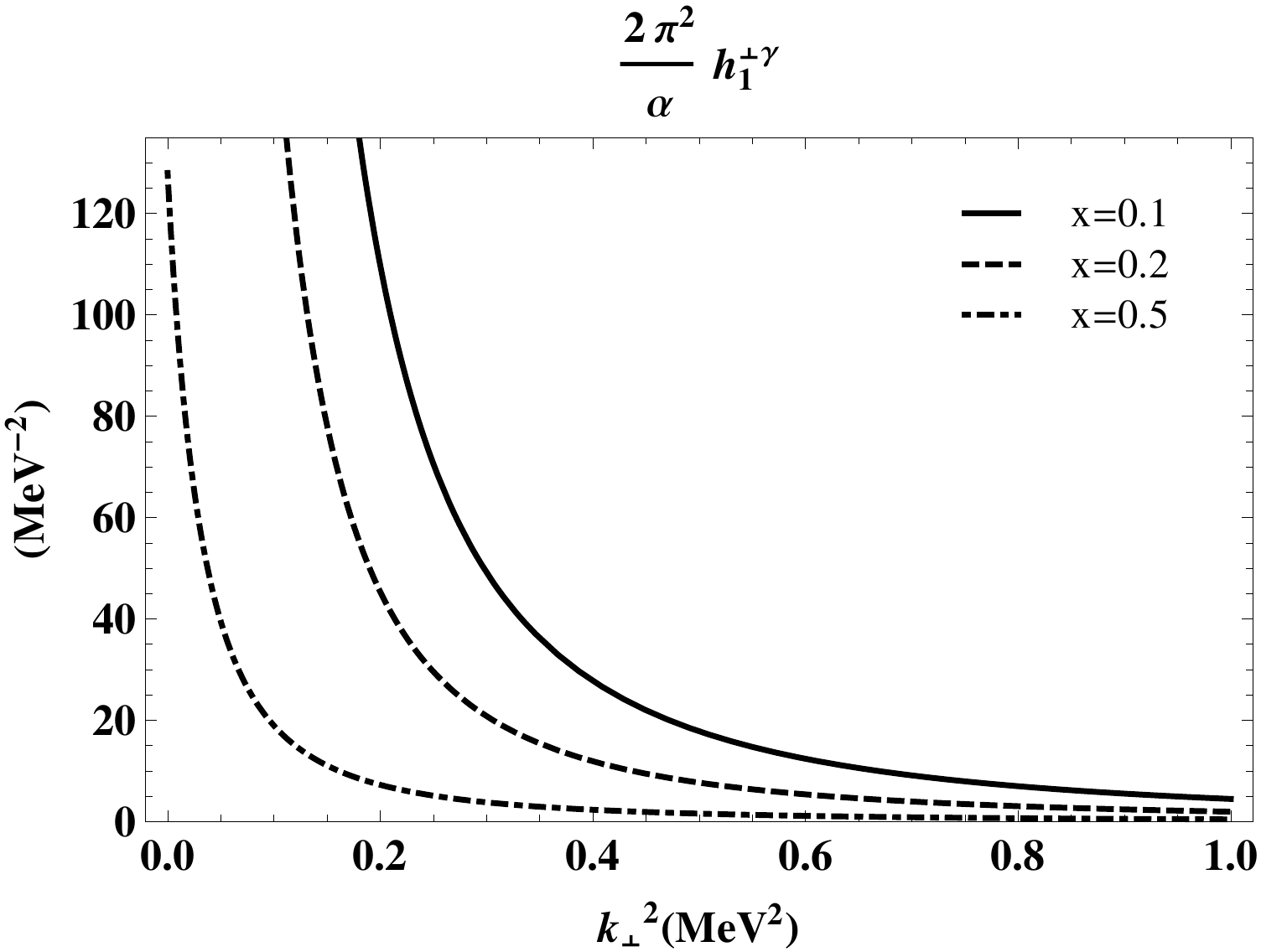}
\hspace{0.5cm}
\includegraphics[width=0.45\textwidth, keepaspectratio]{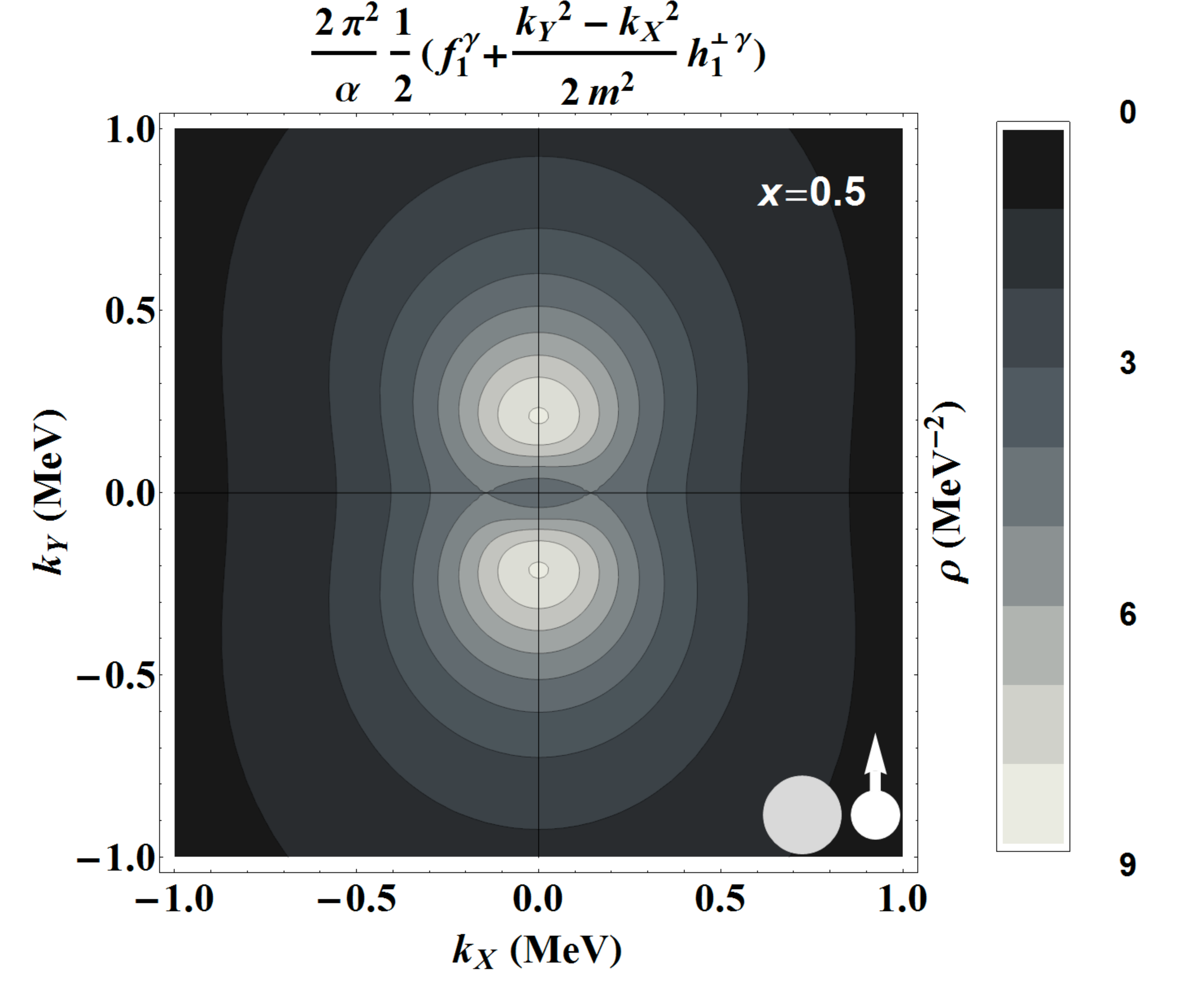}
\caption{\footnotesize{Left panel: photon
TMD $h_{1}^{\perp \gamma}\xkq$, rescaled by a factor $2\pi^2/\alpha$,
as function of  $\pep{k}^2$, and for different values of the longitudinal
momentum fraction $x$: $x=0.1$ (solid curves); $x=0.2$ (dashed curves), and
$x=0.5$ (dashed-dotted curves). Right panel: density plot 
in the transverse momentum plane $k_{x}-k_{y}$ and at fixed value of $x=0.5$
for the combination 
$\big(f_1^{\gamma}+(k_y^2 -k_x^2)/(2m^2) h_1^{\perp \gamma}\big)/2$,
rescaled by a factor $2\pi^2/\alpha$.  The legend in the bottom-right corner
of indicates the corresponding spin configurations: the grey empty disc refers
to the unpolarized dressed electron, the white disc with arrow indicate a
photon with linear polarization along the $y$ direction.
}}
\label{figh1perp}
\end{figure}

We avoid showing plots of all the photon TMDs, but we take the chance to make
a few observations. The functions $f_1^{\gamma}$ and $g_{1T}^{\gamma}$ are
similar to their electron counterparts with the replacement 
$x \leftrightarrow (1-x)$. The helicity distribution has no straightforward
relation to the electron helicity distribution. 
At variance with the
electron case, the combination $f^{- \gamma}$ 
contains contributions only from P waves,
while $f^{+ \gamma}$ contains contributions from both P and S waves.
The other nonvanishing photon TMD is the so-called Boer-Mulders function
$h_{1}^{\perp \gamma}$, describing linearly polarized photons in an unpolarized
electron~\cite{Boer:1997nt,Mulders:2000sh,Meissner:2007rx}. 
We plot it in Fig.~\ref{figh1perp} (left panel), where we observe 
that it is very large, especially at low $x$. The corresponding function for gluons 
in a proton has been the topic of a few articles in the last 
years~\cite{Boer:2010zf,Boer:2011kf,Pisano:2013cya}, mainly because it does
not require polarized targets and it can be generated through perturbative QCD
corrections~\cite{Catani:2010pd}, 
similarly to what happens in our QED calculation.
In Fig.~\ref{figh1perp} (right panel), we plot the combination 
$\big(f_1^{\gamma}+(k_y^2 -k_x^2)/(2m^2) h_1^{\perp \gamma}\big)/2$
corresponding to 
the density of photons with linear polarization along the $y$ direction in an
unpolarized dressed electron. 
The distribution has two peaks, shifted along the polarization direction 
by approximately 
$\pm 200$ keV at $x=0.5$.

\section{Conclusions}

In this article we analyzed the dressed electron at order $\alpha$ as a system
effectively composed of an electron and a photon. In analogy with QCD parton
distribution functions, we studied all twist-2 polarized
transverse-momentum-dependent
distribution functions (TMDs) 
of the electron and photon in the dressed electron,
which describe the three-dimensional structure of the dressed electron in
momentum space.  

We performed the calculation of electron TMDs in Feynman gauge and light-cone
gauge up to order $\alpha$ and for $x \neq 1$, $\pep{k} \neq \pep{0}$. We  
explicitly checked
the equivalence of the two gauges and the critical role played by Wilson
lines in the gauge-invariant definition of TMDs. This result was already
known for the QCD case, but the explicit QED calculation is particularly
useful for illustration purposes.
A nontrivial question concerns the contributions of transverse Wilson lines,
which in our QED calculation at order $\alpha$ appear only at the end point
$x=1$. In this case, we showed how the choice of different prescriptions for
the photon propagator is correctly compensated by the contribution of
transverse Wilson lines.

In light-cone gauge, we performed the calculation of TMDs in two different
approaches, namely using Feynman diagrams and using the overlap representation
in terms of light-front wave functions (LFWFs), confirming the
equivalence of the two approaches.

We calculated photon TMDs using their gauge-invariant definition which, in
the QED case, does not require the addition of Wilson lines. The results are
in fact the same for Feynman gauge and light-cone gauge.

The results of the electron TMDs are summarized in Eqs.~\eqref{Tmd1} to
\eqref{Tmd8}. T-odd TMDs ($h_{1}^{\perp e}$ and $f_{1T}^{\perp e}$)
are zero, a result that holds true also at order $\alpha^2$ due to the lack of
contributing diagrams.
Also the T-even TMD $h_{1T}^{\perp e}$ vanishes in QED at order
$\alpha$.  

The results of the photon TMDs are summarized in Eqs.~\eqref{Tmd1f} to
\eqref{Tmd8f}. All T-odd TMDs ($f_{1T}^{\perp\gamma}$,
$h_{1L}^{\perp\gamma}$, $h_{1T}^{\perp\gamma}$, $h_{1T}^{\gamma}$)
are zero. We checked that this is true also at order $\alpha^2$, and we argued
that it is probably true at any order $\alpha^n$.
      
In the last part of the article, we showed several plots that visualize TMDs in
various ways, including multidimensional density plots of the structure of
the dressed electron in momentum space. For instance, we showed that
the density of electrons (and photons, by symmetry) 
in a dressed electron in momentum space, 
averaged over all polarizations, at $x=0.5$, 
looks like a ring-shaped image, with a
radius of about 200 keV. 

We observed that the inclusion of polarization causes several interesting
effects. 
Different combinations of longitudinal polarizations
make it possible to isolate contributions with different projections of orbital
angular momentum. 
Some polarized distributions are not cylindrically symmetric. For instance, 
the density of linearly polarized photons inside the
dressed electron (related to the TMD $h_{1}^{\perp \gamma}$)
has a distribution with two peaks, shifted along the polarization direction 
by approximately 
$\pm 200$ keV at $x=0.5$. The shifts increase towards lower $x$. 
Transverse polarization of the dressed electron causes distortions in the
density distributions of longitudinally polarized electrons, with shifts
of the order of 200 keV at $x=0.5$, decreasing towards lower $x$.

Our analysis can be further extended by including the contributions from the
end points and the treatment of infrared and rapidity divergences. Another
possible direction of study would be to identify which observables are
sensitive to the TMDs of a dressed electron.

\bibliographystyle{myrevtex}
\bibliography{BiblioTMDs} 

\begin{thebibliography}{58}
\expandafter\ifx\csname natexlab\endcsname\relax\def\natexlab#1{#1}\fi
\expandafter\ifx\csname bibnamefont\endcsname\relax
  \def\bibnamefont#1{#1}\fi
\expandafter\ifx\csname bibfnamefont\endcsname\relax
  \def\bibfnamefont#1{#1}\fi
\expandafter\ifx\csname citenamefont\endcsname\relax
  \def\citenamefont#1{#1}\fi
\expandafter\ifx\csname url\endcsname\relax
  \def\url#1{\texttt{#1}}\fi
\expandafter\ifx\csname urlprefix\endcsname\relax\def\urlprefix{URL }\fi
\providecommand{\bibinfo}[2]{#2}
\providecommand{\eprint}[2][]{\url{#2}}

\bibitem[{\citenamefont{Altarelli and Parisi}(1977)}]{Altarelli:1977zs}
\bibinfo{author}{\bibfnamefont{G.}~\bibnamefont{Altarelli}} \bibnamefont{and}
  \bibinfo{author}{\bibfnamefont{G.}~\bibnamefont{Parisi}},
  \bibinfo{journal}{Nucl. Phys.} \textbf{\bibinfo{volume}{B126}},
  \bibinfo{pages}{298} (\bibinfo{year}{1977}).

\bibitem[{\citenamefont{Kuraev and Fadin}(1985)}]{Kuraev:1985hb}
\bibinfo{author}{\bibfnamefont{E.~A.} \bibnamefont{Kuraev}} \bibnamefont{and}
  \bibinfo{author}{\bibfnamefont{V.~S.} \bibnamefont{Fadin}},
  \bibinfo{journal}{Sov. J. Nucl. Phys.} \textbf{\bibinfo{volume}{41}},
  \bibinfo{pages}{466} (\bibinfo{year}{1985}), \bibinfo{note}{[Yad.
  Fiz.41,733(1985)]}.

\bibitem[{\citenamefont{Montagna et~al.}(1991)\citenamefont{Montagna,
  Nicrosini, and Trentadue}}]{Montagna:1991ku}
\bibinfo{author}{\bibfnamefont{G.}~\bibnamefont{Montagna}},
  \bibinfo{author}{\bibfnamefont{O.}~\bibnamefont{Nicrosini}},
  \bibnamefont{and}
  \bibinfo{author}{\bibfnamefont{L.}~\bibnamefont{Trentadue}},
  \bibinfo{journal}{Nucl. Phys.} \textbf{\bibinfo{volume}{B357}},
  \bibinfo{pages}{390} (\bibinfo{year}{1991}).

\bibitem[{\citenamefont{Meissner et~al.}(2009)\citenamefont{Meissner, Metz, and
  Schlegel}}]{Meissner:2009ww}
\bibinfo{author}{\bibfnamefont{S.}~\bibnamefont{Meissner}},
  \bibinfo{author}{\bibfnamefont{A.}~\bibnamefont{Metz}}, \bibnamefont{and}
  \bibinfo{author}{\bibfnamefont{M.}~\bibnamefont{Schlegel}},
  \bibinfo{journal}{JHEP} \textbf{\bibinfo{volume}{08}}, \bibinfo{pages}{056}
  (\bibinfo{year}{2009}).

\bibitem[{\citenamefont{Lorc\'e and Pasquini}(2013)}]{Lorce:2013pza}
\bibinfo{author}{\bibfnamefont{C.}~\bibnamefont{Lorc\'e}} \bibnamefont{and}
  \bibinfo{author}{\bibfnamefont{B.}~\bibnamefont{Pasquini}},
  \bibinfo{journal}{JHEP} \textbf{\bibinfo{volume}{1309}}, \bibinfo{pages}{138}
  (\bibinfo{year}{2013}).

\bibitem[{\citenamefont{Lorc\'e et~al.}(2011)\citenamefont{Lorc\'e, Pasquini,
  and Vanderhaeghen}}]{Lorce:2011dv}
\bibinfo{author}{\bibfnamefont{C.}~\bibnamefont{Lorc\'e}},
  \bibinfo{author}{\bibfnamefont{B.}~\bibnamefont{Pasquini}}, \bibnamefont{and}
  \bibinfo{author}{\bibfnamefont{M.}~\bibnamefont{Vanderhaeghen}},
  \bibinfo{journal}{JHEP} \textbf{\bibinfo{volume}{05}}, \bibinfo{pages}{041}
  (\bibinfo{year}{2011}).

\bibitem[{\citenamefont{Hoyer and Kurki}(2010)}]{Hoyer:2009sg}
\bibinfo{author}{\bibfnamefont{P.}~\bibnamefont{Hoyer}} \bibnamefont{and}
  \bibinfo{author}{\bibfnamefont{S.}~\bibnamefont{Kurki}},
  \bibinfo{journal}{Phys. Rev.} \textbf{\bibinfo{volume}{D81}},
  \bibinfo{pages}{013002} (\bibinfo{year}{2010}).

\bibitem[{\citenamefont{Miller}(2014)}]{Miller:2014vla}
\bibinfo{author}{\bibfnamefont{G.~A.} \bibnamefont{Miller}},
  \bibinfo{journal}{Phys. Rev.} \textbf{\bibinfo{volume}{D90}},
  \bibinfo{pages}{113001} (\bibinfo{year}{2014}).

\bibitem[{\citenamefont{Brodsky
  et~al.}(2001{\natexlab{a}})\citenamefont{Brodsky, Hwang, Ma, and
  Schmidt}}]{Brodsky:2000ii}
\bibinfo{author}{\bibfnamefont{S.~J.} \bibnamefont{Brodsky}},
  \bibinfo{author}{\bibfnamefont{D.~S.} \bibnamefont{Hwang}},
  \bibinfo{author}{\bibfnamefont{B.-Q.} \bibnamefont{Ma}}, \bibnamefont{and}
  \bibinfo{author}{\bibfnamefont{I.}~\bibnamefont{Schmidt}},
  \bibinfo{journal}{Nucl. Phys.} \textbf{\bibinfo{volume}{B593}},
  \bibinfo{pages}{311} (\bibinfo{year}{2001}{\natexlab{a}}).

\bibitem[{\citenamefont{Burkardt and BC}(2009)}]{Burkardt:2008ua}
\bibinfo{author}{\bibfnamefont{M.}~\bibnamefont{Burkardt}} \bibnamefont{and}
  \bibinfo{author}{\bibfnamefont{H.}~\bibnamefont{BC}}, \bibinfo{journal}{Phys.
  Rev.} \textbf{\bibinfo{volume}{D79}}, \bibinfo{pages}{071501}
  (\bibinfo{year}{2009}).

\bibitem[{\citenamefont{Liu and Ma}(2015)}]{Liu:2014fxa}
\bibinfo{author}{\bibfnamefont{T.}~\bibnamefont{Liu}} \bibnamefont{and}
  \bibinfo{author}{\bibfnamefont{B.-Q.} \bibnamefont{Ma}},
  \bibinfo{journal}{Phys. Rev.} \textbf{\bibinfo{volume}{D91}},
  \bibinfo{pages}{017501} (\bibinfo{year}{2015}).

\bibitem[{\citenamefont{Ji and Yuan}(2002)}]{Ji:2002aa}
\bibinfo{author}{\bibfnamefont{X.-d.} \bibnamefont{Ji}} \bibnamefont{and}
  \bibinfo{author}{\bibfnamefont{F.}~\bibnamefont{Yuan}},
  \bibinfo{journal}{Phys. Lett.} \textbf{\bibinfo{volume}{B543}},
  \bibinfo{pages}{66} (\bibinfo{year}{2002}).

\bibitem[{\citenamefont{Belitsky et~al.}(2003)\citenamefont{Belitsky, Ji, and
  Yuan}}]{Belitsky:2002sm}
\bibinfo{author}{\bibfnamefont{A.~V.} \bibnamefont{Belitsky}},
  \bibinfo{author}{\bibfnamefont{X.}~\bibnamefont{Ji}}, \bibnamefont{and}
  \bibinfo{author}{\bibfnamefont{F.}~\bibnamefont{Yuan}},
  \bibinfo{journal}{Nucl.Phys.} \textbf{\bibinfo{volume}{B656}},
  \bibinfo{pages}{165} (\bibinfo{year}{2003}).

\bibitem[{\citenamefont{Boer et~al.}(2003)\citenamefont{Boer, Mulders, and
  Pijlman}}]{Boer:2003cm}
\bibinfo{author}{\bibfnamefont{D.}~\bibnamefont{Boer}},
  \bibinfo{author}{\bibfnamefont{P.}~\bibnamefont{Mulders}}, \bibnamefont{and}
  \bibinfo{author}{\bibfnamefont{F.}~\bibnamefont{Pijlman}},
  \bibinfo{journal}{Nucl.Phys.} \textbf{\bibinfo{volume}{B667}},
  \bibinfo{pages}{201} (\bibinfo{year}{2003}).

\bibitem[{\citenamefont{Ji et~al.}(2005)\citenamefont{Ji, Ma, and
  Yuan}}]{Ji:2004wu}
\bibinfo{author}{\bibfnamefont{X.}~\bibnamefont{Ji}},
  \bibinfo{author}{\bibfnamefont{J.-P.} \bibnamefont{Ma}}, \bibnamefont{and}
  \bibinfo{author}{\bibfnamefont{F.}~\bibnamefont{Yuan}},
  \bibinfo{journal}{Phys. Rev.} \textbf{\bibinfo{volume}{D71}},
  \bibinfo{pages}{034005} (\bibinfo{year}{2005}).

\bibitem[{\citenamefont{Cherednikov et~al.}(2014)\citenamefont{Cherednikov,
  Mertens, and Van~der Veken}}]{Cherednikov:2014mua}
\bibinfo{author}{\bibfnamefont{I.~O.} \bibnamefont{Cherednikov}},
  \bibinfo{author}{\bibfnamefont{T.}~\bibnamefont{Mertens}}, \bibnamefont{and}
  \bibinfo{author}{\bibfnamefont{F.~F.} \bibnamefont{Van~der Veken}},
  \emph{\bibinfo{title}{{Wilson lines in quantum field theory}}}, vol.
  \bibinfo{volume}{24. 24} of \emph{\bibinfo{series}{De Gruyter Studies in
  Mathematical Physics}} (\bibinfo{publisher}{de Gruyter},
  \bibinfo{address}{Berlin}, \bibinfo{year}{2014}).

\bibitem[{\citenamefont{Brodsky et~al.}(2002)\citenamefont{Brodsky, Hwang, and
  Schmidt}}]{Brodsky:2002cx}
\bibinfo{author}{\bibfnamefont{S.~J.} \bibnamefont{Brodsky}},
  \bibinfo{author}{\bibfnamefont{D.~S.} \bibnamefont{Hwang}}, \bibnamefont{and}
  \bibinfo{author}{\bibfnamefont{I.}~\bibnamefont{Schmidt}},
  \bibinfo{journal}{Phys. Lett.} \textbf{\bibinfo{volume}{B530}},
  \bibinfo{pages}{99} (\bibinfo{year}{2002}).

\bibitem[{\citenamefont{Collins}(2002)}]{Collins:2002kn}
\bibinfo{author}{\bibfnamefont{J.~C.} \bibnamefont{Collins}},
  \bibinfo{journal}{Phys. Lett.} \textbf{\bibinfo{volume}{B536}},
  \bibinfo{pages}{43} (\bibinfo{year}{2002}).

\bibitem[{\citenamefont{Pasquini et~al.}(2008)\citenamefont{Pasquini,
  Cazzaniga, and Boffi}}]{Pasquini:2008ax}
\bibinfo{author}{\bibfnamefont{B.}~\bibnamefont{Pasquini}},
  \bibinfo{author}{\bibfnamefont{S.}~\bibnamefont{Cazzaniga}},
  \bibnamefont{and} \bibinfo{author}{\bibfnamefont{S.}~\bibnamefont{Boffi}},
  \bibinfo{journal}{Phys. Rev.} \textbf{\bibinfo{volume}{D78}},
  \bibinfo{pages}{034025} (\bibinfo{year}{2008}).

\bibitem[{\citenamefont{Brodsky
  et~al.}(2001{\natexlab{b}})\citenamefont{Brodsky, Diehl, and
  Hwang}}]{Brodsky:2000xy}
\bibinfo{author}{\bibfnamefont{S.~J.} \bibnamefont{Brodsky}},
  \bibinfo{author}{\bibfnamefont{M.}~\bibnamefont{Diehl}}, \bibnamefont{and}
  \bibinfo{author}{\bibfnamefont{D.~S.} \bibnamefont{Hwang}},
  \bibinfo{journal}{Nucl. Phys.} \textbf{\bibinfo{volume}{B596}},
  \bibinfo{pages}{99} (\bibinfo{year}{2001}{\natexlab{b}}).

\bibitem[{\citenamefont{Chakrabarti and
  Mukherjee}(2005{\natexlab{a}})}]{Chakrabarti:2004ci}
\bibinfo{author}{\bibfnamefont{D.}~\bibnamefont{Chakrabarti}} \bibnamefont{and}
  \bibinfo{author}{\bibfnamefont{A.}~\bibnamefont{Mukherjee}},
  \bibinfo{journal}{Phys. Rev.} \textbf{\bibinfo{volume}{D71}},
  \bibinfo{pages}{014038} (\bibinfo{year}{2005}{\natexlab{a}}).

\bibitem[{\citenamefont{Chakrabarti and
  Mukherjee}(2005{\natexlab{b}})}]{Chakrabarti:2005zm}
\bibinfo{author}{\bibfnamefont{D.}~\bibnamefont{Chakrabarti}} \bibnamefont{and}
  \bibinfo{author}{\bibfnamefont{A.}~\bibnamefont{Mukherjee}},
  \bibinfo{journal}{Phys. Rev.} \textbf{\bibinfo{volume}{D72}},
  \bibinfo{pages}{034013} (\bibinfo{year}{2005}{\natexlab{b}}).

\bibitem[{\citenamefont{Brodsky et~al.}(2007)\citenamefont{Brodsky,
  Chakrabarti, Harindranath, Mukherjee, and Vary}}]{Brodsky:2006ku}
\bibinfo{author}{\bibfnamefont{S.~J.} \bibnamefont{Brodsky}},
  \bibinfo{author}{\bibfnamefont{D.}~\bibnamefont{Chakrabarti}},
  \bibinfo{author}{\bibfnamefont{A.}~\bibnamefont{Harindranath}},
  \bibinfo{author}{\bibfnamefont{A.}~\bibnamefont{Mukherjee}},
  \bibnamefont{and} \bibinfo{author}{\bibfnamefont{J.~P.} \bibnamefont{Vary}},
  \bibinfo{journal}{Phys. Rev.} \textbf{\bibinfo{volume}{D75}},
  \bibinfo{pages}{014003} (\bibinfo{year}{2007}).

\bibitem[{\citenamefont{Chakrabarti et~al.}(2009)\citenamefont{Chakrabarti,
  Manohar, and Mukherjee}}]{Chakrabarti:2008mw}
\bibinfo{author}{\bibfnamefont{D.}~\bibnamefont{Chakrabarti}},
  \bibinfo{author}{\bibfnamefont{R.}~\bibnamefont{Manohar}}, \bibnamefont{and}
  \bibinfo{author}{\bibfnamefont{A.}~\bibnamefont{Mukherjee}},
  \bibinfo{journal}{Phys. Rev.} \textbf{\bibinfo{volume}{D79}},
  \bibinfo{pages}{034006} (\bibinfo{year}{2009}).

\bibitem[{\citenamefont{Kumar and Dahiya}(2015)}]{Kumar:2015tpa}
\bibinfo{author}{\bibfnamefont{N.}~\bibnamefont{Kumar}} \bibnamefont{and}
  \bibinfo{author}{\bibfnamefont{H.}~\bibnamefont{Dahiya}},
  \bibinfo{journal}{Eur. Phys. J.} \textbf{\bibinfo{volume}{A51}},
  \bibinfo{pages}{19} (\bibinfo{year}{2015}).

\bibitem[{\citenamefont{Soper}(1977)}]{Soper:1977jc}
\bibinfo{author}{\bibfnamefont{D.~E.} \bibnamefont{Soper}},
  \bibinfo{journal}{Phys. Rev.} \textbf{\bibinfo{volume}{D15}},
  \bibinfo{pages}{1141} (\bibinfo{year}{1977}).

\bibitem[{\citenamefont{Collins and
  Soper}(1982{\natexlab{a}})}]{Collins:1982uw}
\bibinfo{author}{\bibfnamefont{J.~C.} \bibnamefont{Collins}} \bibnamefont{and}
  \bibinfo{author}{\bibfnamefont{D.~E.} \bibnamefont{Soper}},
  \bibinfo{journal}{Nucl. Phys.} \textbf{\bibinfo{volume}{B194}},
  \bibinfo{pages}{445} (\bibinfo{year}{1982}{\natexlab{a}}).

\bibitem[{\citenamefont{Mulders and Tangerman}(1996)}]{Mulders:1995dh}
\bibinfo{author}{\bibfnamefont{P.}~\bibnamefont{Mulders}} \bibnamefont{and}
  \bibinfo{author}{\bibfnamefont{R.}~\bibnamefont{Tangerman}},
  \bibinfo{journal}{Nucl.Phys.} \textbf{\bibinfo{volume}{B461}},
  \bibinfo{pages}{197} (\bibinfo{year}{1996}).

\bibitem[{\citenamefont{Bomhof et~al.}(2004)\citenamefont{Bomhof, Mulders, and
  Pijlman}}]{Bomhof:2004aw}
\bibinfo{author}{\bibfnamefont{C.~J.} \bibnamefont{Bomhof}},
  \bibinfo{author}{\bibfnamefont{P.~J.} \bibnamefont{Mulders}},
  \bibnamefont{and} \bibinfo{author}{\bibfnamefont{F.}~\bibnamefont{Pijlman}},
  \bibinfo{journal}{Phys. Lett.} \textbf{\bibinfo{volume}{B596}},
  \bibinfo{pages}{277} (\bibinfo{year}{2004}).

\bibitem[{\citenamefont{Bomhof et~al.}(2006)\citenamefont{Bomhof, Mulders, and
  Pijlman}}]{Bomhof:2006dp}
\bibinfo{author}{\bibfnamefont{C.~J.} \bibnamefont{Bomhof}},
  \bibinfo{author}{\bibfnamefont{P.~J.} \bibnamefont{Mulders}},
  \bibnamefont{and} \bibinfo{author}{\bibfnamefont{F.}~\bibnamefont{Pijlman}},
  \bibinfo{journal}{Eur. Phys. J.} \textbf{\bibinfo{volume}{C47}},
  \bibinfo{pages}{147} (\bibinfo{year}{2006}).

\bibitem[{\citenamefont{Collins and Soper}(1981)}]{Collins:1981uk}
\bibinfo{author}{\bibfnamefont{J.~C.} \bibnamefont{Collins}} \bibnamefont{and}
  \bibinfo{author}{\bibfnamefont{D.~E.} \bibnamefont{Soper}},
  \bibinfo{journal}{Nucl.Phys.} \textbf{\bibinfo{volume}{B193}},
  \bibinfo{pages}{381} (\bibinfo{year}{1981}).

\bibitem[{\citenamefont{Aybat and Rogers}(2011)}]{Aybat:2011zv}
\bibinfo{author}{\bibfnamefont{S.}~\bibnamefont{Aybat}} \bibnamefont{and}
  \bibinfo{author}{\bibfnamefont{T.~C.} \bibnamefont{Rogers}},
  \bibinfo{journal}{Phys. Rev.} \textbf{\bibinfo{volume}{D83}},
  \bibinfo{pages}{114042} (\bibinfo{year}{2011}).

\bibitem[{\citenamefont{Echevarria
  et~al.}(2012{\natexlab{a}})\citenamefont{Echevarria, Idilbi, and
  Scimemi}}]{GarciaEchevarria:2011rb}
\bibinfo{author}{\bibfnamefont{M.~G.} \bibnamefont{Echevarria}},
  \bibinfo{author}{\bibfnamefont{A.}~\bibnamefont{Idilbi}}, \bibnamefont{and}
  \bibinfo{author}{\bibfnamefont{I.}~\bibnamefont{Scimemi}},
  \bibinfo{journal}{JHEP} \textbf{\bibinfo{volume}{1207}}, \bibinfo{pages}{002}
  (\bibinfo{year}{2012}{\natexlab{a}}).

\bibitem[{\citenamefont{Collins}(2011)}]{Collins:2011zzd}
\bibinfo{author}{\bibfnamefont{J.}~\bibnamefont{Collins}},
  \emph{\bibinfo{title}{Foundations of Perturbative {QCD}}}, Cambridge
  Monographs on Particle Physics, Nuclear Physics and Cosmology
  (\bibinfo{publisher}{Cambridge University Press}, \bibinfo{year}{2011}), ISBN
  \bibinfo{isbn}{9780521855334},
  \urlprefix\url{http://books.google.it/books?id=0xGi1KW9vykC}.

\bibitem[{\citenamefont{Echevarria
  et~al.}(2012{\natexlab{b}})\citenamefont{Echevarria, Idilbi, Schafer, and
  Scimemi}}]{Echevarria:2012pw}
\bibinfo{author}{\bibfnamefont{M.~G.} \bibnamefont{Echevarria}},
  \bibinfo{author}{\bibfnamefont{A.}~\bibnamefont{Idilbi}},
  \bibinfo{author}{\bibfnamefont{A.}~\bibnamefont{Schafer}}, \bibnamefont{and}
  \bibinfo{author}{\bibfnamefont{I.}~\bibnamefont{Scimemi}}
  (\bibinfo{year}{2012}{\natexlab{b}}), {arXiv:1208.1281 [hep-ph]}.

\bibitem[{\citenamefont{Echevarria et~al.}(2013)\citenamefont{Echevarria,
  Idilbi, and Scimemi}}]{Echevarria:2012js}
\bibinfo{author}{\bibfnamefont{M.~G.} \bibnamefont{Echevarria}},
  \bibinfo{author}{\bibfnamefont{A.}~\bibnamefont{Idilbi}}, \bibnamefont{and}
  \bibinfo{author}{\bibfnamefont{I.}~\bibnamefont{Scimemi}},
  \bibinfo{journal}{Phys. Lett.} \textbf{\bibinfo{volume}{B726}},
  \bibinfo{pages}{795} (\bibinfo{year}{2013}).

\bibitem[{\citenamefont{Collins and Rogers}(2013)}]{Collins:2012uy}
\bibinfo{author}{\bibfnamefont{J.~C.} \bibnamefont{Collins}} \bibnamefont{and}
  \bibinfo{author}{\bibfnamefont{T.~C.} \bibnamefont{Rogers}},
  \bibinfo{journal}{Phys. Rev.} \textbf{\bibinfo{volume}{D87}},
  \bibinfo{pages}{034018} (\bibinfo{year}{2013}).

\bibitem[{\citenamefont{Goeke et~al.}(2005)\citenamefont{Goeke, Metz, and
  Schlegel}}]{Goeke:2005hb}
\bibinfo{author}{\bibfnamefont{K.}~\bibnamefont{Goeke}},
  \bibinfo{author}{\bibfnamefont{A.}~\bibnamefont{Metz}}, \bibnamefont{and}
  \bibinfo{author}{\bibfnamefont{M.}~\bibnamefont{Schlegel}},
  \bibinfo{journal}{Phys. Lett.} \textbf{\bibinfo{volume}{B618}},
  \bibinfo{pages}{90} (\bibinfo{year}{2005}).

\bibitem[{\citenamefont{Bacchetta et~al.}(2007)\citenamefont{Bacchetta, Diehl,
  Goeke, Metz, Mulders et~al.}}]{Bacchetta:2006tn}
\bibinfo{author}{\bibfnamefont{A.}~\bibnamefont{Bacchetta}},
  \bibinfo{author}{\bibfnamefont{M.}~\bibnamefont{Diehl}},
  \bibinfo{author}{\bibfnamefont{K.}~\bibnamefont{Goeke}},
  \bibinfo{author}{\bibfnamefont{A.}~\bibnamefont{Metz}},
  \bibinfo{author}{\bibfnamefont{P.~J.} \bibnamefont{Mulders}},
  \bibnamefont{et~al.}, \bibinfo{journal}{JHEP}
  \textbf{\bibinfo{volume}{0702}}, \bibinfo{pages}{093} (\bibinfo{year}{2007}).

\bibitem[{\citenamefont{Boer and
  Mulders}(1998{\natexlab{a}})}]{PhysRevD.57.5780}
\bibinfo{author}{\bibfnamefont{D.}~\bibnamefont{Boer}} \bibnamefont{and}
  \bibinfo{author}{\bibfnamefont{P.~J.} \bibnamefont{Mulders}},
  \bibinfo{journal}{Phys. Rev. D} \textbf{\bibinfo{volume}{57}},
  \bibinfo{pages}{5780} (\bibinfo{year}{1998}{\natexlab{a}}).

\bibitem[{\citenamefont{Sivers}(1990)}]{PhysRevD.41.83}
\bibinfo{author}{\bibfnamefont{D.}~\bibnamefont{Sivers}},
  \bibinfo{journal}{Phys. Rev. D} \textbf{\bibinfo{volume}{41}},
  \bibinfo{pages}{83} (\bibinfo{year}{1990}).

\bibitem[{\citenamefont{Collins and
  Soper}(1982{\natexlab{b}})}]{Collins:1981uw}
\bibinfo{author}{\bibfnamefont{J.~C.} \bibnamefont{Collins}} \bibnamefont{and}
  \bibinfo{author}{\bibfnamefont{D.~E.} \bibnamefont{Soper}},
  \bibinfo{journal}{Nucl.Phys.} \textbf{\bibinfo{volume}{B194}},
  \bibinfo{pages}{445} (\bibinfo{year}{1982}{\natexlab{b}}).

\bibitem[{\citenamefont{Brodsky et~al.}(1998)\citenamefont{Brodsky, Pauli, and
  Pinsky}}]{Brodsky:1997de}
\bibinfo{author}{\bibfnamefont{S.~J.} \bibnamefont{Brodsky}},
  \bibinfo{author}{\bibfnamefont{H.-C.} \bibnamefont{Pauli}}, \bibnamefont{and}
  \bibinfo{author}{\bibfnamefont{S.~S.} \bibnamefont{Pinsky}},
  \bibinfo{journal}{Phys. Rept.} \textbf{\bibinfo{volume}{301}},
  \bibinfo{pages}{299} (\bibinfo{year}{1998}).

\bibitem[{\citenamefont{Brodsky and Lepage}(1989)}]{brodskynot}
\bibinfo{author}{\bibfnamefont{S.~J.} \bibnamefont{Brodsky}} \bibnamefont{and}
  \bibinfo{author}{\bibfnamefont{G.}~\bibnamefont{Lepage}}, in
  \emph{\bibinfo{booktitle}{Perturbative Quantum Chromodynamics}}, edited by
  \bibinfo{editor}{\bibfnamefont{A.}~\bibnamefont{Mueller}}
  (\bibinfo{publisher}{World Scientific}, \bibinfo{address}{Singapore},
  \bibinfo{year}{1989}).

\bibitem[{\citenamefont{Lorc\'e and Pasquini}(2011)}]{Lorce:2011zta}
\bibinfo{author}{\bibfnamefont{C.}~\bibnamefont{Lorc\'e}} \bibnamefont{and}
  \bibinfo{author}{\bibfnamefont{B.}~\bibnamefont{Pasquini}},
  \bibinfo{journal}{Phys.Rev.} \textbf{\bibinfo{volume}{D84}},
  \bibinfo{pages}{034039} (\bibinfo{year}{2011}).

\bibitem[{\citenamefont{Bacchetta et~al.}(2000)\citenamefont{Bacchetta,
  Boglione, Henneman, and Mulders}}]{Bacchetta:1999kz}
\bibinfo{author}{\bibfnamefont{A.}~\bibnamefont{Bacchetta}},
  \bibinfo{author}{\bibfnamefont{M.}~\bibnamefont{Boglione}},
  \bibinfo{author}{\bibfnamefont{A.}~\bibnamefont{Henneman}}, \bibnamefont{and}
  \bibinfo{author}{\bibfnamefont{P.~J.} \bibnamefont{Mulders}},
  \bibinfo{journal}{Phys. Rev. Lett.} \textbf{\bibinfo{volume}{85}},
  \bibinfo{pages}{712} (\bibinfo{year}{2000}).

\bibitem[{\citenamefont{Mulders and Rodrigues}(2001)}]{Mulders:2000sh}
\bibinfo{author}{\bibfnamefont{P.~J.} \bibnamefont{Mulders}} \bibnamefont{and}
  \bibinfo{author}{\bibfnamefont{J.}~\bibnamefont{Rodrigues}},
  \bibinfo{journal}{Phys. Rev.} \textbf{\bibinfo{volume}{D63}},
  \bibinfo{pages}{094021} (\bibinfo{year}{2001}).

\bibitem[{\citenamefont{Goeke et~al.}(2006)\citenamefont{Goeke, Meissner, Metz,
  and Schlegel}}]{Goeke:2006ef}
\bibinfo{author}{\bibfnamefont{K.}~\bibnamefont{Goeke}},
  \bibinfo{author}{\bibfnamefont{S.}~\bibnamefont{Meissner}},
  \bibinfo{author}{\bibfnamefont{A.}~\bibnamefont{Metz}}, \bibnamefont{and}
  \bibinfo{author}{\bibfnamefont{M.}~\bibnamefont{Schlegel}},
  \bibinfo{journal}{Phys. Lett.} \textbf{\bibinfo{volume}{B637}},
  \bibinfo{pages}{241} (\bibinfo{year}{2006}).

\bibitem[{\citenamefont{Meissner et~al.}(2007)\citenamefont{Meissner, Metz, and
  Goeke}}]{Meissner:2007rx}
\bibinfo{author}{\bibfnamefont{S.}~\bibnamefont{Meissner}},
  \bibinfo{author}{\bibfnamefont{A.}~\bibnamefont{Metz}}, \bibnamefont{and}
  \bibinfo{author}{\bibfnamefont{K.}~\bibnamefont{Goeke}},
  \bibinfo{journal}{Phys. Rev.} \textbf{\bibinfo{volume}{D76}},
  \bibinfo{pages}{034002} (\bibinfo{year}{2007}).

\bibitem[{\citenamefont{Collins and Rogers}(2008)}]{Collins:2008sg}
\bibinfo{author}{\bibfnamefont{J.~C.} \bibnamefont{Collins}} \bibnamefont{and}
  \bibinfo{author}{\bibfnamefont{T.~C.} \bibnamefont{Rogers}},
  \bibinfo{journal}{Phys. Rev.} \textbf{\bibinfo{volume}{D78}},
  \bibinfo{pages}{054012} (\bibinfo{year}{2008}).

\bibitem[{\citenamefont{Burkardt}(2004{\natexlab{a}})}]{Burkardt:2004ur}
\bibinfo{author}{\bibfnamefont{M.}~\bibnamefont{Burkardt}},
  \bibinfo{journal}{Phys. Rev.} \textbf{\bibinfo{volume}{D69}},
  \bibinfo{pages}{091501} (\bibinfo{year}{2004}{\natexlab{a}}).

\bibitem[{\citenamefont{Burkardt}(2004{\natexlab{b}})}]{Burkardt:2003yg}
\bibinfo{author}{\bibfnamefont{M.}~\bibnamefont{Burkardt}},
  \bibinfo{journal}{Phys. Rev.} \textbf{\bibinfo{volume}{D69}},
  \bibinfo{pages}{057501} (\bibinfo{year}{2004}{\natexlab{b}}).

\bibitem[{\citenamefont{Diehl and Hagler}(2005)}]{Diehl:2005jf}
\bibinfo{author}{\bibfnamefont{M.}~\bibnamefont{Diehl}} \bibnamefont{and}
  \bibinfo{author}{\bibfnamefont{P.}~\bibnamefont{Hagler}},
  \bibinfo{journal}{Eur. Phys. J.} \textbf{\bibinfo{volume}{C44}},
  \bibinfo{pages}{87} (\bibinfo{year}{2005}).

\bibitem[{\citenamefont{Boer and Mulders}(1998{\natexlab{b}})}]{Boer:1997nt}
\bibinfo{author}{\bibfnamefont{D.}~\bibnamefont{Boer}} \bibnamefont{and}
  \bibinfo{author}{\bibfnamefont{P.}~\bibnamefont{Mulders}},
  \bibinfo{journal}{Phys.Rev.} \textbf{\bibinfo{volume}{D57}},
  \bibinfo{pages}{5780} (\bibinfo{year}{1998}{\natexlab{b}}).

\bibitem[{\citenamefont{Boer et~al.}(2011)\citenamefont{Boer, Brodsky, Mulders,
  and Pisano}}]{Boer:2010zf}
\bibinfo{author}{\bibfnamefont{D.}~\bibnamefont{Boer}},
  \bibinfo{author}{\bibfnamefont{S.~J.} \bibnamefont{Brodsky}},
  \bibinfo{author}{\bibfnamefont{P.~J.} \bibnamefont{Mulders}},
  \bibnamefont{and} \bibinfo{author}{\bibfnamefont{C.}~\bibnamefont{Pisano}},
  \bibinfo{journal}{Phys. Rev. Lett.} \textbf{\bibinfo{volume}{106}},
  \bibinfo{pages}{132001} (\bibinfo{year}{2011}).

\bibitem[{\citenamefont{Boer et~al.}(2012)\citenamefont{Boer, den Dunnen,
  Pisano, Schlegel, and Vogelsang}}]{Boer:2011kf}
\bibinfo{author}{\bibfnamefont{D.}~\bibnamefont{Boer}},
  \bibinfo{author}{\bibfnamefont{W.~J.} \bibnamefont{den Dunnen}},
  \bibinfo{author}{\bibfnamefont{C.}~\bibnamefont{Pisano}},
  \bibinfo{author}{\bibfnamefont{M.}~\bibnamefont{Schlegel}}, \bibnamefont{and}
  \bibinfo{author}{\bibfnamefont{W.}~\bibnamefont{Vogelsang}},
  \bibinfo{journal}{Phys. Rev. Lett.} \textbf{\bibinfo{volume}{108}},
  \bibinfo{pages}{032002} (\bibinfo{year}{2012}).

\bibitem[{\citenamefont{Pisano et~al.}(2013)\citenamefont{Pisano, Boer,
  Brodsky, Buffing, and Mulders}}]{Pisano:2013cya}
\bibinfo{author}{\bibfnamefont{C.}~\bibnamefont{Pisano}},
  \bibinfo{author}{\bibfnamefont{D.}~\bibnamefont{Boer}},
  \bibinfo{author}{\bibfnamefont{S.~J.} \bibnamefont{Brodsky}},
  \bibinfo{author}{\bibfnamefont{M.~G.} \bibnamefont{Buffing}},
  \bibnamefont{and} \bibinfo{author}{\bibfnamefont{P.~J.}
  \bibnamefont{Mulders}}, \bibinfo{journal}{JHEP}
  \textbf{\bibinfo{volume}{1310}}, \bibinfo{pages}{024} (\bibinfo{year}{2013}).

\bibitem[{\citenamefont{Catani and Grazzini}(2011)}]{Catani:2010pd}
\bibinfo{author}{\bibfnamefont{S.}~\bibnamefont{Catani}} \bibnamefont{and}
  \bibinfo{author}{\bibfnamefont{M.}~\bibnamefont{Grazzini}},
  \bibinfo{journal}{Nucl. Phys.} \textbf{\bibinfo{volume}{B845}},
  \bibinfo{pages}{297} (\bibinfo{year}{2011}).

\end{thebibliography}
\end{document}